%% file: main.tex
\documentclass[12pt]{article}
\usepackage{amsmath}
\usepackage{amssymb,bbm}
\usepackage{amsfonts}
\usepackage{latexsym}

\usepackage{amsthm}
\usepackage{geometry} 
\usepackage{graphicx}
\usepackage{verbatim}
\usepackage[colorlinks=true]{hyperref} 
\usepackage{cite}
\usepackage{bm} 
\usepackage[export]{adjustbox}

\usepackage{subfigure}
\usepackage{graphicx}

\usepackage{xcolor}
\usepackage{color}
\usepackage{comment}
\usepackage{gensymb}
\usepackage{bigints}
\usepackage[lined,boxed,commentsnumbered, ruled]{algorithm2e}
\usepackage{pifont}

\usepackage{tikz}
\usetikzlibrary{arrows}
\usetikzlibrary{fit}
\usetikzlibrary{shapes.geometric}
\usetikzlibrary{patterns}

\usepackage{caption}
\usepackage{pgfplots}
\pgfplotsset{compat=1.8}
\usepgfplotslibrary{statistics}

\usepgfplotslibrary{colorbrewer}
\pgfplotsset{compat = 1.15, cycle list/Set1-8} 
\usetikzlibrary{pgfplots.statistics, pgfplots.colorbrewer} 
\usepackage{pgfplotstable}
\usepackage{filecontents}
\usepgfplotslibrary{statistics}
\pgfplotsset{compat=1.8}
\pgfplotstableset{
  every head row/.style={before row=\toprule,after row=\midrule},
  every last row/.style={after row=\bottomrule},
  fixed,precision=5,
}

\pgfmathdeclarefunction{fpumod}{2}{%
    \pgfmathfloatdivide{#1}{#2}%
    \pgfmathfloatint{\pgfmathresult}%
    \pgfmathfloatmultiply{\pgfmathresult}{#2}%
    \pgfmathfloatsubtract{#1}{\pgfmathresult}%
    \pgfmathfloatifapproxequalrel{\pgfmathresult}{#2}{\def\pgfmathresult{3}}{}%
}
\usepackage{booktabs}
\usepackage{array}
\usepackage{colortbl}

\def\lddots{\mathinner{\mkern1mu\raise1pt\hbox{.}\mkern2mu  
\raise4pt\hbox{.}\mkern2mu\raise7pt\vbox{\kern7pt\hbox{.}}\mkern1mu}}
\makeatletter
\def\numberbysection{\@addtoreset{equation}{section}
 \def\theequation{\thesection.\arabic{equation}}}
\makeatother  
  
\numberbysection

\newcommand{\be}{\begin{eqnarray}}  
\newcommand{\ee}{\end{eqnarray}}

 \title{\bf Statistical modelling and Bayesian inversion for a Compton imaging system: application to radioactive source localisation.}
\author{\textsf{C\'ecilia Tarpau$^{1,2,*}$,}
\textsf{Ming Fang$^3$,}
\textsf{Konstantinos C. Zygalakis$^{2,4}$,}\\
\textsf{Marcelo Pereyra$^{1,2}$,}
\textsf{Angela Di Fulvio$^3$,}
\textsf{Yoann Altmann$^5$}
\\
\textit{\scriptsize $^1$ School of Mathematical and Computer Sciences, Heriot-Watt University, Edinburgh EH14 4AS, UK} 
\\
\textit{\scriptsize $^2$ Maxwell Institute for Mathematical Sciences, Bayes Centre, University of Edinburgh, Edinburgh EH8 9BT, UK} 
\\
\textit{\scriptsize $^3$ Department of Nuclear, Plasma, and Radiological Engineering, University of Illinois Urbana-Champaign, Urbana IL 61801, USA}
\\ 
\textit{\scriptsize $^4$ School of Mathematics, University of Edinburgh, Edinburgh EH9 3FD, UK}
\\ 
\textit{\scriptsize $^5$ School of Engineering and Physical Sciences, Heriot-Watt University, Edinburgh EH14 4AS, UK}
\\
    {\footnotesize $^{*}$ Corresponding author} \\}
\date{}

\footnotetext[1]{Email: {\tt c.tarpau@hw.ac.uk}}

\begin{document}

\definecolor{rouge}{HTML}{e6194B}
\definecolor{orange}{HTML}{f58231}
\definecolor{grey}{HTML}{a9a9a9}
\definecolor{magenta}{HTML}{f032e6}
\definecolor{purple}{HTML}{911eb4}
\definecolor{blue}{HTML}{4363d8}
\definecolor{cyan}{HTML}{42d4f4}
\definecolor{green}{HTML}{3cb44b}

\maketitle
\begin{abstract}
    This paper presents a statistical forward model for a Compton imaging system, called Compton imager. This system, under development at the University of Illinois Urbana Champaign, is a variant of Compton cameras with a single type of sensors which can simultaneously act as scatterers and absorbers. This imager is convenient for imaging situations requiring a wide field of view. The proposed statistical forward model is then used to solve the inverse problem of estimating the location and energy of point-like sources from observed data.  This inverse problem is formulated and solved in a Bayesian framework by using a Metropolis within Gibbs algorithm for the estimation of the location, and an expectation-maximization algorithm for the estimation of the energy. This approach leads to more accurate estimation when compared with the deterministic  standard back-projection approach, with the additional benefit of uncertainty quantification in the low photon imaging setting.  

\end{abstract}


\section{Introduction}

\begin{figure}[!ht]
    \centering
    \includegraphics[width = 0.5\textwidth]{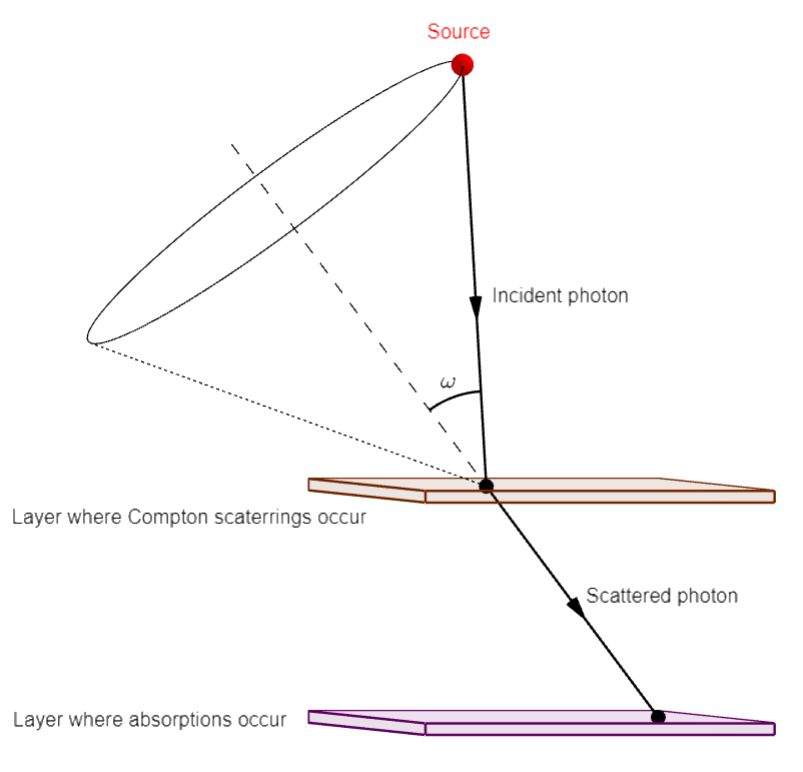}
    \caption{ Working principle of a Compton camera. The locations and the amount of energy lost by the incoming photon are recorded by each layer. From this information, the position of the source lies on a conical surface of semi-aperture angle $\omega$ \eqref{eq:Compton_formula}.}
    \label{fig:WP_Compton_camera}
\end{figure}

Compton imaging plays a central role in radiation detection and analysis \cite{parajuli2022development, terzioglu2018compton, frandes2016image}, and has important applications in modern astrophysics, cosmology \cite{schonfelder1973telescope, schonfelder1993instrument, todd1974proposed, takahashi2004hard, takahashi2014astro}, nuclear safety \cite{sweeney2014compton, steinberger2020imaging, mukhopadhyay2020modern, vetter2018gamma, al2019passive}, environmental radiation monitoring \cite{sato2018remote, tomono2017first} and medical imaging \cite{stichelbaut2003verification, min2006prompt}. Traditionally, Compton cameras are constructed by using two layers of sensors. The sensors in the first layer interact with incoming photons via Compton scattering \cite{ChoppinGregory2013RaNC}; i.e., a photon interacts with a charged particle within the sensor in a manner that results in a change of direction and a decrease of energy. The scattered photon is then absorbed by one of the sensors in the second layer. Both sensors record the location of the interaction and the amount of energy that the photon has lost as a result of the interaction. From this information and the physics of Compton scattering, it is possible to partially determine the location of the source that generated the income photon, up to a conical surface. This is illustrated in Figure \ref{fig:WP_Compton_camera}, where a photon emitted by the source highlighted in colour red interacts with the two layers of the Compton camera. From the location and energy loss related to these interactions, it is possible to determine that the source is located somewhere on the depicted conical surface, whose apex coincides with the location of the first interaction. The \emph{direction-of-arrival} of the source can be accurately estimated by detecting additional photons and analysing the intersection of the resulting conical surfaces, with the accuracy of the estimates depending strongly on the number of photons detected and the level of measurement noise.

Several approaches for inversion in Compton Cameras have been proposed in the literature, from simple back-projection techniques to iterative reconstruction algorithms implementing maximum likelihood as well as Bayesian inference strategies. Back-projection techniques \cite{rohe_backprojection, wilderman1998fast} are highly computationally efficient but they can deliver solutions that suffer from blur distortions. Such distortions can be mitigated through filtered back-projection schemes \cite{mundy2011accelerated, lee2016adaptation, basko1997analytical, basko1997fully, parra2000reconstruction, tomitani2002image, hirasawa2003analytical, xu2006filtered, shy2020filtered}, whose objective is to implement suitable filters to reduce blurring on reconstructions. Moreover, many reconstruction methods rely on maximum likelihood estimation computed by using an expectation-maximisation (EM) algorithm. These operate predominantly through a list mode in which photon detection events are considered sequentially \cite{wilderman2001improved, YuemengFengthesis, maxim2015probabilistic, yabu2021tomographic, tornga2009three}, or alternatively through bin mode that relies on a quantization of the space \cite{ikeda2014bin}. In particular, list-mode EM (LM-EM) has become the most widely used reconstruction technique in the context of Compton cameras, with several improvements available to accelerate its computational efficiency and reconstruction quality. For example, Ordered-Subset EM (OS-EM) \cite{sakai2020improved, hudson1994accelerated, kim2010fully} implements LM-EM with data batching in order to reduce computing times, and the reconstruction quality of LM-EM techniques can be improved by leveraging prior information \cite{green1990use}. Furthermore, regarding the use of Monte Carlo methods for Compton Camera inversion, we note the Stochastic Origin Ensemble methods (SOE) \cite{andreyev2009stochastic, andreyev2011fast, andreyev2016resolution} which rely on Markov chain Monte Carlo (MCMC) sampling. The comparisons reported in \cite{andreyev2009stochastic, andreyev2011fast, andreyev2016resolution} suggest that SOE can deliver solutions of comparable accuracy to LM-EM at a reduced computational cost.

\begin{figure}[!ht]
    \centering
    \subfigure[]{\includegraphics[width = 0.45\linewidth]{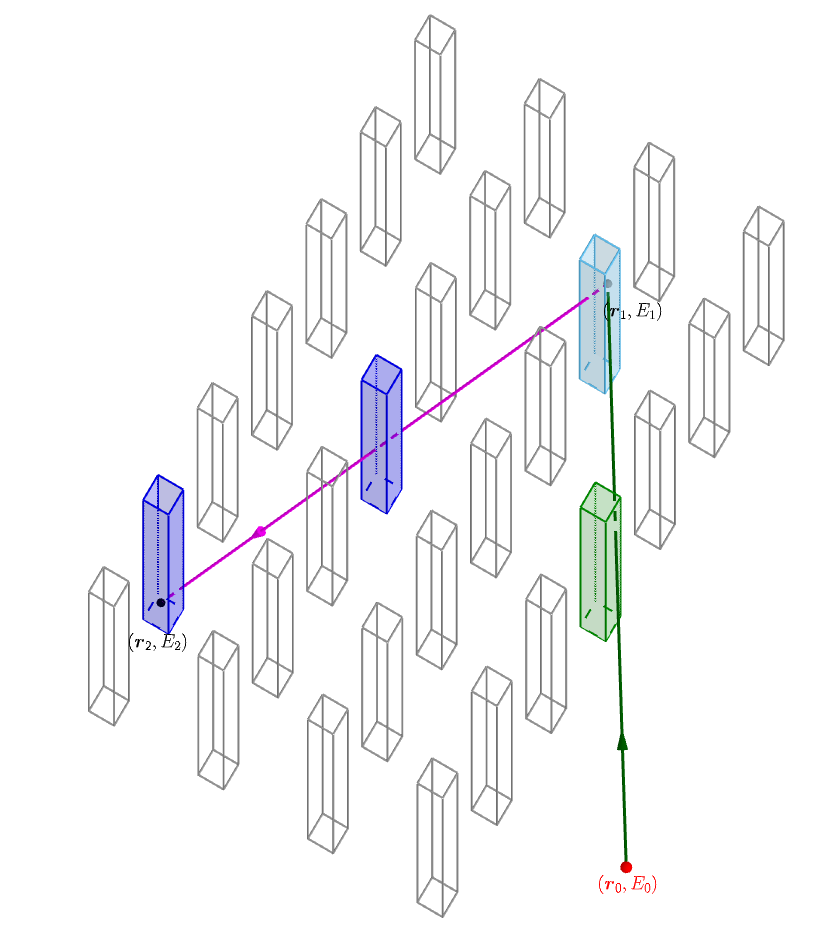}}\hfill
    \subfigure[]{\includegraphics[width = 0.45\linewidth]{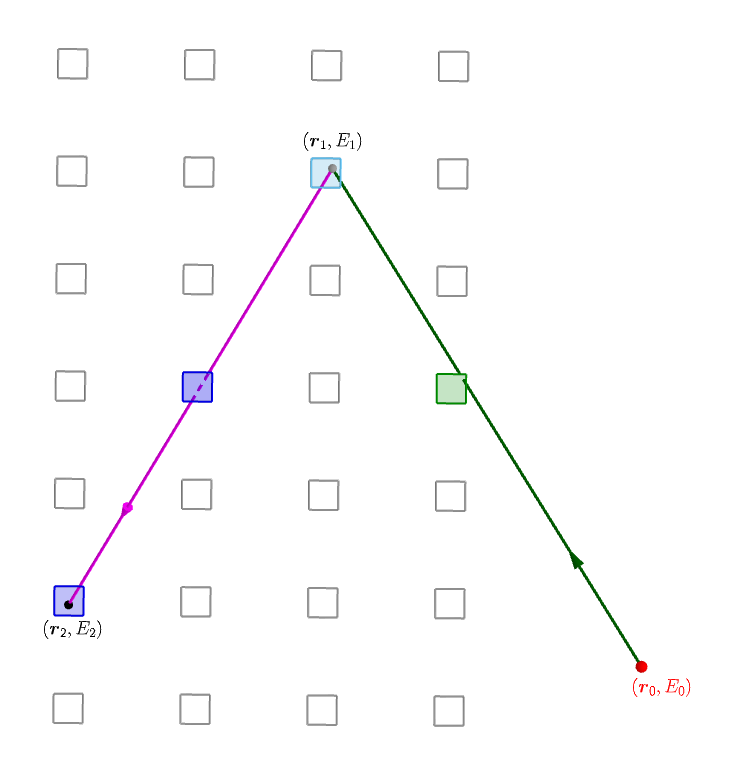}}
     \caption{ Compton imager: setup and working principle. An incoming photon of energy $E_0$ is emitted by a source (in red) at position $\boldsymbol{r}_0$. The photon travels in a first sensor in green before undergoing scattered at position $\boldsymbol{r}_1$ in the sensor in light blue according to the path represented by the green array. The energy deposited during the first interaction is denoted $E_1$. The scattered photon travels in the two sensors in dark blue. Its path is depicted by the pink array. The second interaction position happens at the position $\boldsymbol{r}_2$ with a deposited energy $E_2$. (a) Side view. (b) Top view.}
    \label{fig:CI}
\end{figure}

Recently,  variants of conventional Compton cameras have become of interest to allow imaging with a wide field of view \cite{vetter2018gamma,
al2019passive}. The main difference here stems from the fact that the sensors used are now able to both scatter and absorb photons. Such a system is currently being developed at the University of Illinois Urbana-Champaign. In particular, sensors are organised in a two dimensional array such as to allow imaging high-energy sources in applications that require a full field of view spanning a complete sphere. This is illustrated in Figure \ref{fig:CI}, which shows the setup of this  Compton imaging system and an example of a photon path. We henceforth refer to this variant of the Compton camera as the Compton imager (CI).

In this  paper, we are interested in solving  the problem of estimating the location and energy of point-like sources from the interactions  (data) recorded by a CI, taking into account the different sources of measurement noise as well as the presence of background noise. The fact that  the sensors used in the CI are  now able to both scatter and absorb photon means that the previous modelling approaches \cite{parajuli2022development, terzioglu2018compton, frandes2016image}  are not directly applicable. We thus develop a novel forward model that is capable of dealing with the new features of the CI. This in turn allows us to solve our inverse problem by a hierarchical Bayesian approach. In particular, the forward model is used to construct the likelihood and is combined with appropriate priors to give rise to the  joint posterior distribution for the energy and location. The expectation maximization algorithm \cite{moon1996expectation} is then used  to compute the minimum mean square error (MMSE) estimator for the underlying energy, while a Metropolis within Gibbs approach \cite{brooks2011handbook} is  employed to compute the corresponding MMSE estimator for the location of the point-like sources. This statistical approach is  then used to estimate the energy and the location of a known number of point like sources using simulated data obtained from a Monte Carlo $N$-particle transport code \cite{TechReport_2022_LANL_LA-UR-22-32951Rev.1_JoseyClarkEtAl, TechReport_2022_LANL_LA-UR-22-32851Rev.1_BullKuleszaEtAl, TechReport_2022_LANL_LA-UR-22-30006Rev.1_KuleszaAdamsEtAl, TechReport_2023_LANL_LA-UR-22-33103Rev.1_RisingArmstrongEtAl}.




The remainder of this paper is organized as follows. In Section \ref{ci_setup}, we present in more details the CI instrument considered and its working principle for data acquisition. The proposed statistical forward model is presented in Section \ref{sec:forward_model}. This model is then used to perform Bayesian estimation for the location and the energy of the sources in Section \ref{sec:bayesian_scheme}. The accuracy of the method is thoroughly analysed via numerical experiments in Section \ref{ref:simulation_results}. Conclusions and perspectives for future work are finally reported in Section \ref{sec:concluding}.

\section{Problem statement}
\label{ci_setup}
We first present the CI instrument in more detail, set notation, and formalise the estimation problem that we seek to address.
The considered CI instrument is composed of $L$ scintillation crystals, arranged in a $L_1 \times L_2 = L$ two-dimensional array configuration (in our experiments, we consider an $4\times 7$ array with $L = 28$ sensors as depicted in Figure \ref{fig:CI}). We consider the presence of $K$ point-like sources and aim to determine their position $\left(\boldsymbol{r}_0^{(1)}, \dots, \boldsymbol{r}_0^{(K)}\right)$ and energy $E_0$ from $N\gg K$ incoming photons detected by the CI. Note that the sources are assumed to be mono-energetic and share the same energy $E_0$. More precisely, the photons emitted by the sources interact with the CI as follows: for any $n \in \{1, \dots, N\}$, the $n$th photon produces a list of interactions
\begin{equation}
    I_n = \left(I_{1n}, I_{2n}, I_{3n}, \dots\right),
\end{equation}
in which each $I_{in}$ records the amount of energy lost/deposited $E_{in}$ at position $\boldsymbol{r}_{in}$
\begin{equation}
    I_{in} = (\boldsymbol{r}_{in}, E_{in}).
\end{equation}
It is assumed that the photons can only interact inside the sensors. 

For illustration, Figure \ref{fig:CI} depicts the flight path of a photon originating in a source at location $\boldsymbol{r}_0$ 
and with energy $E_0$, which interacts via Compton scattering with a sensor at the location $\boldsymbol{r}_1$ and deposits $E_1$ energy as a result, and subsequently interacts with a second sensor at location $\boldsymbol{r}_2$. In case of absorption,  this photon looses its remaining energy, hence $E_2 = E_0 - E_1$. In case of second scattering interaction, it follows that $E_2 < E_0 - E_1$.  Notice that in addition to the two photon-sensor interactions at $\boldsymbol{r}_1$ and $\boldsymbol{r}_2$, the photon path also crosses two other sensors without any interaction. 

\begin{figure}[!ht]
    \centering
    \subfigure[]{\includegraphics[width = 0.45\linewidth]{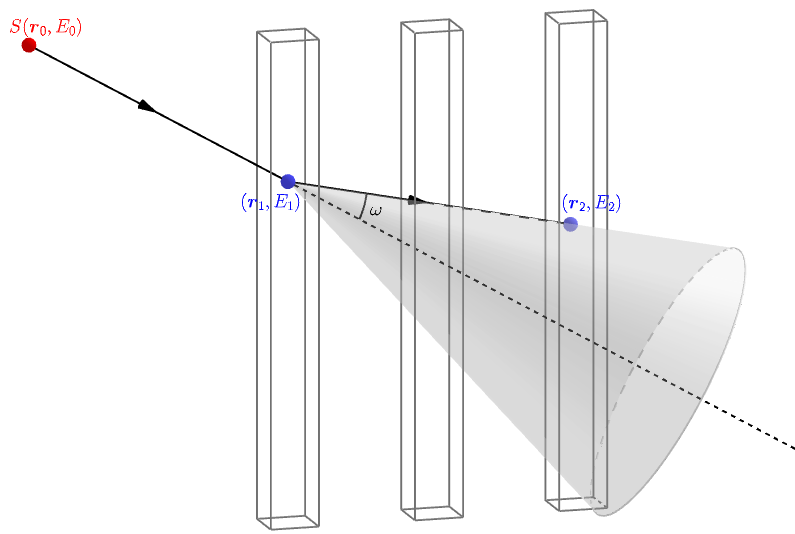}}\hfill
    \subfigure[]{\includegraphics[width = 0.45\linewidth]{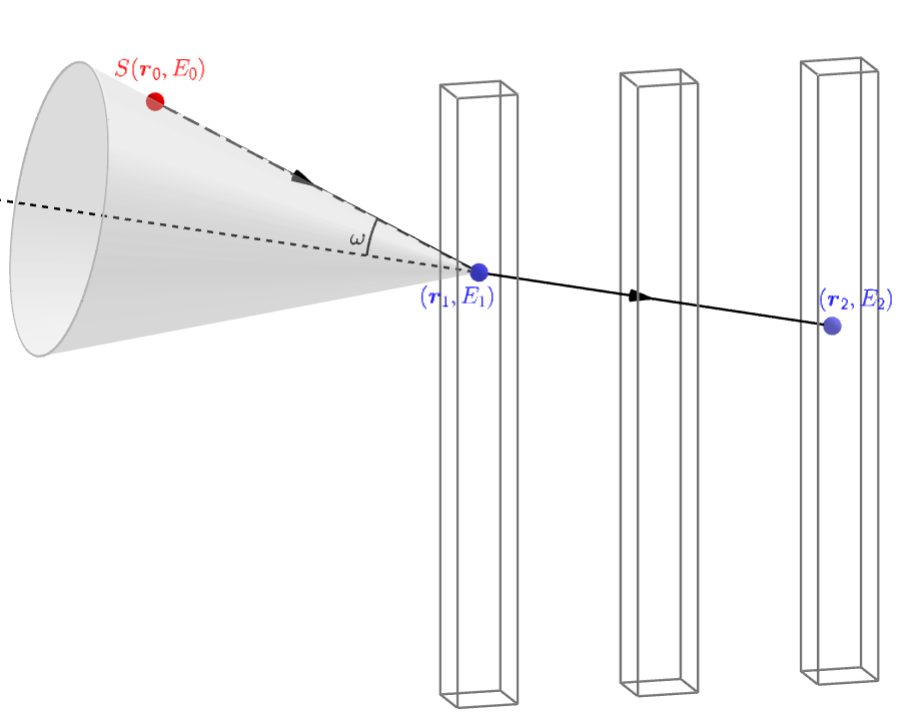}}
    \caption{ Physics of the Compton imager. (a) A point-like source $S$ at position $\boldsymbol{r}_0$ emits a photon of energy $E_0$. This photon interacts with one of the sensor of the imager at position $\boldsymbol{r}_1$. This interaction results in a scattered photon of energy $(E_0-E_1)$, where $E_1$ is the part of energy collected by the sensor at $\boldsymbol{r}_1$. The scattered photon interacts then at position $\boldsymbol{r}_2$. According to the Compton kinematics $\boldsymbol{r}_2$ is on a cone of apex $\boldsymbol{r}_1$, axis $(\boldsymbol{r}_1-\boldsymbol{r}_0)$ and semi-aperture $\omega$, that can be computed via the Compton formula \eqref{eq:Compton_formula}. The sensor records the amount of energy $E_2$ collected at $\boldsymbol{r}_2$, which is equal to $(E_0-E_1)$ if this photon interaction was a photoelectric absorption or less than $(E_0-E_1)$ if it was a Compton interaction. (b) Conversely, given the positions $\boldsymbol{r}_1, \boldsymbol{r}_2$ and energy depositions $E_1, E_2$ of the two interactions and assuming that we know $E_0$, $\boldsymbol{r}_0$ is on the cone of apex $\boldsymbol{r}_1$, axis $(\boldsymbol{r}_1, \boldsymbol{r}_2)$ and semi angle $\omega(E_0, E_1)$.}
    \label{fig:CS}
\end{figure}

While the events with only one interaction are useless for our radiation localization problem (since we cannot describe the conical surface containing the source), the events with two or more interactions could be used. In the present study, we choose to only exploit the positions and the deposited energies of the two first interactions $I_{1n}$ and $I_{2n}$ for each event. In fact, as for the Compton camera setup, the photon interactions recorded as ``event $n$" originate from a source lying on a conical surface whose apex is the position of the Compton interaction $\boldsymbol{r}_{1n}$, and of axis the line passing through the two first interactions sites $(\boldsymbol{r}_{1n}\boldsymbol{r}_{2n})$. The opening angle $\omega$ of that conical surface is given by the Compton formula \cite{ChoppinGregory2013RaNC} 
\begin{equation}
    \omega(E_0, E_1) = \arccos{\left(1-mc^2\left(\frac{1}{E_0-E_1}-\frac{1}{E_0}\right)\right)},
    \label{eq:Compton_formula}
\end{equation}
where $mc^2$ is the energy of an electron at rest (see Figures \ref{fig:CI} and \ref{fig:CS}). The next successive interactions do not give direct additional information to find the position of the source; however, considering the next interactions $(\boldsymbol{r}_{i}, E_{i})_{i\geq3}$ (if they exist) could allow estimating the position $\boldsymbol{r}_{i-2}$ with fewer uncertainties and exploiting them could be an interesting perspective to this work. 

In practice, the measurements delivered by the CI suffer from inaccuracies due to the finite energetic and spatial resolutions of the sensors, as well as from other sources of error related to the identification of the events. The photon interactions detected by the CI are gathered into events (or interaction pairs) by using a time-gating technique. This occasionally gives rise to aberrant events resulting from incorrect pairings. For instance, there are timing errors that lead to incorrect temporal ordering. Also, photons originating from different sources, including background photons, are sometimes incorrectly grouped together. In our inversion algorithm presented later in the paper, we show how we can process these types of aberrant data as outliers. We henceforth denote the noise-corrupted measurements of $I_{in}$, including possibly incorrectly detected events (outliers), by
\begin{equation}
    \tilde{I}_{in} = (\tilde{\boldsymbol{r}}_{in}, \tilde{E}_{in}).
\end{equation}


In the next section, we derive a statistical observation model for data acquisition. An important feature of this model is that it can handle two types of events, i.e., depending on whether the second interaction is a photon absorption or Compton scattering. Furthermore, the physical interactions of photons are fully described at each stage. This allows the development of new inversion algorithms and solving the radiation localization problem in the Bayesian framework.

\section{Observation model for the noise-free Compton imager}
\label{sec:forward_model}
In this section, we formulate a probabilistic model for data acquisition. We consider an arbitrary photon, stemmed from a source of position $\boldsymbol{r}_0$ and energy $E_0$, which undergoes at least two interactions $I_1$ and $I_2$ with the Compton Imager. The first interaction $I_1$ is a Compton scattering, the second interaction $I_2$ is either an absorption or Compton scattering. For notation brevity, we henceforth denote the probabilistic event ``Compton interaction (resp. absorption) occurs at interaction $i$." by $CS_i$ (resp. $A_i$). Also, we omit subscript $n$ and superscript $(k)$ in this section for the sake of clarity. 

For simplicity, we first introduce this observation model assuming a perfect (noise-free) Compton imager and focus on the role of Compton scattering. This noise-free model is extended to noisy measurements in Section \ref{sec:bayesian_scheme}.

\begin{figure}[!ht]
    \centering
    \begin{tikzpicture}[->,>=stealth',shorten >=1pt,auto, node distance=2cm,thick,main node/.style={circle,draw,font=\sffamily\Large\bfseries}, ex node/.style={rectangle,draw,font=\sffamily, , text height=0.5cm, text width=8cm}, ex node 2/.style={rectangle,draw,font=\sffamily, , text height=0.5cm, text width=4cm}]

  \node[ex node] (emission) {Emission of a photon at position $(\boldsymbol{r}_0, E_0)$};

    \node[ex node, below of =emission] (int1) {Reaches a sensor and interacts at position $\boldsymbol{r}_1$};

    \node[ex node, below of =int1] (E1) {$CS_1$ with an energy deposition $E_1$};

    \node[ex node, below of =E1] (int2) {Second interaction at position $r_2$};

    \node[ex node 2, below of = int2, yshift = -0.5cm, xshift=-3.5cm] (A) {$A_2$ with an energy deposition \\ $E_2 \,(= E_0-E_1)$};

    \node[ex node 2, below of = int2, yshift = -0.5cm, xshift=3.5cm] (CS) {$CS_2$ with an energy deposition $E_2 \,(< E_0-E_1)$};
 
    \path[every node/.style={font=\sffamily\small}]
    (emission) edge node [right] {$f(\boldsymbol{r}_1|\boldsymbol{r}_0, E_0)$ \eqref{eq:fr1sr0E0}} (int1)
    (int1) edge node [right] {$f(E_1|\boldsymbol{r}_0, E_0, \boldsymbol{r}_1, CS_1)$ \eqref{eq:fE1|E0,I1=CS}} (E1) 
    (E1) edge node [right] {{$f(\boldsymbol{r}_2|\boldsymbol{r}_0, E_0, \boldsymbol{r}_1, CS_1, E_1)$ \eqref{eq:f_r2sr0r1E1}}} (int2)
    (int2) edge node [left] { \small \scalebox{0.9}{$f(E_2|\boldsymbol{r}_0, E_0, \boldsymbol{r}_1, CS_1, E_1, \boldsymbol{r}_2, A_2)$} \eqref{eq:fE2|E0,E1,I1=CS,I2=A}} (A)
    (int2) edge node [right] {\small \scalebox{0.9}{$f(E_2|\boldsymbol{r}_0, E_0, \boldsymbol{r}_1, CS_1, E_1, \boldsymbol{r}_2, CS_2)$}\eqref{eq:fE2|E0,E1,I1=CS,I2=CS}} (CS);
    \end{tikzpicture}
    \caption{Photon trajectory from its emission (top) to the second interaction (bottom) and associated densities of each stage.}
    \label{fig:scheme_data_acq}
\end{figure}
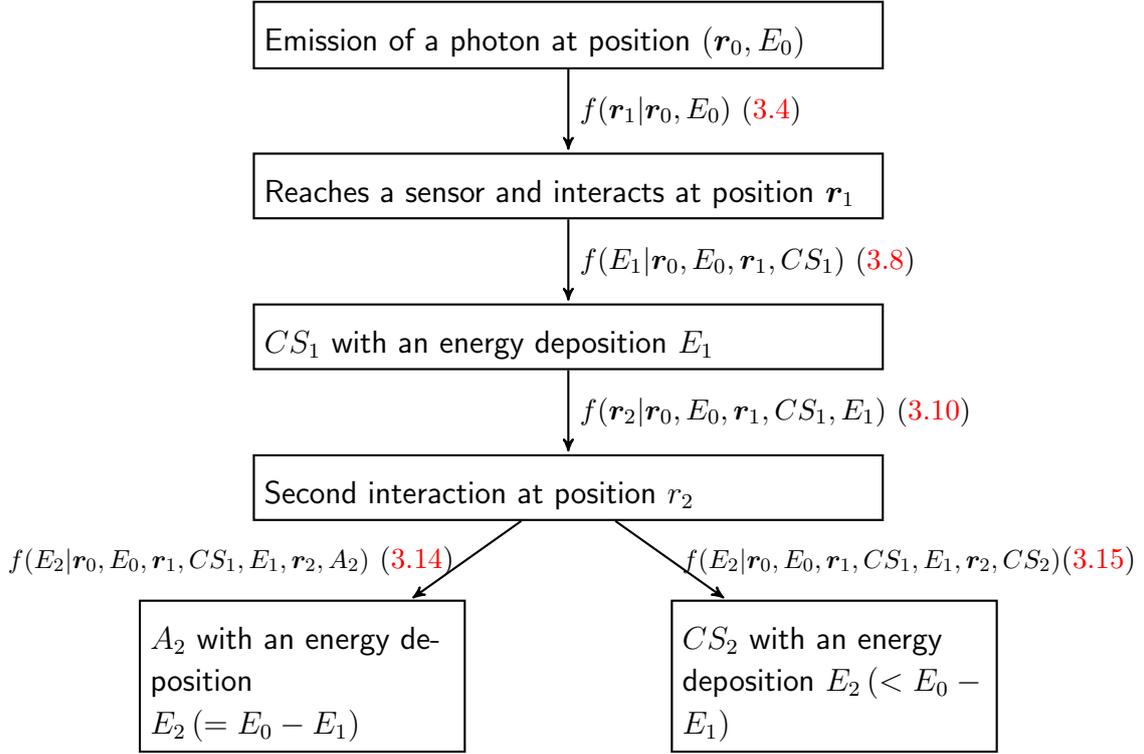

The noise-free proposed observation model rests on the computation of the densities
\begin{equation}
    f(\boldsymbol{r}_1, E_1, \boldsymbol{r}_2, E_2 | \boldsymbol{r}_0, E_0, CS_1, A_2) \text{ and } f(\boldsymbol{r}_1, E_1, \boldsymbol{r}_2, E_2 | \boldsymbol{r}_0, E_0, CS_1, CS_2).
\end{equation}
The observation model is obtained by splitting the photon trajectory into its different stages, from its emission by the source to its second interaction and the two above distributions differ only from the nature of the second interaction. A graphical representation of the different stages in depicted in Figure \ref{fig:scheme_data_acq}. Due to the hierarchical/sequential nature of the forward process, the statistical observation model can be expressed as a product of conditional distributions, associated with each stage of the photon propagation
\begin{multline}
    f( \boldsymbol{r}_1, E_1, \boldsymbol{r}_2, E_2 | \boldsymbol{r}_0, E_0, CS_1, X_2)  \\=\left\{\begin{array}{ll}
         f( \boldsymbol{r}_1, E_1, \boldsymbol{r}_2| \boldsymbol{r}_0, E_0, CS_1) 
    f(E_2 | \boldsymbol{r}_0, E_0, \boldsymbol{r}_1, E_1, CS_1, \boldsymbol{r}_2, A_2) & \text{ if } X_2 = A_2\\
         f( \boldsymbol{r}_1, E_1, \boldsymbol{r}_2| \boldsymbol{r}_0, E_0, CS_1)f(E_2 | \boldsymbol{r}_0, E_0, \boldsymbol{r}_1, E_1, CS_1, \boldsymbol{r}_2, CS_2) & \text{ if } X_2 = CS_2,
    \end{array}
\right.
    \label{eq:likelihood}
\end{multline}
with 
\begin{equation}
    f( \boldsymbol{r}_1, E_1, \boldsymbol{r}_2| \boldsymbol{r}_0, E_0, CS_1) = f( \boldsymbol{r}_1| \boldsymbol{r}_0, E_0)f( E_1| \boldsymbol{r}_0, E_0,  \boldsymbol{r}_1, CS_1) f(\boldsymbol{r}_2|\boldsymbol{r}_0, E_0, \boldsymbol{r}_1, E_1, CS_1). 
\end{equation}

The corresponding densities to each step of the data acquisition process are now derived. 

\paragraph{Computation of $f(\boldsymbol{r}_1|\boldsymbol{r}_0, E_0)$.}
The first step is the first interaction with the Compton imager at position $\boldsymbol{r}_1$. Denoting $\bar{d}_1$ the distance between $\boldsymbol{r}_0$ and $\boldsymbol{r}_1$ and $\boldsymbol{\Theta}_1 = (\boldsymbol{r}_1-\boldsymbol{r}_0)/||\boldsymbol{r}_1-\boldsymbol{r}_0||_2$ the direction of travel of the ray such that $\boldsymbol{r}_1 = \boldsymbol{r_0} + \bar{d}_1\boldsymbol{\Theta}_1$, leads to 
\begin{equation}
    f(\boldsymbol{r}_1|\boldsymbol{r}_0, E_0) = f(\bar{d}_1, \boldsymbol{\Theta}_1|\boldsymbol{r}_0, E_0) = f(\boldsymbol{\Theta}_1|\boldsymbol{r}_0, E_0) f(\bar{d}_1| \boldsymbol{r}_0, E_0, \boldsymbol{\Theta}_1).
        \label{eq:fr1sr0E0}
\end{equation}
$f(\boldsymbol{\Theta}_1|\boldsymbol{r}_0) $ refers to the probability of having an interaction in the direction $\boldsymbol{\Theta}_1$, that can be expressed from the Beer law
\begin{equation}
    f(\boldsymbol{\Theta}_1|\boldsymbol{r}_0, E_0) = \frac{\displaystyle\int_0^{d_{1,\text{max}}(\boldsymbol{r}_0, \boldsymbol{\Theta}_1)} \exp\left( -\mu_{E_0}x\right)dx}{\displaystyle\int_{\boldsymbol{\Theta}} \int_0^{d_{1,\text{max}}(\boldsymbol{r}_0, \boldsymbol{\Theta})} \exp\left( -\mu_{E_0}x\right)dx d\boldsymbol{\Theta}} = \frac{1 - \exp\left( -\mu_{E_0}d_{1,\text{max}}(\boldsymbol{r}_0, \boldsymbol{\Theta}_1)\right)}{\displaystyle\int_{\boldsymbol{\Theta}} \left(1 - \exp\left( -\mu_{E_0}d_{1,\text{max}}(\boldsymbol{r}_0, \boldsymbol{\Theta}_1)\right) \right)d\boldsymbol{\Theta}}, 
    \label{eq:solid_angle}
\end{equation}
where $d_{1,\text{max}}(\boldsymbol{r}_0, \boldsymbol{\Theta}_1)$ is the maximal distance a photon can travel inside the detectors from position $\boldsymbol{r}_0$ in direction $\boldsymbol{\Theta}_1$ and $\mu_{E}$ is the linear attenuation coefficient of the material of the sensor at energy $E$. 
Note that this density distribution is difficult to evaluate analytically. In our simulations, we will use a rejection-sampling scheme (see Appendix \ref{sec:calculation_solid_angle} for details).
Then, assuming the direction of the incoming photon is known and interactions only happen inside sensors, the probability for this photon to travel a distance $\bar{d}_1$ corresponds to the Beer law applied to the effective distance $d_1$ travelled by the photon inside the detectors between $\boldsymbol{r}_0$ and $\boldsymbol{r}_1$
\begin{equation}
   \displaystyle f(\bar{d}_1|\boldsymbol{r}_0, E_0, \boldsymbol{\Theta}_1 ) = \displaystyle f(d_1|\boldsymbol{r}_0, E_0, \boldsymbol{\Theta}_1 ) = \frac{\mu_{E_0} \exp\left(-\mu_{E_0}d_1(\boldsymbol{r}_0, \bar{d}_1, \boldsymbol{\Theta}_1)\right)}{\displaystyle 1-\mu_{E_0} \exp\left(-\mu_{E_0}d_{1,\text{max}}(\boldsymbol{r}_0, \boldsymbol{\Theta}_1)\right)},
    \label{eq:beer_d1}
\end{equation}
and $d_1\in \, ]0, d_{1,\text{max}}(\boldsymbol{r}_0, \boldsymbol{\Theta}_1)[$.  The combination of \eqref{eq:solid_angle} and \eqref{eq:beer_d1} gives \eqref{eq:fr1sr0E0}.

\paragraph{Computation of $f(E_1|\boldsymbol{r}_0, E_0, \boldsymbol{r}_1, CS_1)$.} At the position $\boldsymbol{r}_1$, a Compton interaction occurs and a part $E_1$ of the energy of the photon is deposited. The corresponding probability $ f(E_1|\boldsymbol{r}_0, E_0, \boldsymbol{r}_1, CS_1)$ does not depend on position of the event, since the array of detectors consists of a single material.  It follows that

\begin{align}
     f(E_1|\boldsymbol{r}_0, E_0, \boldsymbol{r}_1, CS_1) 
        &= f(E_1|E_0, CS_1).
        \label{eq:f_E1sr0r1}
\end{align}
The Klein-Nishina formula \cite{klein_nishina} leads to
\begin{equation}
    f(E_1|E_0, CS_1) =
    \frac{\displaystyle \varphi_{E_0}(E_1)}{ \displaystyle\int_{E_{1min,cs}}^{E_{1max,cs}} \varphi_{E_0}(x)\,dx},
    \label{eq:fE1|E0,I1=CS}
\end{equation}
where 
\begin{equation}
    \varphi_{E_0}(x) = \left(\frac{E_0-x}{E_0}\right)^2\left(\frac{E_0-x}{E_0} + \frac{x}{E_0-x} + \left(1-\frac{mc^2}{E_0}\left(\frac{x}{E_0-x}\right)\right)^2\right)\sqrt{1-\left(1-\frac{mc^2}{E_0}\left(\frac{x}{E_0-x}\right)\right)}.
\end{equation}

\noindent Extra explanation and the explicit formulation of the denominator used for calculations is derived in Appendix \ref{app:f(E_1sCS_E0)}. The domain of integration is set according to the Compton formula \eqref{eq:Compton_formula}, hence $E_{1min,cs} = 0$ and $E_{1max,cs} = E_0 - E_0/(1+2E_0/mc^2)$. Note that once $E_0$ is known, the denominator here is a constant. 

\paragraph{Computation of $f(\boldsymbol{r}_2|\boldsymbol{r}_0, E_0, \boldsymbol{r}_1, E_1, CS_1)$.}
As a third step, the scattered photon reaches position $\boldsymbol{r}_2$ before having a second interaction. As above, we denote by $\bar{d_2}$ the distance travelled by the photon on a direction $\boldsymbol{\Theta}_2 = (\boldsymbol{r}_2-\boldsymbol{r}_1)/||\boldsymbol{r}_2-\boldsymbol{r}_1||_2$ such that $\boldsymbol{r}_2 = \boldsymbol{r_1} + \bar{d}_2\boldsymbol{\Theta}_2$. It follows that 
\begin{align}
    f(\boldsymbol{r}_2|\boldsymbol{r}_0, E_0, \boldsymbol{r}_1, E_1, CS_1)  &= f(\bar{d}_2, \boldsymbol{\Theta}_2|\boldsymbol{r}_0, E_0, \boldsymbol{r}_1, CS_1, E_1) \nonumber\\&= f(\boldsymbol{\Theta}_2|\boldsymbol{r}_0, E_0, \boldsymbol{r}_1, E_1, CS_1) f(\bar{d}_2|\boldsymbol{r}_0, E_0, \boldsymbol{r}_1, E_1, CS_1, \boldsymbol{\Theta}_2).
    \label{eq:f_r2sr0r1E1}
\end{align}
From the Compton kinematics, the vector from $\boldsymbol{r}_1$ to $\boldsymbol{r}_2$ belongs to the $2\pi$-directional space whose shape is a conical surface of semi aperture angle $\omega(E_0, E_1)$ and apex $\boldsymbol{r}_1$, hence 
\begin{equation}
    f(\boldsymbol{\Theta}_2|\boldsymbol{r}_0, E_0, \boldsymbol{r}_1, E_1, CS_1) = \frac{1}{2\pi}\delta(\omega(E_0, E_1) - \arccos(\boldsymbol{\Theta}_1^T\boldsymbol{\Theta}_2)),
    \label{eq:dirac_compton_kinematics}
\end{equation}
where $\delta(\cdot)$ is the Dirac delta distribution (see Figure \ref{fig:CS}). Note that, for numerical computation, the  Dirac delta distribution is approximated using a Gaussian distribution
\begin{equation}
\delta(x) \approx \sqrt{\frac{a}{\pi}} \exp{(-ax^2)},
\label{eq:delta_approx}
\end{equation}
where $a\in \mathbb{R}^+$ is a user-defined parameter.
A similar reasoning as for \eqref{eq:beer_d1} leads to  
\begin{align}
    f(\bar{d}_2| \boldsymbol{r}_0, E_0, \boldsymbol{r}_1, E_1, CS_1,\boldsymbol{\Theta}_2) &= f(d_2| \boldsymbol{r}_0, E_0, \boldsymbol{r}_1, E_1, CS_1,\boldsymbol{\Theta}_2) \nonumber \\&= \frac{\mu_{E_0-E_1} \exp\left(-\mu_{E_0-E_1}d_2(\boldsymbol{r}_1, \bar{d}_2, \boldsymbol{\Theta}_2)\right)}{\displaystyle 1-\mu_{E_0-E_1} \exp\left(-\mu_{E_0-E_1}d_{2,\text{max}}(\boldsymbol{r}_1, \boldsymbol{\Theta}_2)\right)},
    \label{eq:beer_d2}
\end{align}
where $d_2\in \, ]0, d_{2,\text{max}}(\boldsymbol{r}_1, \boldsymbol{\Theta}_2)[$ is the effective distance travelled by the photon inside the detectors between $\boldsymbol{r}_1$ and $\boldsymbol{r}_2$ and $d_{2,\text{max}}$ is the maximal distance the photon can travel inside the detectors from $\boldsymbol{r}_1$ in the direction $\boldsymbol{\Theta}_2$. The combination of \eqref{eq:dirac_compton_kinematics} and \eqref{eq:beer_d2} leads to \eqref{eq:f_r2sr0r1E1}.

\paragraph{Computation of $f(E_2|\boldsymbol{r}_0, E_0, \boldsymbol{r}_1, E_1, CS_1, \boldsymbol{r}_2, CS_2)$ and $f(E_2|\boldsymbol{r}_0, E_0, \boldsymbol{r}_1, E_1, CS_1, \boldsymbol{r}_2, A_2)$.} The second interaction is either an absorption or a Compton interaction. Here, as it was the case for the first energy deposition, $f(E_2|\boldsymbol{r}_0, E_0, \boldsymbol{r}_1, E_1, CS_1, \boldsymbol{r}_2, CS_2)$ and $f(E_2|\boldsymbol{r}_0, E_0, \boldsymbol{r}_1, E_1, CS_1, \boldsymbol{r}_2, A_2)$ are not function of the positions $\boldsymbol{r}_0, \boldsymbol{r}_1$ and $\boldsymbol{r}_2$ involved in this problem.  
In case of an absorption, that is $E_2 = E_0 - E_1$, $f(E_2|E_0, E_1, CS_1, A_2)$ is derived using the Dirac delta distribution, leading to 
\begin{equation}
    f(E_2|E_0, E_1, CS_1, A_2) = \delta(E_2-(E_0-E_1)).
    \label{eq:fE2|E0,E1,I1=CS,I2=A}
\end{equation}
For a Compton scattering interaction, $f(E_2|E_0, E_1, CS_1, CS_2) $ is computed in a similar fashion to \eqref{eq:fE1|E0,I1=CS} replacing $E_1$ by $E_2$ and $E_0$ by $(E_0-E_1)$, leading to
\begin{equation}
    f(E_2|E_0, E_1, CS_1, CS_2) =
    \frac{\varphi_{E_0-E_1}(E_2)}{ \displaystyle\int_{E_{2min,cs}}^{E_{2max,cs}} \varphi_{E_0-E_1}(x) \,dx}, 
    \label{eq:fE2|E0,E1,I1=CS,I2=CS}
\end{equation}
with $E_{2min,cs} = 0$ and $E_{2max,cs} = E_0 -E_1 - (E_0-E_1)/(1+2(E_0-E_1)/(mc^2))$. Notice that the denominator here is a function of $E_1$. 

The likelihood $f_{\text{for}}$ \eqref{eq:likelihood} is finally obtained combining \eqref{eq:fr1sr0E0}, \eqref{eq:f_E1sr0r1}, \eqref{eq:f_r2sr0r1E1} and \eqref{eq:fE2|E0,E1,I1=CS,I2=A} or \eqref{eq:fE2|E0,E1,I1=CS,I2=CS}. 

\section{Bayesian estimation algorithms for source localization}
\label{sec:bayesian_scheme}
 
This section discusses a Bayesian approach to estimate positions and energy of fixed sources. More precisely, the problem is divided into the following tasks:
\begin{enumerate}
    \item the estimation of the position of the source(s), assuming the energy of the incoming photons and the nature of the interactions (scattering/absorption) is known.
    \item the joint estimation of the energy of the incoming photons and the classification of the events according to their nature (e.g. two successive Compton interactions or Compton scattering followed by absorption),
\end{enumerate}


In the numerical simulations we report in Section \ref{ref:simulation_results} for the source localization problem (assuming the source energy is known), we consider cases with limited number of noisy events, observed over a short period, ultimately targeting tracking of moving sources. We investigate a Bayesian estimation algorithm based on a Metropolis-within-Gibbs scheme \cite{brooks2011handbook} for this task. While the model presented in Section \ref{sec:forward_model} could be embedded in any EM-based scheme \cite{moon1996expectation} such as those mentioned in the introduction, back-projection techniques and EM-algorithms give equivalent results, and often suffer from (local) convergence issues for such a low-photon imaging situation. 


We discuss our EM algorithm and our Monte Carlo sampling based method in the next paragraphs, starting by the source localization estimation problem in Section \ref{subsec:est_position_source}. The joint estimation of the sources energy and the nature of the second interaction, which is simpler and can be addressed by maximum likelihood estimation, will be described in Section \ref{sec:estimation_E0}. 

\subsection{Estimation of the position of the sources}
\label{subsec:est_position_source}
We consider a set of $N$ noisy events $\{\tilde{\boldsymbol{r}}_{1n}, \tilde{E}_{1n}, \tilde{\boldsymbol{r}}_{2n}, \tilde{E}_{2n}\}_{n=1}^N$. The energy $E_0$ and the nature of the second interactions $\{X_{2n}\}$ are assumed to be either known or estimated using the method described in Section \ref{sec:estimation_E0}. The algorithm here assumes that $K$, the number of sources, is known and aims at estimating the position of these $K$ source(s) $\{\boldsymbol{r}_0^{(k)}\}_{k=1}^K$.

The use of the forward model presented in Section  \ref{sec:forward_model} requires extra knowledge due to measurement noise (e.g., the relationship between $\tilde{E}_{in}$ and $E_{in}$) and the presence of outliers/spurious data. For this purpose, we introduce latent variables which are part of an extended model, discuss the corresponding posterior distribution in Section \ref{subsec:extended_model} and then present the Gibbs sampler in Section \ref{sec:gibbs}. 

\subsubsection{Statistical observation model for noisy measurements and outliers}
\label{subsec:extended_model}
To deal with the finite spatial and energy resolutions of the sensors and the possible presence of outliers, we extend the model and include additional latent variables.

We introduce first $\{\boldsymbol{Z}_n\}_{n=1}^N = \{{\boldsymbol{r}}_{1n}, {E}_{1n}, {\boldsymbol{r}}_{2n}, {E}_{2n}\}_{n=1}^N$ standing for the true (unknown) positions and energy depositions of the interactions $\{\tilde{\boldsymbol{Z}}_n\}_{n=1}^N = \{\tilde{\boldsymbol{r}}_{1n}, \tilde{E}_{1n}, \tilde{\boldsymbol{r}}_{2n}, \tilde{E}_{2n}\}_{n=1}^N$. To keep the derivation simple, we assumed that the actual interaction positions and deposited energies are corrupted by truncated Gaussian noise of corresponding hidden standard deviations denoted by $\sigma_{x,y}$, $\sigma_z$ and $\sigma_E$. According to our observations, we have chosen to estimate these variances to make the algorithm more stable. We have also observed that the $x$ and $y$-coordinates of the positions are corrupted by similar noise levels, hence we consider a single variance parameter for those two dimensions. The prior distributions of standard deviations $f(\sigma_{xy})$, $f(\sigma_{z})$ and $f(\sigma_{E})$ are assumed to be uniform. For each event $n$, the density modelling data uncertainty is 
\begin{multline}
    f\left(\tilde{\boldsymbol{Z}}_n| E_0,\boldsymbol{Z}_n, \sigma_{xy}, \sigma_z, \sigma_E\right) = \\f\left(\tilde{\boldsymbol{r}}_{1n} | \boldsymbol{r}_{1n}, \sigma_{xy}, \sigma_z \right) 
    f(\tilde{E}_{1n}| E_{1n}, \sigma_E) f(\tilde{\boldsymbol{r}}_{2n}|\boldsymbol{r}_{2n}, \sigma_{xy}, \sigma_z)f(\tilde{E}_{2n}| E_{1n}, E_{2n}, \sigma_E). 
    \label{eq:model_uncertainty}
\end{multline}
Spatial uncertainties of the detectors $f\left(\tilde{\boldsymbol{r}}_{1} | \boldsymbol{r}_{1n}, \sigma_{xy}, \sigma_z \right) $ and $f(\tilde{\boldsymbol{r}}_{2n}|\boldsymbol{r}_{2n}, \sigma_{xy}, \sigma_z)$ are modelled using truncated Gaussian distributions of mean the true value $\boldsymbol{r}_{in}$ and the hidden standard deviations are the spatial resolutions $\sigma_{x,y}, \sigma_{z}$ of the sensors. The intervals of the distributions are determined by the boundaries of the sensor where the interaction took place. It follows that for $i = \{1,2\}$, 
\begin{equation}
     f(\tilde{\boldsymbol{r}}_{in}| \boldsymbol{r}_{in}, \sigma_{xy}, \sigma_z) = f(\tilde{x}_{in}| x_{in}, \sigma_{xy}) f(\tilde{y}_{in}| y_{in}, \sigma_{xy}) f(\tilde{z}_{in}| z_{in}, \sigma_{z}),
     \label{eq:rnoise_s_r}
\end{equation}
where $(x_{in}, y_{in}, z_{in})$ and $(\tilde{x}_{in}, \tilde{y}_{in}, \tilde{z}_{in})$ are the respective Cartesian coordinates of $\boldsymbol{r}_{in}$ and $\tilde{\boldsymbol{r}}_{in}$. 
For instance, $ f(\tilde{x}_{in}| x_{in}, \sigma_{xy})$ is defined by
\begin{equation}
    f(\tilde{x}_{in}| x_{in}, \sigma_{xy}) = \frac{1}{\displaystyle{\sigma_{xy}\sqrt{2\pi}}} \frac{\exp\left( - \frac{\displaystyle (\tilde{x}_{in}-x_{in})^2}{\displaystyle 2\sigma_{xy}^2} \right)}{\displaystyle \left(\Phi\left(\frac{x_{in,\text{max}}-x_{in}}{\sigma_{xy}}\right) - \Phi\left(\frac{x_{in,\text{min}}-x_{in}}{\sigma_{xy}}\right)\right)},
\end{equation}
with $[x_{in,\text{min}},x_{in,\text{max}}]$ is the $x$-domain of the sensor including position $\boldsymbol{r}_{in}$. Function $\Phi$ is the cumulative distribution function of the standard normal distribution
\begin{equation}
    \Phi(x) = \frac{1}{2}\left(1 + \text{erf}\left(\frac{x}{\sqrt{2}}\right)\right),
\end{equation}
and $\text{erf}(\cdot)$ stands for the error function \cite{andrews1998special}. The densities $f(\tilde{y}_{in}| y_{in}, \sigma_{xy})$ and $f(\tilde{z}_{in}| z_{in}, \sigma_{z})$ are defined accordingly. 

\noindent Energy uncertainties of detectors are modelled using truncated Gaussian distributions centred at the true value $E_{in}$ and of standard deviation $\sigma_E$. We only need to constrain the noisy energy depositions to be positive,  i.e. for each $i = \{1,2\}$
\begin{equation}
    f(\tilde{E}_{in}| E_{in}, \sigma_E) = \frac{1}{\displaystyle{\sigma_E\sqrt{2\pi}}}\,\frac{\displaystyle\exp\left(- \frac{\displaystyle (\tilde{E}_{in}-E_{in})^2}{\displaystyle 2\sigma_E^2} \right)}{\displaystyle 1-\Phi \left(-\frac{E_{in}}{\sigma_E}\right)}.
    \label{eq:noisy_E}
\end{equation}

Moreover, it is assumed that no prior knowledge on the position of the sources is available. Consequently, the $K$ sources to track are supposed to be independent and a-priori uniformly distributed
\begin{equation}
    f(\boldsymbol{r}_0^{(1)}, \dots, \boldsymbol{r}_0^{(K)}) = \prod_{k = 1}^K f(\boldsymbol{r}_0^{(k)}) \text{ where } f(\boldsymbol{r}_0^{(k)}) = \frac{\sin(\theta_0^{(k)})}{4\pi},
    \label{eq:prior_r0}
\end{equation}
and $\theta_0^{(k)}$ is colatitude of $\boldsymbol{r}_0^{(k)}$.
To assign each of the $N$ events to the relevant source and account for potential outliers, we propose to add $N$ additional virtual source positions, denoted $\{\boldsymbol{r}_{0n}\}_{n\in \{1, \dots, N\}}$, acting as if there were $N$ sources, each associated with an event.  
We also include the unknown relative intensities of the $K$ sources, denoted by $\{{w}_0^{(k)}\}_{k=1}^K$ and satisfying $\sum_{k=1}^K {w}_0^{(k)}<1$, that will also be estimated. The weights $\{w_0^{(k)}\}_{k=1}^K$ are assigned a Dirichlet distribution, 
\begin{equation}
    f\left( \{w_0^{(k)}\}_{k=1}^K\right) = \frac{1}{B(\alpha_0, \alpha_1, \dots, \alpha_K)} \left(\prod_{k = 1}^K {w_0^{(k)}}^{\alpha_k-1}\right) \left(1-\sum_{k = 1}^K w_0^{(k)}\right)^{\alpha_0-1}
    \label{eq:prior_w}
\end{equation}
where $\{\alpha_k\}_{k=1}^K$ and $\alpha_0$ are the respective concentration parameters of $\{w_0^{(k)}\}_{k=1}^K$ and $\left(1- \allowbreak \sum_{k = 1}^K w_0^{(k)}\right)$. $B(\alpha_0, \alpha_1, \dots, \alpha_K)$ is the normalization constant and can be written in terms of the Gamma function $\Gamma$, i.e. $B(\alpha_0, \alpha_1, \dots, \alpha_K) = \prod_{k=0}^{K} \Gamma({\alpha_k})/\Gamma\left( \sum_{k=0}^K \alpha_k\right). $

\noindent When the number of outliers is expected to be much smaller than the number of true events, as it will be the case in our experiments, $\alpha_0$ is assigned to a small value. Then, $\{\alpha_k\}_{k=1}^K$ are set by the user, according to the expected proportion of photons emitted by each source.  

\noindent Finally, the density distribution of the $n$-th virtual source conditioned to the $K$ sources and their relative intensities is defined as a weighted sum of $K+1$ terms 
\begin{equation}
    f(\boldsymbol{r}_{0n}|\{\boldsymbol{r}_0^{(k)}, w_0^{(k)}\}_{k=1}^K) = \sum_{k = 1}^K w_0^{(k)} \frac{\kappa}{\sinh{\kappa}} \exp{\left( \kappa \,\boldsymbol{r}_{0n}\cdot \boldsymbol{r}_0^{(k)}\right)} + \left(1-\sum_{k = 1}^K w_0^{(k)} \right)\frac{\sin(\theta_n)}{4\pi}, 
    \label{eq:prior_r0n}
\end{equation}
where $\kappa$ is the concentration parameter and $\cdot$ refers to the inner product. The $K$ first terms of \eqref{eq:prior_r0n} promote clustering of the virtual source positions to the right $\boldsymbol{r}_0^{(k)}$. The higher $\kappa$ is, the more the virtual sources $\boldsymbol{r}_{0n}$ are enforced to get close to one $\boldsymbol{r}_0^{(k)}$. The last term gives the possibility for an event to be an outlier. In such a case, it is assumed that this event is emitted by an arbitrary source on the sphere. The joint prior distribution assigned to the source positions and their relative intensities is obtained from the combination of \eqref{eq:prior_r0}, \eqref{eq:prior_w} and \eqref{eq:prior_r0n}, 

\begin{equation}
    f\left(\{\boldsymbol{r}_0^{(k)}, w_0^{(k)}\}_{k=1}^K, \boldsymbol{r}_{0n}\right) = \\f\left(\boldsymbol{r}_{0n}|\{\boldsymbol{r}_0^{(k)}, w_0^{(k)}\}\right) f\left(\{\boldsymbol{r}_0^{(k)}\}\right)f\left(\{ w_0^{(k)}\}\right). 
    \label{eq:jprior_pos_sources}
\end{equation}
 
 We finally illustrate the dependency of the measurements $ \{\tilde{\boldsymbol{Z}}_n\}_{n=1}^N$, the localization of the $K$ sources $\{\boldsymbol{r}_0^{(k)}\}_{k=1}^K$ with the introduced variables of the extended model $\{\boldsymbol{r}_0^{(n)}, \boldsymbol{Z}_n\}_{n=1}^N$ and $\{w_0^{(k)}\}_{k=1}^K$ (and their hyperparameters) in Figure \ref{fig:hierarchical_model}. Following Bayes' theorem and exploiting independence between variables, the joint posterior distribution $f_\text{jpos}$ to estimate results from the combination of \eqref{eq:model_uncertainty}, \eqref{eq:likelihood} and \eqref{eq:jprior_pos_sources}
\begin{equation} 
f_\text{jpos} \propto \prod_{n = 1}^N \left(f(\tilde{\boldsymbol{Z}}_n| E_0, \boldsymbol{Z}_n, \sigma_{xy}, \sigma_z, \sigma_E)f_{\text{for}}\left(\boldsymbol{Z}_n|\boldsymbol{r}_{0n}, E_0, CS_1, X_2\right) f( \{\boldsymbol{r}_0^{(k)},w_0^{(k)}\}_{k =1}^K, \boldsymbol{r}_{0n})\right). 
\label{eq:jposdist}
\end{equation}

\input{hierarchical_model.tex}

\subsubsection{Metropolis-within-Gibbs sampler}
\label{sec:gibbs}
We perform Bayesian computation for the proposed Bayesian model \eqref{eq:jposdist} by using a Metropolis-within-Gibbs sampling Markov chain Monte Carlo scheme. At each iteration of this algorithm, the $N$ events of interest $\{\tilde{\boldsymbol{Z}}_{n}\}$ are processed sequentially to generate new values for $\boldsymbol{r}_{1n}$, $\boldsymbol{r}_{2n}$, $E_{1n}$, $E_{2n}$ and $\boldsymbol{r}_{0n}$. This part of the algorithm can be done in parallel, since the events are conditionally independent. Then, $\boldsymbol{r}_{0}^{(k)}$ and $w^{(k)}$ are sampled from the set of $\{\boldsymbol{r}_{0n}\}_{n\in[|1, N|]}$. 
Finally, the values of the standard deviations $\sigma_{xy,z,E}$ are updated. Algorithm \ref{algo:gibbs} summarizes the proposed procedure. Each sampling step was carried out using a Metropolis-Hastings sampling scheme, and the distributions involved are detailed in Appendix \ref{sec:Gibbs_details}.

\IncMargin{1em}
    \begin{algorithm}[!ht]
        \SetAlgoLined
        \SetKwHangingKw{KwInit}{Initialisation:} 
        \KwData{Set of $N$ events $\{\tilde{\boldsymbol{Z}}_n\}_{n=1}^N$ and the nature of the second interaction $\{X_{2n} = CS_{2n} \cup A_{2n}\}_{n=1}^N$, Energy of emitted rays $E_0$, Number of iterations $T$}
        \KwResult{Position of the $K$ sources $\{\boldsymbol{r}_0^{(k)}\}_{k=1}^K$}
        \KwInit{Set initial values for $\{\boldsymbol{r}_0^{(k,0)}, w_0^{(k,0)},\alpha_k \}_{k=1}^K, \alpha_0, \{\boldsymbol{r}_{0n}^{(0)}, \boldsymbol{Z}_n^{(0)}\}_{n=1}^N, \sigma_{xy}^{(0)}, \sigma_z^{(0)}, \sigma_E^{(0)}$}\
        
        \For{$t = 1$ \KwTo $T$}{
            \For {$n = 1$ \KwTo $N$}{
             \tcc{This for loop can be performed in parallel.}
            Sample $\boldsymbol{r}_{1n}^{(t)} \sim f(\boldsymbol{r}_{1n} | \boldsymbol{r}_{0n}^{(t-1)}, E_0, CS_{1n}, E_{1n}^{(t-1)},  \boldsymbol{r}_{2n}^{(t-1)}, X_{2n}, E_{2n}^{(t-1)}, \tilde{\boldsymbol{r}}_{1n}, \sigma_{xy}^{(t-1)}, \sigma_{z}^{(t-1)})$

            Sample $\boldsymbol{r}_{2n}^{(t)} \sim f(\boldsymbol{r}_{2n} | \boldsymbol{r}_{0n}^{(t-1)}, E_0, \boldsymbol{r}_{1n}^{(t)}, CS_{1n}, E_{1n}^{(t-1)}, X_{2n}, E_{2n}^{(t-1)}, \tilde{\boldsymbol{r}}_{2n}, \sigma_{xy}^{(t-1)}, \sigma_{z}^{(t-1)})$

            Sample $E_{1n}^{(t)}, E_{2n}^{(t)}\sim f(E_{1n}, E_{2n}| \boldsymbol{r}_{0n}^{(t-1)}, E_0, \boldsymbol{r}_{1n}^{(t)}, CS_{1n}, \tilde{E}_{1n}, \boldsymbol{r}_{2n}^{(t)}, X_{2n}, \tilde{E}_{2n}, \sigma_{E}^{(t-1)})$ 
            
            Sample $\boldsymbol{r}_{0n}^{(t)} \sim f(\boldsymbol{r}_{0n}|\{\boldsymbol{r}_0^{(k, t-1)}, w_0^{(k, t-1)}\}_{k=1}^K,\boldsymbol{r}_{1n}^{(t)},  CS_{1n},{E}_{1n}^{(t)}, \boldsymbol{r}_{2n}^{(t)}, X_{2n}, {E}_{2n}^{(t)})$

            }
            \For {$k = 1$ \KwTo $K$}{
             Sample $\boldsymbol{r}_0^{(k, t)} \sim f(\boldsymbol{r}_0^{(k)} | \{\boldsymbol{r}_{0n}^{(t)}\}_{n=1}^N, w_0^{(k,t-1)}, \{\boldsymbol{r}_0^{(k',t)}, w_0^{(k',t)}\}_{k'=1}^{k-1}, \{\boldsymbol{r}_0^{(k',t-1)}, w_0^{(k',t-1)}\}_{k'=k+1}^{K})$
             
             Sample  $w_0^{(k, t)} \sim f(w_0^{(k)} | \{\boldsymbol{r}_{0n}^{(t)}\}_{n=1}^N, \boldsymbol{r}_0^{(k,t)}, \{\boldsymbol{r}_0^{(k',t)}, w_0^{(k',t)}\}_{k'=1}^{k-1}, \{\boldsymbol{r}_0^{(k',t-1)}, w_0^{(k',t-1)}\}_{k'=k+1}^{K})$
             }
    Sample $\sigma_{xy}^{(t)} \sim f(\sigma_{xy}|\{\boldsymbol{r}^{(t)}_{1n}, \tilde{\boldsymbol{r}}_{1n}, \boldsymbol{r}^{(t)}_{2n},\tilde{\boldsymbol{r}}_{2n}\}_{n=1}^N)$
    
    Sample $\sigma_z^{(t)} \sim f(\sigma_z|\{\boldsymbol{r}^{(t)}_{1n}, \tilde{\boldsymbol{r}}_{1n}, \boldsymbol{r}^{(t)}_{2n},\tilde{\boldsymbol{r}}_{2n}\}_{n=1}^N)$

    Sample $\sigma_E^{(t)} \sim f(\sigma_E|\{E^{(t)}_{1n}, \tilde{E}_{1n}, E^{(t)}_{2n},\tilde{E}_{2n}\}_{n=1}^N)$
    
}\
    \caption{Full Bayesian method for the estimation of the position of the sources $\mathbf{r_0}^{(k)}$ from experimental measurements}
    \label{algo:gibbs}
    \end{algorithm}
\DecMargin{1em}

\subsection{Estimation of the energy of the incoming photons and identification of the nature of the interactions}
\label{sec:estimation_E0}
This section discusses the energy estimation algorithm. The proposed algorithm takes only into account the corrupted versions of the energies $\{\tilde{E}_{1n}, \tilde{E}_{2n}\}_{n=1}^N$ and the attenuation coefficients of the material of the sensor. The knowledge of the position of the source(s) as well as the positions of the interactions are not necessary. The estimation of $E_0$ and the nature of the events are posed in terms of a maximum-likelihood estimation problem associated to the sum of the energy depositions, denoted in the following $\{\tilde{E}_n = \tilde{E}_{1n}+\tilde{E}_{2n}\}_{n=1}^N$. The log-likelihood of interest is presented in paragraph \ref{subsubsec:log_lik_EM} and then embedded in the expectation-maximization (EM) algorithm proposed in \ref{sec:EM_algorithm}.

\subsubsection{Log-likelihood of the model} \label{subsubsec:log_lik_EM}
The log-likelihood to maximize involves three sets of variables; (1) $\mathcal{X} = \{ \tilde{E}_n\}_{n=1}^N$, i.e., the set of the sum of the measured energies, (2) the set of latent variables consisting of the nature of the second interactions $\mathcal{Y} = \{X_{2n}\}_{n=1}^N$ and (3) the set of unknown parameters $\boldsymbol{\theta}$ which are optimized during the maximization step $\boldsymbol{\theta}= (p_A, p_{CS}, E_0, \sigma^2)$. In $\boldsymbol{\theta}$, $p_A = f(A_{2n}|E_0, CS_{1n})$ (resp. $p_{CS}= f(CS_{2n}|E_0, CS_{1n})$) is the probability that the second interaction of event $n$ is an absorption (resp. Compton scattering) and $\sigma$ is the standard deviation of the Gaussian distribution modelling noise on $\mathcal{X}$. This leads to
\begin{equation}
    \log \mathcal{L}(\mathcal{X}, \mathcal{Y};\,\boldsymbol{\theta})
    = \sum_{n=1}^N \log\left(f\left(\tilde{E}_n|E_0, CS_{1n}, X_{2n}, \sigma^2 \right)f\left(X_{2n}|E_0, CS_{1n}\right)\right).
    \label{eq:log-likelihood}
\end{equation}
Probabilities $p_A$ and $p_{CS}$ can be either computed analytically or estimated via the EM algorithm. In order to tackle the potential sensitivity of real sensors, we have chosen to estimate these values from the data, together with $E_0$. For the sake of completeness, their analytical derivations are presented in Appendix \ref{app:est_pA_pCS}. 

The numerical computation of $f(\tilde{E}_n|E_0, CS_{1n}, X_{2n}, \sigma^2)$ involves the marginalization of extra hidden variables, that is $\{{E}_n\}_{n=1}^N$, which correspond to the true (but unknown) sums of the energy depositions. The noise affecting energy measurements is modelled as white Gaussian noise with standard deviation $\sigma$, leading to
\begin{align}
 f(\tilde{E_n}|E_n, \sigma^2) = \frac{1}{\sigma\sqrt{2\pi}} \exp{\left(-\frac{1}{2\sigma^2}(\tilde{E}_n-E_n)^2\right)}.
 \label{eq:noisy_E_EM}
\end{align}
It follows that
\begin{equation}
    f(\tilde{E}_n|E_0, CS_{1n}, X_{2n}, \sigma^2) = \frac{1}{\sigma\sqrt{2\pi}}\int   \exp{\left(-\frac{1}{2\sigma^2}(\tilde{E}_n-E_n)^2\right)} f(E_n|E_0, CS_{1n}, X_{2n}) dE_n,
    \label{eq:f_En_noisy}
\end{equation}
and $f(E_n|E_0, CS_{1n}, X_{2n})$ is defined according to the data acquisition model defined in Section \ref{sec:forward_model}. In fact, when the second interaction is an absorption, the sum of the energy depositions $E_n$ is equal to $E_0$, and this is modelled using a Dirac delta distribution
\begin{equation}
    f(E_n|E_0, CS_{1n}, A_{2n}) = \delta(E_n-E_0).
    \label{eq:f_En_A2n}
\end{equation}
When the second interaction is a Compton scattering, then the derivation of the corresponding distribution involves the marginalisation of the energy deposition at first interaction $E_{1n}$
\begin{equation}
    f(E_n|E_0, CS_{1n}, CS_{2n}) = \int_{E_{1nmin,cs}}^{E_{1nmax,cs}} f(E_n|E_0, E_{1n}, CS_{1n}, CS_{2n})f(E_{1n}|E_0, CS_{2n})dE_{1n},
    \label{eq:f_En_C2n}
\end{equation}
where $E_{1nmin,cs} = 0$, $E_{1nmax,cs} = E_0 - E_0/(1+2E_0/mc^2)$, $f(E_{1n}|E_0, CS_{2n})$ is obtained using \eqref{eq:fE1|E0,I1=CS} and $f(E_n|E_0, E_{1n}, CS_{1n}, CS_{2n})f(E_{1n}|E_0, CS_{2n})$ is obtained from the Klein-Nishina formula
\begin{equation}
     f(E_n|E_0, E_{1n}, CS_{1n}, X_{2n}) =
         \frac{\displaystyle\varphi_{E_0-E_{1n}}(E-E_{1n})}{\displaystyle\int_{E_{min,cs}}^{E_{max,cs}} \varphi_{E_0-E_{1n}}(x-E_{1n}) \,dx}
         \label{eq:f_En_E1nmarg_CS2n}
\end{equation}
with $E_{min,cs} = E_{1n}$ and $E_{max,cs} = E_0 - (E_0-E_{1n})/(1 +2(E_0-E_{1n})/(mc^2))$.  

We present now the EM algorithm, estimating the energy of the source(s) and the nature of the interactions.

\subsubsection{EM algorithm}
In the expectation step, the expectations of the unknown parameters $\boldsymbol{\theta}$ conditioned on their current estimate $\boldsymbol{\theta}^{(c)}$ and the observations $\mathcal{X}$ are computed. In the maximization step, a new estimate of the parameters is provided. 
\label{sec:EM_algorithm}
\paragraph{E-step}
For the E-step, the expected value of the log-likelihood $Q(\boldsymbol{\theta}, \boldsymbol{\theta}^{(c)})$ \eqref{eq:log-likelihood} conditioned on the observed data and the versions of the parameters at iteration $(c)$ is computed: 
\begin{equation}
    \mathcal{Q}(\boldsymbol{\theta}, \boldsymbol{\theta}^{(c)}) = E(\log \mathcal{L}(\mathcal{X}, \mathcal{Y}; \boldsymbol{\theta})|\mathcal{X}, \boldsymbol{\theta}^{(c)}) = \sum_{n = 1}^N \sum_{X = \{CS,A\}} t_{nX}^{(c)}\log(p_{X} f(\tilde{E}_n|E_0, CS_{1n}, X_{2n}, \sigma^2) ), 
    \label{eq:Q}
\end{equation}
with 
\begin{equation}
    t_{nX}^{(c)}= \frac{ p_X^{(c)}f(\tilde{E}_n|E_0, CS_{1n}, X_{2n}, \sigma^2)}{p_{CS}^{(c)}f(\tilde{E}_n|E_0, CS_{1n}, CS_{2n}, \sigma^2) + p_A^{(c)} f(\tilde{E}_n|E_0, CS_{1n}, A_{2n}, \sigma^2)}.
\end{equation}

\paragraph{M-step}
The M-step consists of maximizing $Q(\boldsymbol{\theta}, \boldsymbol{\theta}^{(c)})$ \eqref{eq:Q} over $\boldsymbol{\theta}$ to obtain $\boldsymbol{\theta}^{(c+1)}$, that is,
\begin{equation}
    \boldsymbol{\theta}^{(c+1)} = \arg \underset{\boldsymbol{\theta}}{\max} \,Q(\boldsymbol{\theta}, \boldsymbol{\theta}^{(c)}). 
\end{equation}
The expressions of $p_{CS}^{(c+1)}$ and $p_{A}^{(c+1)}$ can be obtained in closed form, i.e. $p_{X}^{(c+1)} = (1/N) \sum_{n=1}^N t_{nX}^{(c)}$,
while $E_0^{(c+1)}$ and $\sigma^{(c+1)}$ are computed using a grid search. The integrals \eqref{eq:f_En_noisy}, \eqref{eq:f_En_C2n} and \eqref{eq:f_En_E1nmarg_CS2n} are calculated numerically using the trapezoidal rule.
The performance of the algorithm will be discussed in Section \ref{ref:simulation_results} below.  

\section{Simulation results}
\label{ref:simulation_results}
\subsection{Experimental setup}
\begin{figure}[!ht]
    \centering
    \includegraphics[width = 0.8\textwidth]{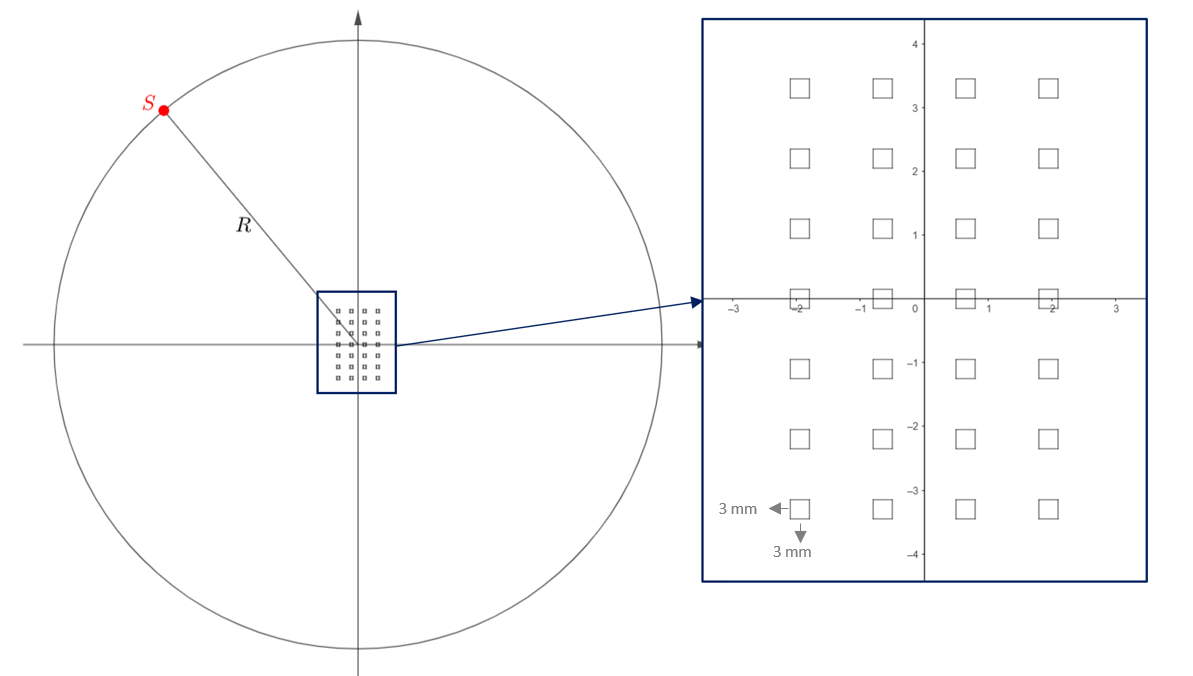}
    \caption{Experimental setup with $K = 1$ sources - Top view}
    \label{fig:CI_top}
\end{figure}

The instrument considered is made of $28$ sensors of size $3\times3\times 50$ mm$^3$ arranged in an $4\times7$ array, centred at positions $(X_i, Y_j, 0)$ where $X_i = -19.5 + 13i$ mm, $i\in \{0, \dots, 3\}$ and $Y_j = -33 + 11j$ mm, $j \in \{0, \dots, 6\}$ (see top view in Figure \ref{fig:CI_top}). These sensors consist of LYSO scintillation crystals (Lu$^{1.9}$Y$^{0.1}$SiO$^5$) and their corresponding attenuation coefficients have been generated using the NIST database \cite{nist_database}. 

We consider a known number of $K$ Cs-137 sources of energy $E_0 = 0.6617$ MeV. In the low-photon imaging experiments proposed here, a small number of noisy events is recorded and in that period, both Compton imager and source are supposed to be fixed to each other. As a consequence, we can only determine the direction-of-arrival of the incoming photons, as it is not possible to determine the distance to static sources. The sources are thus assumed to be placed on a sphere centred at the origin of the coordinates system and of known radius $R = 300$ mm. 

The simulations results presented in the next paragraph originate from simulated data using the Monte Carlo N-Particle (MCNP) code \cite{TechReport_2022_LANL_LA-UR-22-32951Rev.1_JoseyClarkEtAl, TechReport_2022_LANL_LA-UR-22-32851Rev.1_BullKuleszaEtAl, TechReport_2022_LANL_LA-UR-22-30006Rev.1_KuleszaAdamsEtAl, TechReport_2023_LANL_LA-UR-22-33103Rev.1_RisingArmstrongEtAl}. Some effects modelling realistic measurements of energy depositions on the sensors have already been included in the used simulations and the energy resolution is assumed to be $\Delta_E \approx 0.1$MeV. The measured positions of the interactions obtained from the MCNP code are however quite accurate and can be considered as noiseless; hence as a post-processing step, Gaussian noise is added on the measurements to obtain $\Delta_{xy} = 3$ mm resolution on $x$ and $y$-coordinates and $\Delta_{z} = 5$ mm on the third coordinate. These values are coherent with the expected level of noise of the true system and in terms of standard deviations, correspond to $\sigma_{xy} = 0.43$mm,  $\sigma_z = 0.72$mm and $\sigma_E \approx 0.029$MeV. We consider sets of $N$ noisy measurements $\{\tilde{\boldsymbol{Z}}_n\}_{n=1}^N$ whose nature of the second interaction $X_{2n}$ and primary energy $E_0$ are supposed to be unknown. 

The selection of hyper-parameters and initialisation values are reported in Appendix \ref{sec:Gibbs_details}.

\subsection{Estimation of the energy of the source(s)}
We first evaluate the performance of the proposed EM algorithm to estimate the source(s) energy. For illustration purposes, we considered a set of $N=2000$ events. More precisely, the only quantities of interest here are the set of the sums of the energy depositions $\{\tilde{E}_n\}_{n=1}^N$. The distribution of this set of events is depicted in red in Figure \ref{fig:res_EM}. The nature of the second interaction is also known from the MCNP code, and for the considered set of events, it follows that $p_{CS} = 0.1615$ (and $p_{A} = 1-p_{CS} = 0.8385$).

\begin{figure}[ht!]
\centering
\begin{tikzpicture}
\begin{axis}[
    xlabel={$E$},
    ylabel={$f(E|E_0, p_A, p_{CS}, \sigma, I_1=CS)$},
    xtick = {0},
    extra x ticks={0.6617},
    extra x tick style={grid=major, ticklabel pos=bottom},
    extra x tick labels={$E_0$},
    legend pos=outer north east,
    ]
  \addplot[red,fill=red!90!black,opacity=0.5, line width=1pt] table [x=E,y=fEexp] {table_Etheo_E_expe.txt};
  \addplot[black, line width=0.8pt] table [x=E,y=fEtheo] {table_Etheo_E_expe.txt};
  \addlegendentry{Experimental}
  \addlegendentry{Theory}
  \draw [stealth-] (250,25) -- (250,65) node [above]{\footnotesize{CS-CS contribution}};
  \draw [stealth-] (575,150) -- (490,150) node [left]{\footnotesize{CS-A contribution}};
  \node[rectangle, dashed,
    draw = blue, thick,
    minimum width = 3.25cm, 
 minimum height = 0.8cm] (r.southwest) at (255,10) {};
 
\end{axis}
\end{tikzpicture}
\caption{Red: Distribution of a set of the sums of energy depositions. Black: Corresponding theoretical distribution obtained using the estimated values $E_{0est}$, $\sigma_{est}$, $p_{CSest}$, and $p_{Aest}$ from the EM algorithm.}
\label{fig:res_EM}
\end{figure}

The EM algorithm is performed over ten iterations with grid searches of domains $[0.5, 1]$ MeV and $[10^{-4}, 10^{-1}]$ MeV for $E_0$ and $\sigma$ with respective step-sizes $0.02$ MeV and $0.002$ MeV. The algorithm converges quickly towards the closest values of the grid to the ground truth that is $E_{0est} = 0.6667$ MeV and $\sigma_{est}=0.0286$ MeV in three iterations and remains constant until the algorithm stops. The proportions of CS-CS events and CS-A events are also well estimated, as we obtained $p_{CS est} = 0.1697$ (and $p_{A} = 1-p_{CS} = 0.8303$). The estimated distribution of the sum of the energy depositions $f_{est} =$ $ f(\tilde{E}|E_{0,est}, p_{CSest}, p_{Aest}, \sigma_{est}, CS_{1n})$ can then be calculated as follows
\begin{equation}
    f_{est} = p_{CS,est} f(\tilde{E}|E_{0est}, CS_{1n}, CS_{2n}, \sigma_{est}) + p_{Aest} f(\tilde{E}|E_{0est}, CS_{1n}, A_{2n}, \sigma_{est})
    \label{eq:f_exp_em}
\end{equation}
using \eqref{eq:f_En_noisy}. For the example considered, $f_{est}$ is represented by the black line on Figure \ref{fig:res_EM}. Then, for each event, the nature of the second interaction can be estimated. In the present example, only one event over the whole set has been misclassified. This corresponds to an event whose sum of the energy depositions is equal to $0.9564$MeV, which is clearly out of the range of the rest of the energy depositions of the set and thus corresponds to an outlier. 

Finally, the performance of the EM algorithm was evaluated on smaller sets of data, from sets with ten events. The EM algorithm gives already similar estimations for $E_0$ and $\sigma$ from sets with 10 events (for low fractions of outliers) and the events are also generally well classified according to the nature of their second interaction.

\subsection{Estimation of the position of the source(s)}
The results obtained from the Gibbs sampler (Algorithm  \ref{algo:gibbs}) are presented in this paragraph. It is assumed that the EM algorithm has been used first to estimate $E_0$ and the nature of the second interaction for the considered set of events.  First, results from data emitted from one source are discussed. Several positions for the source are considered in order to evaluate the performance and the accuracy of the algorithm according to the localisation on the sphere. In each experiment, ten events including potential outliers $(N=10)$ are processed. Then, results for the two-source localization problem are presented. In this case, the Gibbs sampler deals with $N=20$ events including potential outliers. The sources are assumed to have the same intensity, thus about ten events are issued from photons emitted by each source. 

In both one and two-source localisation problems, 10000 iterations including a burn-in period of 2000 iterations were performed per experiment. Each experiment is repeated 50 times using different data to compute summary statistics of the performance of the proposed estimators.

\subsubsection{Experiments with one source to localise}
The objective of these experiments is to evaluate the performance of the algorithm according to the position of the source on the sphere. Experiments were carried out for several positions $\boldsymbol{r}_0(\alpha, \beta)$ localised on a quarter of the sphere, and $\alpha$ and $\beta$ respectively stand for the longitude and the latitude on the sphere. More precisely, simulations are performed at positions $(0^\circ, 0^\circ)$, $(0^\circ, 30^\circ \text{N})$, $(0^\circ, 60^\circ \text{N})$, $(30^\circ\text{E}, 0^\circ)$, $(60^\circ\text{E}, 0^\circ)$, $(90^\circ\text{E}, 0^\circ)$, $(90^\circ\text{E}, 30^\circ\text{N})$, $(90^\circ\text{E}, 60^\circ\text{N})$, $(120^\circ\text{E}, 0^\circ)$ and  $(150^\circ\text{E}, 0^\circ)$ and presented in the next paragraph using the color code of Table \ref{table:one_source_color_Code}. The expected errors on the rest of the sphere can then be deduced from the proposed experiments by leveraging the symmetries of the imager. 

\begin{table}[h!]
\centering
\begin{tabular}{|c|p{0.12\linewidth}|p{0.12\linewidth}|p{0.12\linewidth}|p{0.12\linewidth}|p{0.12\linewidth}|p{0.12\linewidth}|} 
\hline
Color & {\cellcolor{rouge}}&{\cellcolor{orange}}&{\cellcolor{yellow}}&{\cellcolor{lime}}&{\cellcolor{green}}\\
\hline
Location & $(0^\circ, 0^\circ)$& $(0^\circ, 30^\circ \text{N})$& $(0^\circ, 60^\circ \text{N})$& $(30^\circ\text{E}, 0^\circ)$& $(60^\circ\text{E}, 0^\circ)$\\
\hline
Color & {\cellcolor{cyan}}&{\cellcolor{blue}}&{\cellcolor{purple}}&{\cellcolor{magenta}}&{\cellcolor{gray}}\\
\hline
Location & $(90^\circ\text{E}, 0^\circ)$& $(90^\circ\text{E}, 30^\circ\text{N})$& $(90^\circ\text{E}, 60^\circ\text{N})$& $(120^\circ\text{E}, 0^\circ)$ &$(150^\circ\text{E}, 0^\circ)$\\
\hline
\end{tabular}
\caption{Color code used for the experiments with one source.}
\label{table:one_source_color_Code}
\end{table}

\input{figures_worst_best_1source}

At the end of each experiment, the last $8000$ samples for $\boldsymbol{r}_0$ are used to compute summary statistics. The distribution of these samples is calculated by kernel density estimation. In order to measure the group direction, we also compute the spherical mean $\widehat{\mu}(\alpha, \beta)$ of these samples
\begin{equation}
    \widehat{\mu}(\alpha, \beta) = R \, \frac{\sum_{t=t_{in}}^T \vec{u}^{(t)}(\alpha, \beta)}{\left| \sum_{t=t_{in}}^T \vec{u}^{(t)}(\alpha, \beta)\right|}, 
\end{equation}
where $ \vec{u}^{(t)}$ is the unit direction of the position $\boldsymbol{r}_0^{(t)}$ at iteration $(t)$. $t$ refers to the considered iterations and in the present experiments, $t\in [t_{in},T] = [2001, 10000]$. The bias of the solutions  from the Gibbs sampler can be measured by calculating the geodesic distance between the true position of the source and the obtained spherical mean. 

\paragraph{Simulation results.}

Figures \ref{fig:res_best_worst1} and \ref{fig:res_best_worst2} contains two instances of the obtained distributions from the Gibbs sampler for each considered position for the true sources. The true positions of the sources are represented by stars $\boldsymbol{\star}$ on the different plots, and the region in colour corresponds to the distribution of the accepted samples $\{\boldsymbol{r}_0^{(t)}\}$. In order to give an overview on the whole set of results, it has been chosen to depict on the left-hand side the $5$-th best obtained result and on the right-hand side the $5$-th worst obtained result, in terms of the geodesic distance $d(\boldsymbol{r}_0^{true}, \hat{\mu})$ between the mean of the distributions $\hat{\mu}$ and the true position of the source $\boldsymbol{r}_0^{true}$. The mean of the distribution is represented on each plot by a point {\Large$\cdot$}. The plots of these figures contain also the result obtained by back-projection (BP), depicted with crosses $\boldsymbol{\times}$. The BP result corresponds to the point of highest intensity on the back-projection image. 

\input{boite_moustache.tex}

Furthermore, some metrics regarding the distributions of these sets of geodesic distances have been reported in Figure \ref{fig:boite_moustache} with box-and-whisker plots. These box-plots represent a data summary based on the following values: (1) the median (shown by the line dividing the box into two parts) is the mid-point of the set of distances, (2) the first quartile $Q_1$ (shown by the left line of the box) is the median of the lower half of the set, (3) the third quartile $Q_3$ (shown by the right line of the box) is the median value of the higher half of the set, (4) the minimum $Q_0$ (shown at the end of the left whisker) is the lowest data point excluding potential divergent results and (5) the maximum $Q_4$ (shown at the end of the right whisker) is the highest point excluding potential divergent results. The minimum $Q_0$ and maximum $Q_4$ are calculated as follows 
\begin{equation}
    Q_0 = \text{max}(\text{min}(d(\boldsymbol{r}_0^{true}, \hat{\mu})), Q_1 - 1.5 \,\text{IQR}),
\end{equation}
\begin{equation}
    Q_4 = \text{min}(\text{max}(d(\boldsymbol{r}_0^{true}, \hat{\mu})), Q_3 + 1.5 \,\text{IQR}),
\end{equation}
where $\text{min}(d(\boldsymbol{r}_0^{true}, \hat{\mu}))$ and $\text{max}(d(\boldsymbol{r}_0^{true}, \hat{\mu}))$ are the minimum and maximum value of the related set of geodesic distances and $\text{IQR}$ stands for the inter-quartile range, that is $\text{IQR}=Q_3-Q_1$. Box-plots are depicted at each position of the source for both Gibbs and BP results for comparison; with filled colored boxes for the Gibbs sampler and boxes with hatches for BP.  

Moreover, $\alpha$-confidence regions were estimated. The $\alpha$-regions correspond to the part of the distributions of the samples $\{\boldsymbol{r}_0^{(t)}\}$ includes the spherical mean of the samples plus or minus $\alpha/2$. The mean observed credible level for a source localised randomly on the sphere is reported in Figure \ref{fig:conf_mean}, for $\alpha = \{0, 10, 20, \ldots, 100\}\%$. The radiation localisation algorithm can be considered as accurate if the observed credible level (in red) follows the theoretical level (in black). It is not possible to perform similar statistics from the BP results, since BP results consist of single values and not regions. 

\begin{figure}[ht!]
\centering 
  \scalebox{1}{\begin{tikzpicture}
\pgfplotsset{
    xmin = 0,
    xmax = 1,
    width=0.8\textwidth,
    height=0.4\textwidth,
    }
\begin{axis}[
    xtick = {0, 0.2, 0.4, 0.6, 0.8, 1},
    grid= major ,
    xlabel = Credible level $(1-\alpha)$,
    ylabel = Observed credible level,
    width=0.8\textwidth,
    legend pos=outer north east,
    legend entries={ Empirical , Theoretical}
]
  \addplot[red, mark =*] table [x=xm,y=mean_sphere] {intervalle_conf/mean_conf.txt};
  \addplot[black] table [x=xm,y=xm] {intervalle_conf/mean_conf.txt};
\end{axis}
\end{tikzpicture}}   
\caption{Observed credible level for a source localised randomly on the sphere (in red). The black line corresponds to the theoretical level.   
}
\label{fig:conf_mean}
\end{figure}

\begin{figure}[!ht]
    \centering
    \includegraphics[width = \linewidth, trim={27cm 7cm 3cm 2cm}, clip]{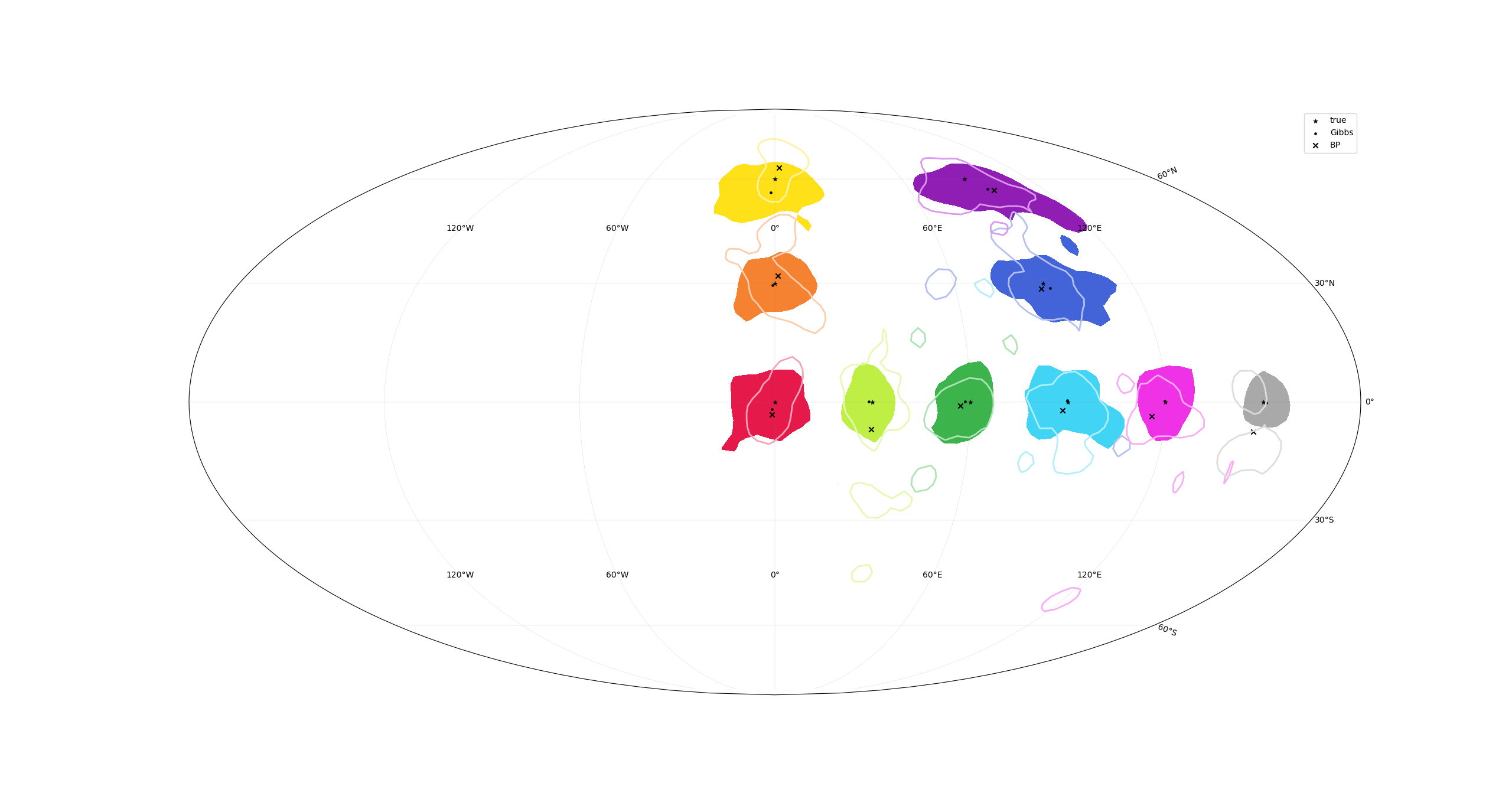}
    \caption{Distribution of the obtained spherical means over the 50 simulations for each source position. Each colour refers to a particular position for the source, according to the same colour code used in Figures \ref{fig:res_best_worst1}, \ref{fig:res_best_worst2}, \ref{fig:boite_moustache} and \ref{fig:conf_mean}. The coloured filled regions are the distributions of the means obtained from the full Bayesian algorithm. The contour areas correspond to the distributions of the means obtained by BP.}
    \label{fig:res_one_source}
\end{figure}

Finally, the distributions of the obtained spherical means over the 50 simulations for each source position are reported on Figure \ref{fig:res_one_source} for the Gibbs sampler and BP.  

\paragraph{Discussions.} The examples of distributions depicted on Figures \ref{fig:res_best_worst1} and \ref{fig:res_best_worst2} show first the convergence of the Gibbs sampler on quite small regions. Even in the worst case results, the regions are quite close to the true source position. The observed bias seems more important when the source to localise is close to the pole and this is confirmed by the observed statistics of Figures \ref{fig:boite_moustache}, \ref{fig:conf_mean} and \ref{fig:res_one_source}. The Gibbs results seems to be more consistent over the whole set of simulations, with box-plot lengths (Figure \ref{fig:boite_moustache})  in the same range while those from BP results are more variable. This is also underlined by the obtained distributions of the means (see Figure \ref{fig:res_one_source}) where extra small spots break away from the main distribution regions on BP results. Finally, the measured uncertainty of the Gibbs results remains satisfactory, since it overall follows the theoretical credible level (Figure \ref{fig:conf_mean}).  

\subsubsection{Experiments with two sources to be localised}

\begin{figure}
    \centering
    \begin{tabular}{cc}
     \includegraphics[width =0.48\linewidth, trim={9.5cm 6cm 1cm 6cm}, clip]{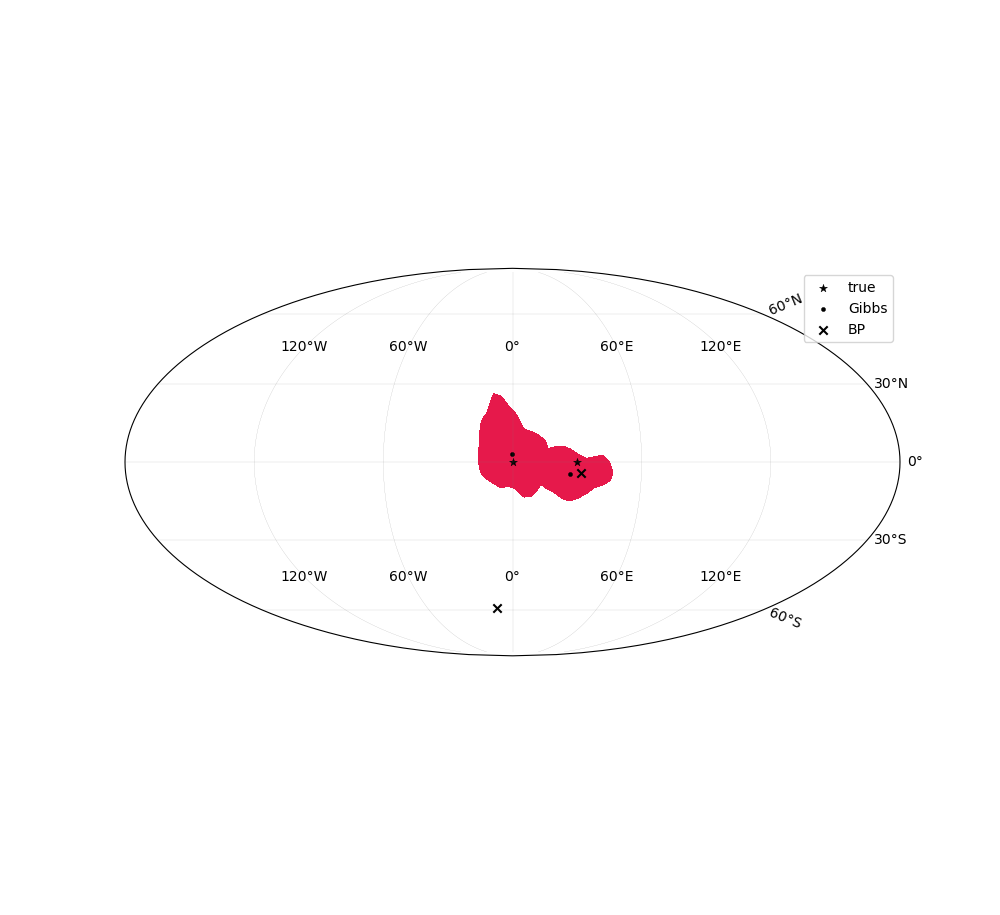} & \includegraphics[width =0.48\linewidth, trim={9.5cm 6cm 1cm 6cm}, clip]{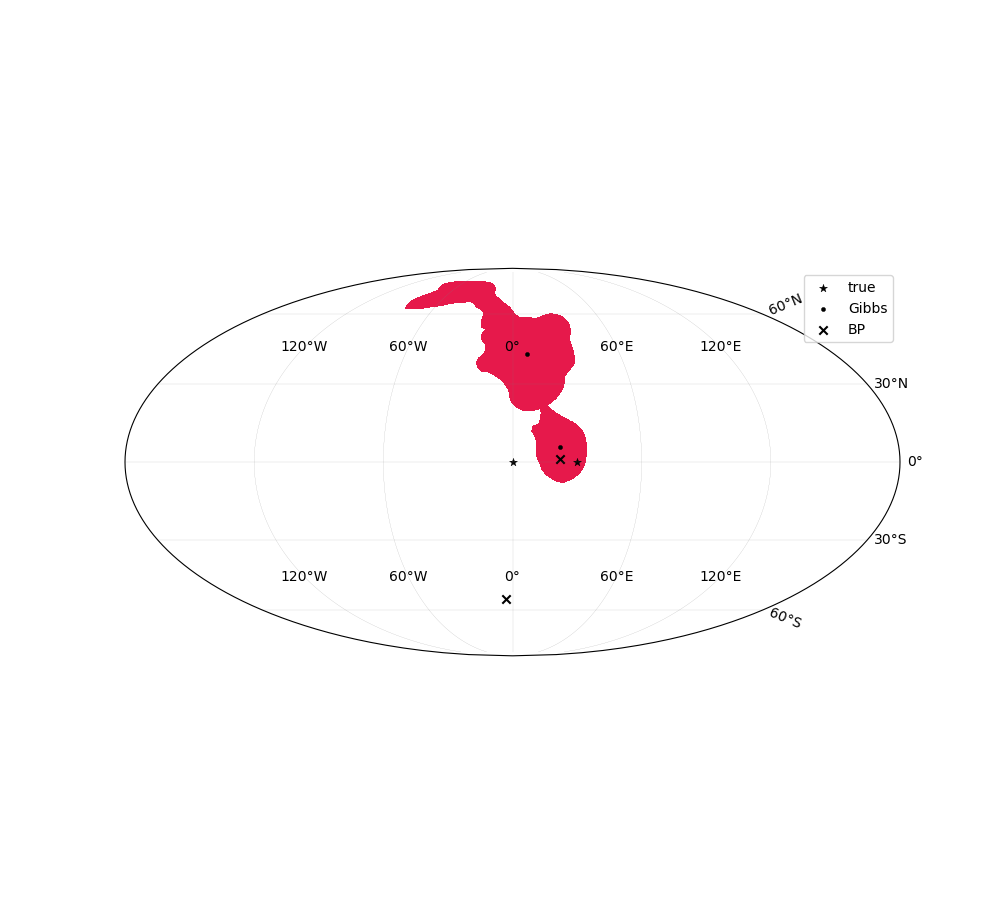}\\
     \multicolumn{2}{c}{\footnotesize (a) Sources at $(0\degree, 0\degree)$ and $(0\degree, 30\degree\text{E})$}\\
      \includegraphics[width =0.48\linewidth, trim={9.5cm 6cm 1cm 6cm}, clip]{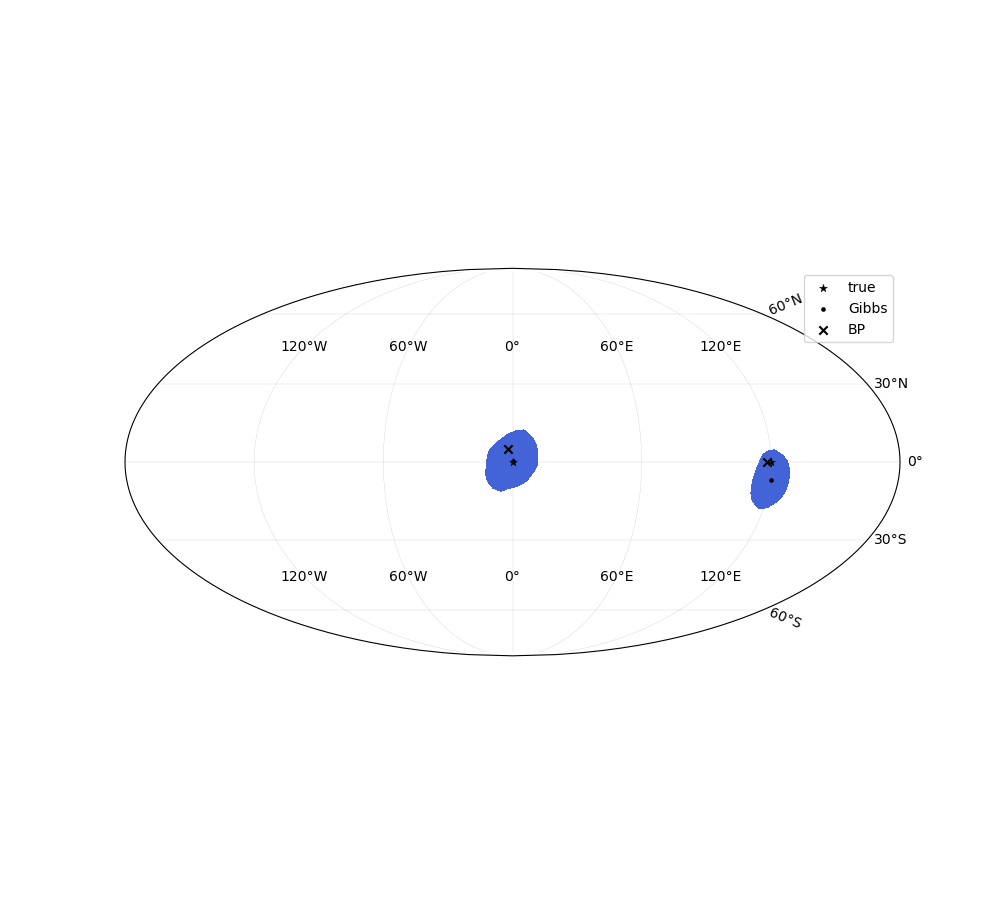} & \includegraphics[width =0.48\linewidth, trim={9.5cm 6cm 1cm 6cm}, clip]{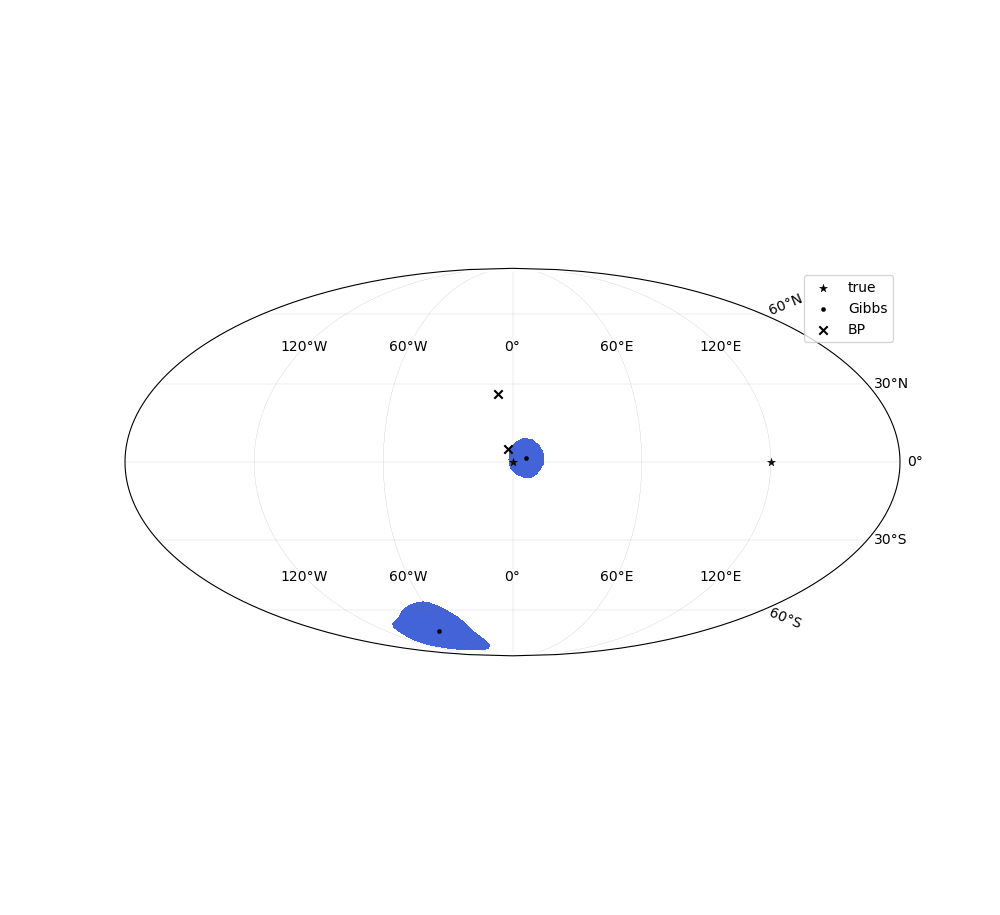}\\
       \multicolumn{2}{c}{\footnotesize (b) Sources at $(0\degree, 0\degree)$ and $(0\degree, 120\degree\text{E})$}\\
    \end{tabular}
     \caption{The distributions of the $5$th best and the $5$th worst result among the obtained distributions are respectively depicted on the left and right columns. }
    \label{fig:res2sources_best_worst}
\end{figure}

\paragraph{Simulation results.} Two experiments for the two-source localization case are now presented. The first situation involves two sources quite distant from each other (e.g. $(0^\circ, 0^\circ)$ and $(0^\circ, 120^\circ\text{E})$), and the second considers two sources close (e.g. $(0^\circ, 0^\circ)$ and $(0^\circ, 30^\circ\text{E})$). At each experiment, the output of the Gibbs consists of two chains of 8000 samples representing the estimated positions of the sources $\{\boldsymbol{r}_0^{(1,t)}, \boldsymbol{r}_0^{(2,t)}\}_{t=t_{in}}^T$. However, since the sources are supposed to share the same intensity, the problem is perfectly symmetric, and it is not possible to know which source each set of sample will estimate. Furthermore, the chains of the samples can swap during iterations, especially when the sources are close to each other. In order to de-entangle the sources, the two Markov chains obtained from the Gibbs sampler are post-processed performing a K-means clustering. Figure \ref{fig:res2sources_best_worst} gives an overview of the obtained results, with the $5$-th best result (in terms of geodesic distance) on the left-hand side column and the $5$-th worst result on the right-hand side. The distributions of the experiments where the two sources are close are depicted in red, and those corresponding where the sources are distant are in blue. The true positions of the sources are represented by stars, the mean of the distributions by points, and the BP result is depicted using crosses. Here, the BP results correspond to the two (sufficiently distant) points of highest intensity on the back-projection image. 


\input{boite_moustache_far_v2}

\begin{figure}[!ht]
    \centering
\begin{tabular}{cc}
    \includegraphics[width =0.49\linewidth, trim={9.5cm 6cm 1cm 6cm}, clip]{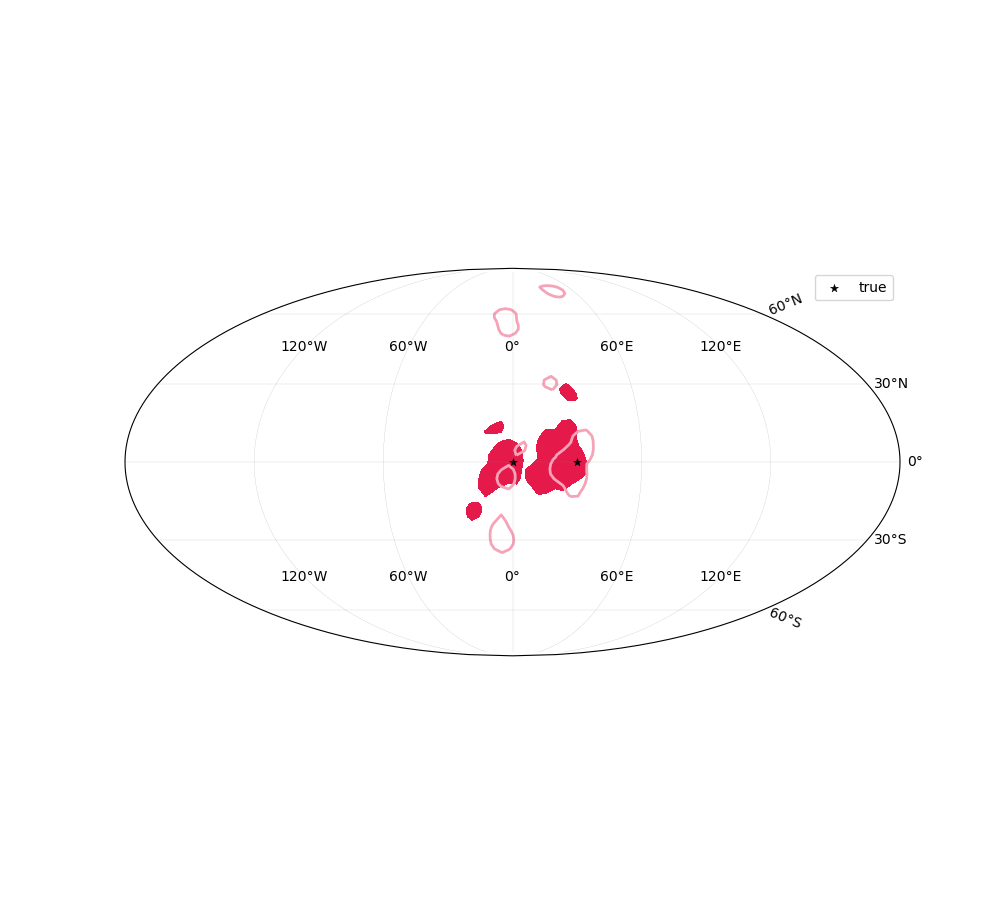} & \includegraphics[width =0.49\linewidth, trim={9.5cm 6cm 1cm 6cm}, clip]{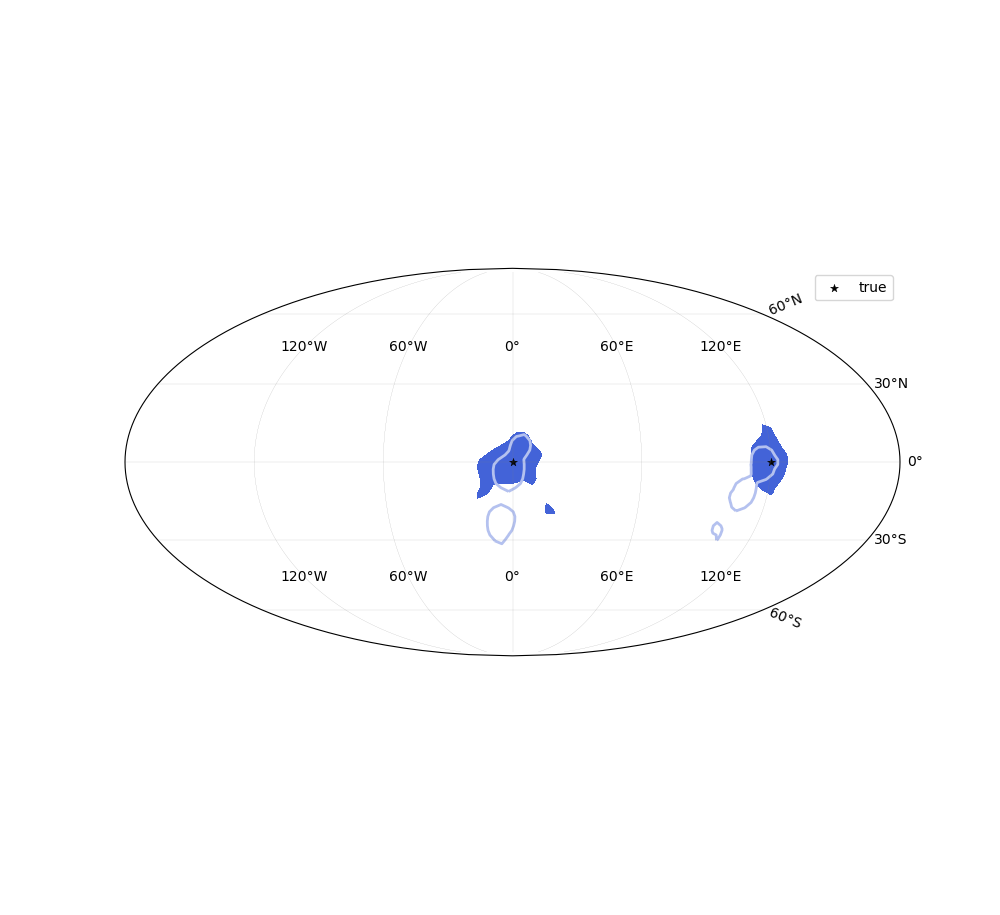}\\{\footnotesize (a)} & {\footnotesize (b)}
    \end{tabular}
     \caption{ Distributions of the results obtained for the two source localization problem. The coloured filled regions are the distributions of the means obtained from the full Bayesian algorithm.  The contour regions are the distributions of the results obtained from the BP algorithm. $\boldsymbol{\star}$ stands for the true positions of the source. (a) Case where the two sources are close to each other ($(0^\circ, 0^\circ)$  and $(0^\circ,30^\circ \text{E})$) (b) Case where the two sources are distant ($(0^\circ,0^\circ)$ and $(0^\circ,120^\circ \text{E})$).}
    \label{fig:res2sources}
\end{figure}

The distributions of the geodesic distances between the mean of the samples and the true positions of the sources are reported in Figure \ref{fig:boite_moustache_2sources} with box-and-whisker plots, using filled coloured boxes for the Gibbs sampler and boxes with hatches for BP. The distributions of the spherical means obtained from the Gibbs sampler are reported on Figure \ref{fig:res2sources}. For comparison purposes, the distributions of the results obtained from BP are also presented on the same figure.



Finally, the obtained means over the 50 simulations are gathered all together to form the distributions presented on Figure \ref{fig:res2sources}. These distributions are compared with the distributions of the positions obtained from the BP algorithm. 

\paragraph{Discussions.} The results obtained from our experiments with two sources are more challenging to draw conclusions from. In most experiments, the Gibbs sampler is effective, and the accepted samples are close to the position of the true sources, even if the initialization (which corresponds to the BP result) is far from the true solutions (see the results presented in the column of the left-hand side of Figure \ref{fig:res2sources_best_worst}).  Nevertheless, in some experiments, some chains of the Gibbs sampler do not converge to the rights positions of the sources, as shown in the results of the right-hand side column of Figure \ref{fig:res2sources_best_worst}. In the presented experiments, one chain is close to one of the sources to be localized, while the other diverged far away. Letting the algorithm run for more iterations could have produced a better result, but the choice of 10,000 iterations was made for computational reasons.  

The results obtained from the Gibbs sampler are on average better and more consistent than those computed from BP, as illustrated by the distribution of the geodesic distance to the true positions in Figure \ref{fig:boite_moustache_2sources}. The domains obtained by BP are always either twice as wide as those obtained with the Gibbs or in a narrower domain, but whose values are greater than most of the errors made by the Gibbs sampler. Furthermore, Figure \ref{fig:res2sources} shows that, while the obtained distributions of the means values are quite well concentrated for the Gibbs algorithm, the distributions obtained from the BP results consist of smaller regions, sometimes far from the true positions of the sources, especially with the two sources to localize are close. The Gibbs sampler is thus more reliable than the BP algorithm. 

These positive results need to be nuanced, however, by the discrepancies observed in some cases and mentioned earlier in this paragraph.

\section{Concluding remarks}
\label{sec:concluding}
This work presented a statistical forward observation model for a variant of the Compton Camera, called Compton Imager. This instrument, whose objective is to detect radioactive sources and determine their energy, is currently under development at the University of Illinois Urbana Champaign. 

The energy estimation problem was formulated in terms of a maximum likelihood estimation problem. The simulations showed excellent results in a reduced computational time. The presented algorithm supposed sources sharing the same energy. A natural  extension of this algorithm could be considered to involve mono-energetic sources of various energies. 

For the localization problem, the developed forward model can be embedded in any EM scheme to perform Bayesian inversion. In the planned experiments to be carried out using this system, it is assumed that the reconstruction method can only access a few numbers of photons, and traditional EM algorithms as well as back projection techniques used to perform poorly in such situations. Another type of algorithm based on a Gibbs sampler was instead investigated in this work. Numerical simulations were performed first with one source to localize to evaluate the performance of the algorithm. The obtained results proved the convergence of the chains to the true positions, and more accurately than the back-projection algorithm. The algorithm was then put to the test with the problem of locating two sources, which is a particular challenge when the sources to be located are close together. The obtained result were encouraging, however, failed to converge in some experiments. One possible explanation could be that the Gibbs sampler needs more iterations to converge. Parts of the forward model which involve the computation of distances are indeed computationally expensive. This drawback was partly solved in the proposed implementation with the use of look-up-tables, other distance calculations need nevertheless to be updated at each iteration. Some calculation methods which approximate the computation of these distances, as well as surrogate forward models were experimented in the Gibbs scheme, however this resulted in larger biases for the obtained distributions. Future works will include the investigation of less computationally expensive algorithms to perform this Bayesian inversion.   

\section*{Acknowledgements} 
This work was supported by the UK Research and Innovation (UKRI) Engineering and Physical Sciences Research Council (EP- SRC) grants EP/V006134/1 , EP/V006177/1 and EP/T007346/1, the UK Royal Academy of Engineering under the Research Fellowship Scheme (RF201617/16/31) and by the Department of Energy National Nuclear Security Administration through the Nuclear Science and Security Consortium under Award Number DE-NA0003996.

\appendix

\section{Numerical computation of \texorpdfstring{$f(\mathbf{\Theta}_1|\mathbf{r}_0, E_0)$}{}}
\label{sec:calculation_solid_angle}
We consider a source of position $\boldsymbol{r}_0$ emitting radiation at energy $E_0$. The objective is to compute numerically $f(\boldsymbol{\Theta}_1|\mathbf{r}_0, E_0)$ \eqref{eq:solid_angle} modelling the probability for an emitted photon in direction $\boldsymbol{\Theta}_1$ to have an interaction with one of the sensors
\begin{equation*}
    f(\boldsymbol{\Theta}_1|\boldsymbol{r}_0, E_0)  = \frac{1 - \exp\left( -\mu_{E_0}d_{1,\text{max}}(\boldsymbol{r}_0, \boldsymbol{\Theta}_1)\right)}{\displaystyle\int_{\boldsymbol{\Theta}}\left(1 - \exp\left( -\mu_{E_0}d_{1,\text{max}}(\boldsymbol{r}_0, \boldsymbol{\Theta}_1)\right) \right)d\boldsymbol{\Theta}}.
\end{equation*}
A rejection-sampling algorithm is considered. The idea is to generate uniformly numerous unit vectors $\boldsymbol{\Theta}_{1\ell}^\star$ of origin $\boldsymbol{r}_0$ so that a line in that direction would intercept at least one of the sensors of the Compton imager. Generating those vectors uniformly in every direction would lead in reality to extra time-consuming computations, we chose instead to generate uniformly these vectors in the smallest cone that encompasses the whole imager. The generated vectors which do not pass at least through one sensor are cancelled during the rejection step of the algorithm. For those which reach at least one sensor, the maximal distance $d_{1\ell,lim}(\boldsymbol{\Theta}_{1\ell}^\star)$ is then computed. This distance corresponds to the maximal distance that can be travelled inside the detectors in that direction.
Then, a potential travelling distance for the photon $d_{1\ell}$ is generated according to the Beer law, picking a uniform value $v\in[0,1)$
\begin{equation}
    d_{1\ell} = - \frac{\ln(1-v)}{\mu_{E_0}}.
\end{equation}

The proposed direction is finally accepted if $d_{1\ell}$ is less than $d_{1\ell,lim}(\boldsymbol{\Theta}_{1\ell}^\star)$. Algorithm \ref{algo:rejection_sampling} summarizes the main steps of this rejection sampling algorithm. 

\IncMargin{1em}
    \begin{algorithm}[!ht]
        \SetAlgoLined
       Generate uniformly $L$ numerous unit vectors $\boldsymbol{\Theta}_{1\ell}^\star, \ell \in \{1, \dots, L\}$ of origin $ \boldsymbol{r}_0$ whose direction may correspond to an interaction with the Compton imager. 
       
        \For{$\ell = 1$ \KwTo $L$}{
       Compute the limit distance $d_{1\ell,lim}(\boldsymbol{\Theta}_{1\ell}^\star)$ that is possible to travel inside the detectors in the direction of $\boldsymbol{\Theta}_{1\ell}^\star$. 

       Pick a value $d_1$ according the Beer Law and the attenuation coefficient of the crystal. 

       Accept the proposed direction  $\boldsymbol{\Theta}_{1\ell}^\star$ if $d_1<d_{1\ell, lim}(\boldsymbol{\Theta}_{1\ell}^\star)$, otherwise reject the proposal.}
    \caption{Rejection sampling algorithm for the computation of $f(\boldsymbol{\Theta}_1|\boldsymbol{r}_0, E_0)$ }
    \label{algo:rejection_sampling}
    \end{algorithm}
\DecMargin{1em}

From the set of accepted direction samples, a spherical kernel density estimation \cite{HandleyKDE} is then performed to obtain the distribution of interest. According to our experiments, this rejection sampling algorithm is time-consuming and represents around $9$ seconds per source position. It has been chosen to pre-compute the distributions for a set of $2563$ source positions uniformly placed on the sphere and used a geodesic nearest neighbour interpolation during the sampling of the posterior distribution. The computation time was reduced to $\sim 0.7$ seconds while the maximal error observed was around $5\%$. 

\section{Derivation of \texorpdfstring{$f(E_1|E_0, CS_1)$}{}}
\label{app:f(E_1sCS_E0)}
Let $\omega(E_1, E_0)$ be the scattering angle corresponding to an energy deposition of $E_1$ at first interaction. Its value is computed using the Compton formula \eqref{eq:Compton_formula}. The probability for a photon of initial energy $E_0$ to be Compton scattered with an angle $\omega$ is given by 
\begin{equation}
    f(\omega(E_1, E_0)|E_0, CS_1)=\frac{\displaystyle\int_{\phi=0}^{2\pi}\left.\frac{d\sigma}{d\Omega}\right|_{\omega(E_0, E_1)}\sin(\omega(E_0, E_1)) d\phi}{\displaystyle\int_{\theta=0}^{\pi}\int_{\phi=0}^{2\pi}\left.\frac{d\sigma}{d\Omega}\right|_{\theta}\sin(\theta)d\phi d\theta },
    \label{eq:f_E1sCS-E0}
\end{equation}
where $\left.\frac{d\sigma}{d\Omega}\right|_{\theta}$ is the differential Compton cross section at scattering angle $\theta$ \cite{klein1929streuung}
\begin{equation}
    \left.\frac{d\sigma}{d\Omega}\right|_{\theta} = \frac{r_e^2}{2}\lambda(\theta)^2 \left[\lambda(\theta)+\frac{1}{\lambda(\theta)}-\sin(\theta)^2\right],
\end{equation}
where $r_e$ is the classical electron radius  ($\sim2.8179$~fm) and $\lambda(\theta)$ is  the ratio of photon energy after and before the collision
\begin{equation}
    \lambda(\theta) = \left(1+\frac{\displaystyle E_0}{\displaystyle m_ec^2}(1-\cos(\theta))\right)^{-1}.
\end{equation}
With the change of variables 
\begin{equation*}
    \theta \longleftarrow \arccos{\left( 1 - \frac{mc^2}{E_0}\left(\frac{E}{E_0-E}\right)\right)}, 
\end{equation*}
the obtained expression of $f(E_1|E_0, CS_1)$ is \eqref{eq:fE1|E0,I1=CS}. The integral of the denominator has also an explicit derivation, which corresponds to 
\begin{equation}
    F(E_0 - E_0/(1+2E_0/mc^2)) - F(0),
\end{equation}
where the function $F$ is defined as
\begin{equation}
    F: E\longrightarrow \frac{mc^2}{E_0^2}\left( -\frac{E^2}{2E_0} + \left(1+\frac{mc^2}{E_0}\right)^2 E + \left(2\left(1+\frac{mc^2}{E_0}\right)mc^2-E_0\right)\ln{(E_0-E)}  + \frac{(mc^2)^2}{E_0-E} \right).
\end{equation}

\section{Details about Algorithm \ref{algo:gibbs}}
\label{sec:Gibbs_details}

\subsection{Initialisation}
\subsubsection{Hyper-parameters of the full Bayesian algorithm}
The hyper-parameters of the Dirichlet distribution $\{\alpha_k\}_{k=0}^K$ modelling the prior distribution of the related intensities $\{w_0^{(k)}\}_{k=1}^K$ were set to $(1, 50)$ for simulations with one source and $(1, 50, 50)$ in case of two sources. This models photons mostly incoming from source(s) (of equal intensity) with a very few outliers. The parameter $a$ of the exponential function \eqref{eq:delta_approx} is set to $400$. Finally, the concentration parameter of the Von-Mises distribution resp. $80$. Both $a$ and $\kappa$ have been chosen arbitrarily high to model Dirac delta distributions. 

\subsubsection{Initial values of the random variables}
The uncertainty on position of the Compton Imager is assumed to be $3$ mm on $x$ and $y$-coordinates and $5$ mm on $z$. In terms of energy, we suppose an uncertainty of $0.1$ keV. The initial values $\sigma_{xy}^{(0)}$, $\sigma_z^{(0)}$ and, $\sigma_E^{(0)}$ are set accordingly and the domains of their respective prior distribution are defined to include these values. The first values for $\{\boldsymbol{r}_{1n}^{(0)}, E_{1n}^{(0)}, \boldsymbol{r}_{2n}^{(0)}, E_{2n}^{(0)}\}_{n\in [|1, N|]}$ are then generated using Gaussian distributions of the corresponding standard deviations $\sigma_{xy}, \sigma_{z}, \sigma_{E}$. 

In the one-source localisation problem, $w_0^{(1, t=0)}$ is set to $0.99$.  The initial position for the source $\boldsymbol{r}_0^{(1,t = 0)}$ set to point having the maximal intensity on the back-projection image. The set of $\{\boldsymbol{r}_{0n}^{(t=0)}\}_{n=1}^N$ is then generated according to a Von-Mises distribution of mean $\boldsymbol{r}_0^{(1,t=0)}$ and concentration parameter $100$. 

In the two-sources case, both $w_0^{(1, t=0)}$, $w_0^{(2, t=0)}$ are set to $0.49$ and we used a local maxima algorithm based on mathematical morphology to determine the two \textit{main} peaks of the back-projection image. This gives the positions of $\boldsymbol{r}_0^{(1,t = 0)}$ and $\boldsymbol{r}_0^{(2,t = 0)}$. Then, for each event $n$ and source $(k)$, the likelihood $ f(\tilde{\boldsymbol{Z}_n|\boldsymbol{r}_0^{(k,t = 0)}, E_0, CS_{1n}, X_{2n})}$ \eqref{eq:likelihood} is computed and each event $n$ is assigned to the most probable source $\boldsymbol{r}_0^{(k,t = 0)}$. The initial value of each $\boldsymbol{r}_{0n}$ is then generated according to a Von-Mises distribution of mean equal to the relevant $\boldsymbol{r}_0^{(k,t = 0)}$ and concentration parameter $100$.

\subsection{Sampling steps }
This paragraph presents the sampling steps of Algorithm \ref{algo:gibbs} and the expressions of the sampled conditionals are derived. These are straightforwardly obtained using basics of conditional probability and keeping only the parameters that depend on the variable of interest. Each sampling step is then carried out using a Metropolis Hastings scheme. For the sake of completeness, a generic version of this Monte Carlo technique for sampling an arbitrary random variable $x$ from a density $\pi$ and a proposal Markov kernel $q$ is presented in Algorithm \ref{algo:mh}. Then, $x$ has to be replaced by the sampling variable of interest along with their corresponding density $\pi$ and a proposal Markov kernel $q$. 

\IncMargin{1em}
    \begin{algorithm}[!ht]
        \SetAlgoLined
        \SetKwHangingKw{KwInit}{Initialisation:} 
        \KwInit{Set an initial value $x^{(0)}$, and $T$ the number of iterations}\
        \For{$t = 1$ \KwTo $T$}{
        Generate a candidate $x^\star$ from a proposal density $q(\cdot|x^{(t-1)})$.\,
        
        Compute the acceptance probability 
        $\rho^{(t)} = \min\left(1, \frac{\pi(x^\star)}{\pi(x^{(t-1)})}\frac{q(x^{(t-1)}|x^\star)}{q(x^\star|x^{(t-1)})}\right)$.

        Generate $u^{(t)} \sim \mathcal{U}(0,1)$.
        
        Accept the proposal and take $x^{(t)} = x^\star$ if $u^{(t)} < \rho^{(t)}$, otherwise reject the proposal and set $x^{(t)} = x^{(t-1)}$.
        }
    \caption{Metropolis Hastings algorithm}
    \label{algo:mh}
    \end{algorithm}
\DecMargin{1em}


\paragraph{Sampling $(\boldsymbol{r}_{1n} | \boldsymbol{r}_{0n}, E_0,  CS_{1n}, E_{1n}, \boldsymbol{r}_{2n},  X_{2n}, E_{2n}, \tilde{\boldsymbol{r}}_{1n}, \sigma_{xy}, \sigma_z)$.} The proposal density $q(\cdot|\boldsymbol{r}_{1n}^{(t-1)})$ is a multivariate truncated Gaussian distribution of mean $\boldsymbol{r}_{1n}^{(t-1)}$ to stay in the same sensor. It is also checked that the distance $d_1^\star$ between $\boldsymbol{r}_{0n}$ and the new proposal $\boldsymbol{r}_{1n}^\star$ is between $0$ and $d_{1lim}^\star$ excluded. The coordinates of $\boldsymbol{r}_{1n}^{(t-1)}$ are supposed to be uncorrelated, hence we chose a diagonal correlation matrix, whose non-zero elements evolve during the algorithm to keep the acceptance rate between $40\%$ and $60\%$. The conditional density $\pi(\boldsymbol{r}_{1n})$ is
\begin{equation}
     \pi(\boldsymbol{r}_{1n} ) \propto f( \boldsymbol{r}_{1n}, E_{1n}, \boldsymbol{r}_{2n}, E_{2n}  | \boldsymbol{r}_{0n}, E_0, CS_{1n}, X_{2n}) f(\tilde{\boldsymbol{r}}_{1n}|\boldsymbol{r}_{1n}, \sigma_{xy}. \sigma_z),     
\end{equation}
where $f( \boldsymbol{r}_{1n}, E_{1n}, \boldsymbol{r}_{2n}, E_{2n}  | \boldsymbol{r}_{0n}, E_0, CS_{1n}, X_{2n}) $ is the probabilistic model for data acquisition \eqref{eq:likelihood} and $f(\tilde{\boldsymbol{r}}_{1n}|\boldsymbol{r}_{1n}, \sigma_{xy}. \sigma_z)$ is the distribution modelling noise on positions \eqref{eq:rnoise_s_r}. 

\paragraph{Sampling $(\boldsymbol{r}_{2n} | \boldsymbol{r}_{0n},\boldsymbol{r}_{1n}, E_{1n}, \tilde{\boldsymbol{r}}_{2n})$.} The proposal $q(\cdot|\boldsymbol{r}_{2n}^{(t-1)})$ is defined similarly as $q(\cdot| \boldsymbol{r}_{1n}^{(t-1)})$. It has been checked that the new proposal $\boldsymbol{r}_2^\star$ involves a distance $d_2^\star$ between $\boldsymbol{r}_{1n}^{(t)}$ and $\boldsymbol{r}_{2n}^\star$ is between $0$ and $d_{2lim}^\star$ excluded. The conditional density $\pi(\boldsymbol{r}_{2n} )$ is 
\begin{equation}
     \pi(\boldsymbol{r}_{2n} ) \propto f(\boldsymbol{r}_{2n}|\boldsymbol{r}_{0n}, E_0, \boldsymbol{r}_{1n}, E_{1n}, CS_{1n}) f(\tilde{\boldsymbol{r}}_{2n}|\boldsymbol{r}_{2n},\sigma_{xy}. \sigma_z),
\end{equation}
where $f(\boldsymbol{r}_{2n}|\boldsymbol{r}_{0n}, E_0, \boldsymbol{r}_{1n}, E_{1n}, CS_{1n})$ and $f(\tilde{\boldsymbol{r}}_{2n}|\boldsymbol{r}_{2n},\sigma_{xy}. \sigma_z)$ are respectively defined in \eqref{eq:f_r2sr0r1E1} and \eqref{eq:rnoise_s_r}. 

\paragraph{Sampling $(E_{1n}, E_{2n} | \boldsymbol{r}_{0n}, E_0, \boldsymbol{r}_{1n}, CS_{1n}, \boldsymbol{r}_{2n}, X_{2n}, \tilde{E}_{1n},  \tilde{E}_{2n}, \sigma_E)$.} 
The proposal density for $(E_{1n}, E_{2n})$ is set to fulfill Compton requirements
\begin{equation}
    q(E_{1n}, E_{2n}|E_0, E_{1n}^{(t-1)}, E_{2n}^{(t-1)}, X_{2n}) = q(E_{1n}|E_0, E_{1n}^{(t-1)})\, q(E_{2n}|E_0, E_{1n}, E_{2n}^{(t-1)}, X_{2n}), 
\end{equation}
where $q(E_{1n}|E_0, E_{1n}^{(t-1)})$ is a truncated normal distribution of mean $E_{1n}^{(t-1)}$ lying within the interval $[0, E_0/(1- mc^2/(2E_0))]$. $q(E_{2n}|E_0, E_{1n}, E_{2n}^{(t-1)}, X_{2n})$ depends on the nature of the second interaction. In case of absorption, the corresponding $q(E_{2n}|E_0, E_{1n}, E_{2n}^{(t-1)}, A_{2n})$ is a delta Dirac distribution such that $E_{2n} = E_0-E_{1n}$.
In case of Compton interaction,   $q(E_{2n}|E_0, E_{1n}, E_{2n}^{(t-1)}, CS_{2n})$ is a truncated Gaussian of mean $E_{2n}^{(t-1)}$ lying within the interval $[0, (E_0-E_{1n})/(1- mc^2/(2(E_0-E_{1n})))]$. The variance of the truncated Gaussian distribution(s) evolves during the algorithm in order to have a rate of acceptance between $40$ and $60\%$.  Moreover, the conditional density is 
\begin{equation}
     \pi(E_{1n}, E_{2n} ) \propto f( E_{1n}, \boldsymbol{r}_{2n},  E_{2n}| \boldsymbol{r}_{0n}, E_0,  \boldsymbol{r}_{1n}, CS_{1n}, X_{2n})  f(\tilde{E}_{1n}|E_{1n}, \sigma_E) f(\tilde{E}_{2n}|E_{2n}, \sigma_E),
\end{equation}
where $f(E_{1n}, \boldsymbol{r}_{2n},  E_{2n}| \boldsymbol{r}_{0n}, E_0,  \boldsymbol{r}_{1n}, CS_{1n}, X_{2n})$ is obtained from the combination of \eqref{eq:f_E1sr0r1}, \eqref{eq:f_r2sr0r1E1} and \eqref{eq:fE2|E0,E1,I1=CS,I2=A} or \eqref{eq:fE2|E0,E1,I1=CS,I2=CS}. $f(\tilde{E}_{1n}|E_{1n}, \sigma_E)$ and $f(\tilde{E}_{2n}|E_{2n}, \sigma_E)$ are the distributions modelling noise on energy depositions \eqref{eq:noisy_E}.

\paragraph{Sampling $(\boldsymbol{r}_{0n}|\{\boldsymbol{r}_0^{(k)}, w_0^{(k)}\}_{k=1}^K, E_0, \boldsymbol{r}_{1n}, CS_{1n}, E_{1n}, \boldsymbol{r}_{2n}, X_{2n}, E_{2n})$.} The proposal density $q(\cdot|\boldsymbol{r}_{0n}^{(t-1)})$ is a Von-Mises distribution of mean $\boldsymbol{r}_{0n}^{(t-1)}$. The concentration parameter varies to keep the rate of acceptance between $40$ and $60\%$. 
The conditional density is 
\begin{equation}
    \pi(\boldsymbol{r}_{0n}) \propto f( \boldsymbol{r}_{1n}, E_{1n}, \boldsymbol{r}_{2n}, E_{2n}  | \boldsymbol{r}_{0n}, E_0, CS_{1n}, X_{2n}) f(\boldsymbol{r}_{0n}|\{\boldsymbol{r}_0^{(k)},w_0^{(k)}\}_{k=1}^K), 
\end{equation}
where $f( \boldsymbol{r}_{1n}, E_{1n}, \boldsymbol{r}_{2n}, E_{2n}  | \boldsymbol{r}_{0n}, E_0, CS_{1n}, X_{2n}) $ is the probabilistic model for data acquisition \eqref{eq:likelihood} and $f(\boldsymbol{r}_{0n}|\{\boldsymbol{r}_0^{(k)},w_0^{(k)}\}_{k=1}^K)$ is the prior distribution of the $n$-th virtual source conditioned to the $K$ source positions and their relative intensities defined in \eqref{eq:prior_r0n}. 
\paragraph{Sampling  $(\boldsymbol{r}_0^{(k)}|\{\boldsymbol{r}_0^{(k^\prime)}, w_0^{(k^\prime )}\}_{k^\prime=1, k^\prime \neq k}^{K}, w_0^{(k)})$.} The proposal density $q(\cdot|\boldsymbol{r}_0^{(k, t-1)})$ is also a Von-Mises distribution of mean $\boldsymbol{r}_0^{(k, t-1)}$ and the concentration parameter varies to keep the rate of acceptance between $40$ and $60\%$.  The conditional probability is 
\begin{equation}
    \pi(\boldsymbol{r}_0^{(k)}) \propto f(\boldsymbol{r}_0^{(k)}) \prod_{n=1}^N f(\boldsymbol{r}_{0n}|\{\boldsymbol{r}_0^{(k)},w_0^{(k)}\}_{k=1}^K), 
\end{equation}
where $f(\boldsymbol{r}_0^{(k)})$ is the prior distribution on the $k$-th source defined in \eqref{eq:prior_r0} and $f(\boldsymbol{r}_{0n}|\{\boldsymbol{r}_0^{(k)},w_0^{(k)}\}_{k=1}^K)$ corresponds to \eqref{eq:prior_r0n}.

\paragraph{Sampling  $(w_0^{(k)}|\{\boldsymbol{r}_0^{(k^\prime)}, w_0^{(k^\prime )}\}_{k^\prime=1, k^\prime \neq k}^{K}, \boldsymbol{r}_0^{(k)})$.}
The proposal $q(\cdot|\{w_0^{(k^\prime, t)}\}_{k^\prime=1}^{k-1}, \{w_0^{(k^\prime, t-1)}\}_{k^\prime=k}^{K})$ is a truncated normal distribution of mean $w_0^{(k, t-1)}$ lying between $0$ and $1 - \sum_{k^\prime = 1}^{k-1}w_0^{(k^\prime, t)} - \sum_{k^\prime = k+1}^{K}w_0^{(k^\prime, t-1)}$. The conditional distribution is 
\begin{equation}
    \pi(w_0^{(k)}) \propto {w_0^{(k)}}^{\alpha_k-1} \left(1-\sum_{k^\prime = 1}^K w_0^{(k^\prime)}\right)^{\alpha_0-1} \prod_{n=1}^N f(\boldsymbol{r}_{0n}|\{\boldsymbol{r}_0^{(k)}, w_0^{(k)}\}_{k=1}^K), 
\end{equation}
where $f(\boldsymbol{r}_{0n}|\{\boldsymbol{r}_0^{(k)},w_0^{(k)}\}_{k=1}^K)$ corresponds to \eqref{eq:prior_r0n}.

\paragraph{Sampling $\sigma_{xy} \sim f(\sigma_{xy}|\{\boldsymbol{r}_{1n}, \tilde{\boldsymbol{r}}_{1n}, \boldsymbol{r}_{2n},\tilde{\boldsymbol{r}}_{2n}\}_{n=1}^N)$ and $\sigma_{z} \sim f(\sigma_{z}|\{\boldsymbol{r}_{1n}, \tilde{\boldsymbol{r}}_{1n}, \boldsymbol{r}_{2n},\tilde{\boldsymbol{r}}_{2n}\}_{n=1}^N)$.} The respective proposal densities are Gaussian distributions of means $\sigma_{xy}^{(t-1)}$ and $\sigma_{z}^{(t-1)}$. The conditional distributions are 
\begin{equation}
\pi(\sigma_{xy})\propto \prod_{n=1}^N f(\tilde{\boldsymbol{r}}_{1n}|\boldsymbol{r}_{1n}, \sigma_{xy},\sigma_{z})f(\tilde{\boldsymbol{r}}_{2n}|\boldsymbol{r}_{2n}, \sigma_{xy},\sigma_{z}),
\end{equation}
and 
\begin{equation}
\pi(\sigma_{z})\propto \prod_{n=1}^N f(\tilde{\boldsymbol{r}}_{1n}|\boldsymbol{r}_{1n}, \sigma_{xy}, \sigma_{z})f(\tilde{\boldsymbol{r}}_{2n}|\boldsymbol{r}_{2n}, \sigma_{xy}, \sigma_{z}), 
\end{equation}
where $f(\tilde{\boldsymbol{r}}_{in}|\boldsymbol{r}_{in}, \sigma_{xy},\sigma_{z})$ is defined in \eqref{eq:rnoise_s_r}. 

\paragraph{Sampling $\sigma_{E} \sim f(\sigma_{E}|\{E_{1n}, \tilde{E}_{1n}, E_{2n},\tilde{E}_{2n}\}_{n=1}^N)$.}
Similarly as in the previous paragraph, the proposal density is Gaussian of mean $\sigma_{E}$. The conditional distribution is
\begin{align}
\pi(\sigma_{E}) = \prod_{n=1}^N f(\tilde{E}_{1n}|E_{1n}, \sigma_{E})f(\tilde{E}_{2n}|E_{2n}, \sigma_{E}), 
\end{align}
where $f(\tilde{E}_{in}|E_{in}, \sigma_{E})$ is defined in \eqref{eq:noisy_E}.
\section{Analytical derivation of \texorpdfstring{$p_A$}{pA} and \texorpdfstring{$p_{CS}$}{pCS}}
\label{app:est_pA_pCS}
The analytical expressions of $p_A = f( X_2 =A_2|E_0, CS_1)$ and $p_{CS} = f(X_2 =CS_2|E_0, CS_1)$ are presented in this paragraph. First, the probability 
 $p_A$ is obtained from the marginalisation of $E_1$, i.e.
\begin{equation}
f( X_2 =A|E_0, X_1=CS) = \int_{E_{1min,a}}^{E_{1max,a}}f(X_2 = A|E_0, E_1, X_1=CS)f(E_1|E_0, X_1=CS)\, dE_1, 
\end{equation}
where $f(E_1|E_0, X_1=CS)$ is given in \eqref{eq:f_E1sr0r1} and $f(X_2 = A|E_0, E_1, X_1=CS)$ is the ratio of the absorption and attenuation coefficients of the considered material at energy $E_0-E_1$, respectively denoted $\mu^a_{E_0-E_1}$ and $\mu_{E_0-E_1}$
\begin{equation}
 f(X_2 = A|E_0, E_1, X_1=CS) = \displaystyle \frac{\mu_{E_0-E_1}^a}{\mu_{E_0-E_1}}.
\end{equation}
The probability $p_{CS}$ can also be obtained  via a similar calculation, nevertheless, since Compton scattering and absorption are assumed to be the only two possible interactions, it follows that $p_{CS} = 1 - p_A$.   

 
\bibliographystyle{IEEEtran}
\bibliography{IEEEabrv,biblio}
\end{document}

%% file: hierarchical_model.tex
\begin{figure}[!ht]
    \centering
    \scalebox{0.7}{\begin{tikzpicture}[->,>=stealth',shorten >=1pt,auto,node distance=3cm,thick,main node/.style={circle,draw,font=\sffamily\Large\bfseries}, minimum size = 1.5cm,ex node/.style={rectangle,draw,font=\sffamily\Large\bfseries},block/.style={draw, thick, text width=3cm, minimum height=1.5cm, align=center}, param node/.style={diamond,draw,font=\sffamily\Large\bfseries, minimum size = 2cm},text node/.style={rectangle,font=\sffamily\Large\bfseries},block/.style={draw, thick, text width=3cm, minimum height=1.5cm, align=center}, param node/.style={diamond,draw,font=\sffamily\Large\bfseries, minimum size = 2cm}]

  \node[ex node] (r1n) {$\tilde{\boldsymbol{r}}_{1n}$};
  \node[ex node] (r2n) [right of=r1n] {$\tilde{\boldsymbol{r}}_{2n}$};
  \node[ex node] (E1n) [right of=r2n] {$\tilde{E}_{1n}$};
  \node[ex node] (E2n) [right of=E1n] {$\tilde{E}_{2n}$};
  \node[main node] (r2) [below of=r2n] {$\boldsymbol{r}_{2n}$};
  \node[main node] (r1) [below of=r2] {$\boldsymbol{r}_{1n}$};
  \node[main node] (r0) [below of=r1] {$\boldsymbol{r}_{0n}$};
  \node[main node] (E1) [below of=E1n] {$E_{1n}$};
  \node[main node] (E2) [below of=E2n] {$E_{2n}$};
  \node[main node] (r0k) [below of=r0] {$\boldsymbol{r}_0^{(k)}$};
  \node[main node] (w0k) [right of=r0k] {$w_0^{(k)}$};
  \node[draw, inner xsep=5mm,inner  
     ysep=6mm, below left, fit=(r1n)(r2n)(E1n)(E2n)(r2)(r1)(r0)](f){};
  \node[text node] (n) [left of=r0, anchor = north]{\small $\quad n=\{1,\dots,N\}$};
  
   \node[main node] (sigmaxy) [left of=r1n] 
 {$\sigma_{x,y}$};

 \node[main node] (sigmaz) [below of=sigmaxy] 
 {$\sigma_{z}$};
 
   \node[main node] (sigmaE) [right of=E2n] 
 {$\sigma_{E}$};
    \node[param node] (alphabeta) [below of=w0k] 
 {$\alpha_k$};
 \node[param node] (alpha0) [right of=alphabeta] 
 {$\alpha_0$};
    \node[param node] (kappa) [left of=r0k] 
 {$\kappa$};

 \node[text node] (k) [left of=alphabeta, anchor = north]{\small $\quad k=\{1,\dots,K\}$};

 \node[draw, inner xsep=5mm,inner  
     ysep=6mm, below left, fit=(r0k)(w0k)(alphabeta)](g){};

  \path[every node/.style={font=\sffamily\small}]
    (r1) edge node [left] {} (r2)
    (r0) edge node [left] {} (r1)
    (r0) edge [bend right] node[right] {} (r2)
    (r2) edge node [left] {} (r2n)
    (r1) edge [bend left] node[left] {} (r1n)
    (E1) edge node [left] {} (E1n)
    (E1) edge node [left] {} (r2)
    (E2) edge node [left] {} (E2n)
    (E1) edge node [left] {} (E2)
    (r0k) edge node [left] {} (r0)
    (w0k) edge node [left] {} (r0)
    (sigmaxy) edge node[right] {} (r1n)
    (sigmaxy) edge [bend left] node[right] {} (r2n)
    (sigmaz) edge [bend right] node[right] {} (r2n)
    (sigmaz) edge [bend left] node[right] {} (r1n)
    (sigmaE) edge [bend right] node [left] {} (E1n)
    (sigmaE) edge node[left] {} (E2n)
    (alphabeta) edge node [left] {} (w0k)
    (kappa) edge node [left] {} (r0)
    (alpha0) edge node [left] {} (w0k);

\end{tikzpicture}}
    \caption{Hierarchical model between variables. From the measurements $\{\tilde{\boldsymbol{r}}_{1n}, \tilde{E}_{1n}, \tilde{\boldsymbol{r}}_{2n}, \tilde{E}_{2n}\}_{n=1}^N$, the true positions and energy depositions $\{{\boldsymbol{r}}_{1n}, {E}_{1n}, {\boldsymbol{r}}_{2n}, {E}_{2n}\}_{n=1}^N$ and the corresponding hidden standard deviations $\sigma_{xy}$ and $\sigma_z$ are estimated, assuming measurements are corrupted by Gaussian noises. $N$ additional virtual source positions, acting as if there were $N$ sources, are also introduced to deal with potential outliers and assign the events to the relevant source $\boldsymbol{r}_0^{(k)}$. The relative intensities of sources $\{w_0^{(k)}\}_{k=1}^K$ are also estimated. Note that $E_0$ is not depicted here since it is considered as a known value for this algorithm. Circular nodes represent the variables to estimate. Measurements are in rectangular nodes. Diamond shaped nodes contain the fixed hyperparameters $\{\alpha_k\}_{k=0}^K$ and $\kappa$ that will be set according to prior knowledge about the experiment.  }
    \label{fig:hierarchical_model}
\end{figure}

%% file: figures_worst_best_1source.tex
\begin{figure}[!ht]
    \centering
    \begin{tabular}{cc}
            \includegraphics[width = 0.4\linewidth, trim={9.5cm 10.3cm 1cm 7.7cm},clip, cframe=gray]{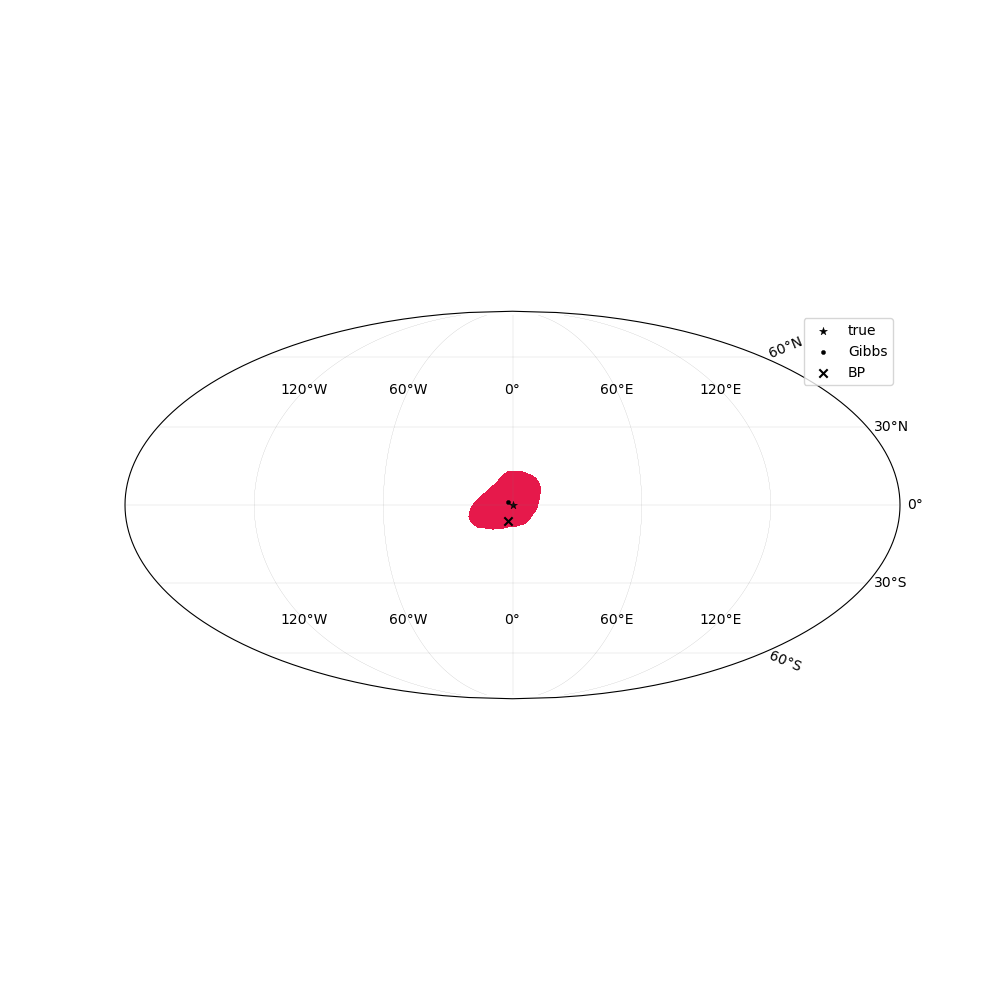} &  \includegraphics[width = 0.4\linewidth,trim={9.5cm 10.3cm 1cm 7.7cm},clip, cframe=gray]{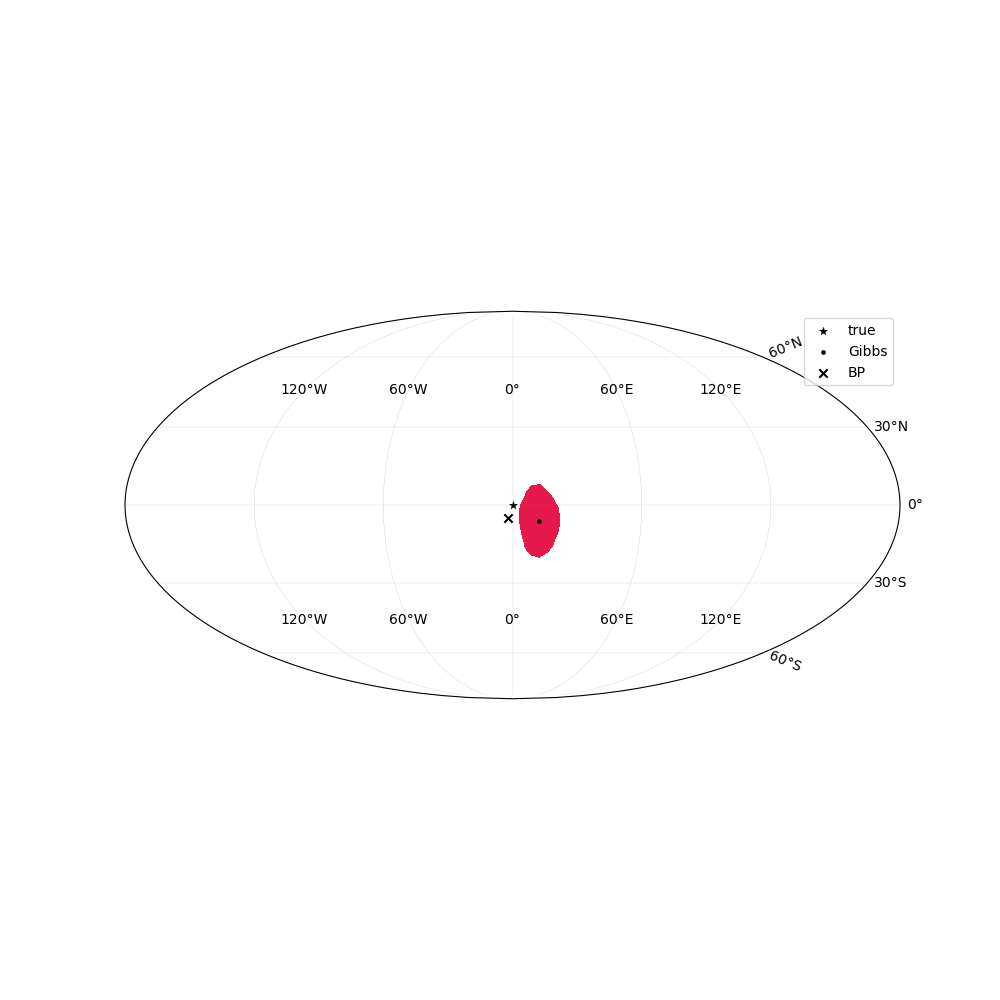}\\ 
            \multicolumn{2}{c}{\footnotesize (a) Source at longitude $0\degree$, latitude $0\degree$}\\ \vspace{-0.2cm} \\
            \includegraphics[width = 0.4\linewidth, trim={9.5cm 10.3cm 1cm 7.7cm},clip, cframe=gray]{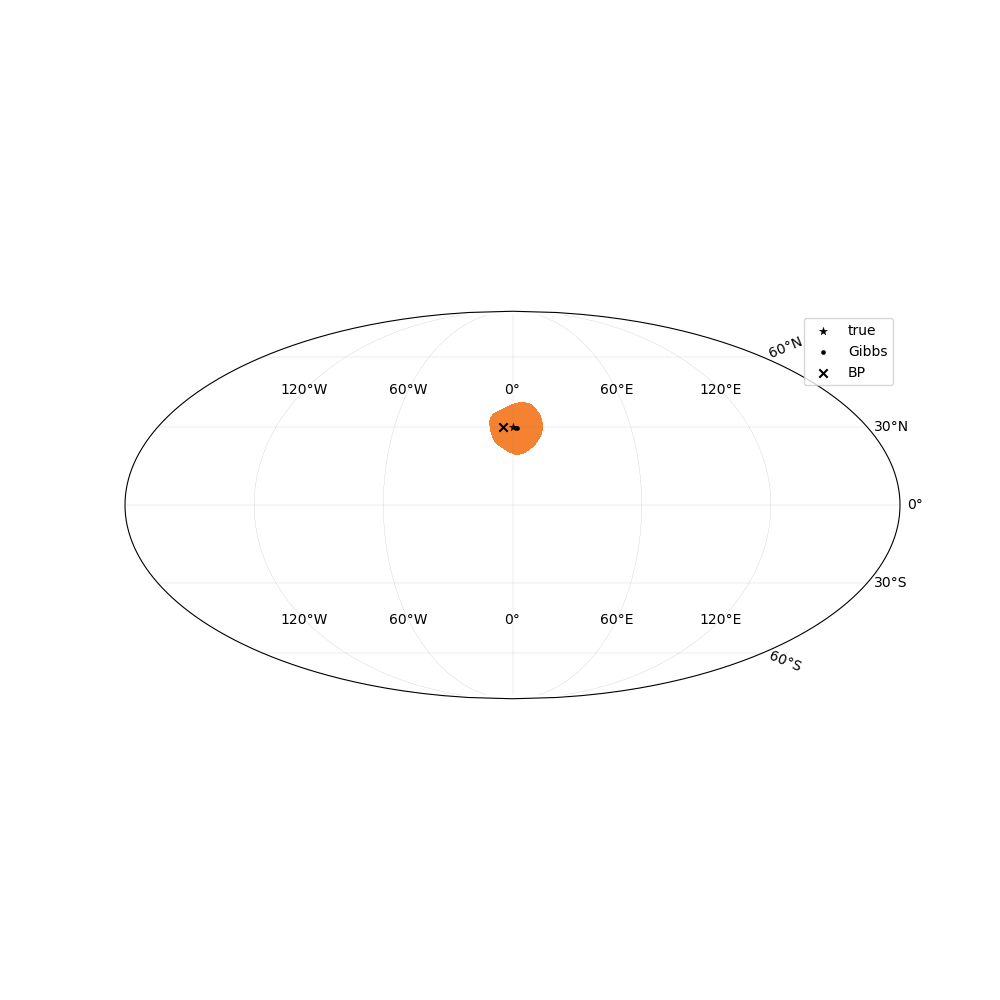} &  \includegraphics[width = 0.4\linewidth,trim={9.5cm 10.3cm 1cm 7.7cm},clip, cframe=gray]{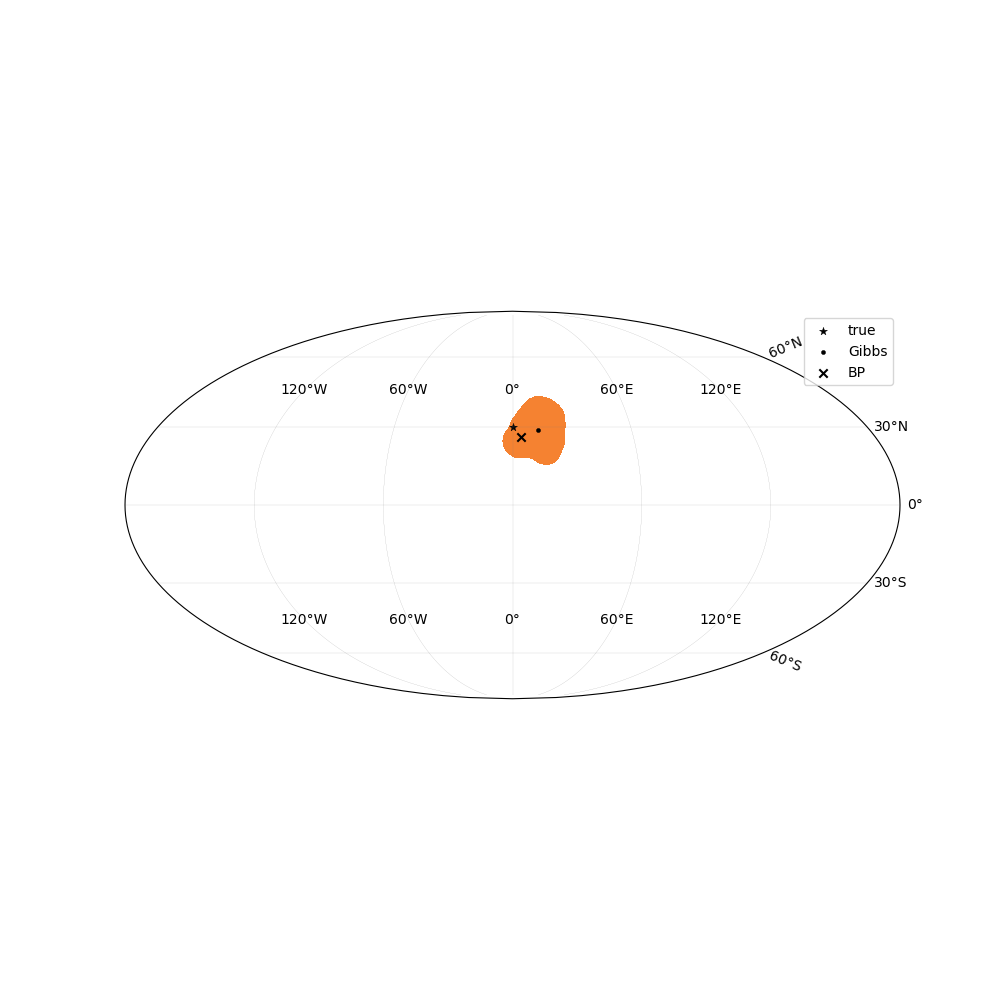}\\
            \multicolumn{2}{c}{\footnotesize (b) Source at longitude $0\degree$, latitude $30\degree$N}\\ \vspace{-0.2cm} \\
            \includegraphics[width = 0.4\linewidth, trim={9.5cm 10.3cm 1cm 7.7cm},clip, cframe=gray]{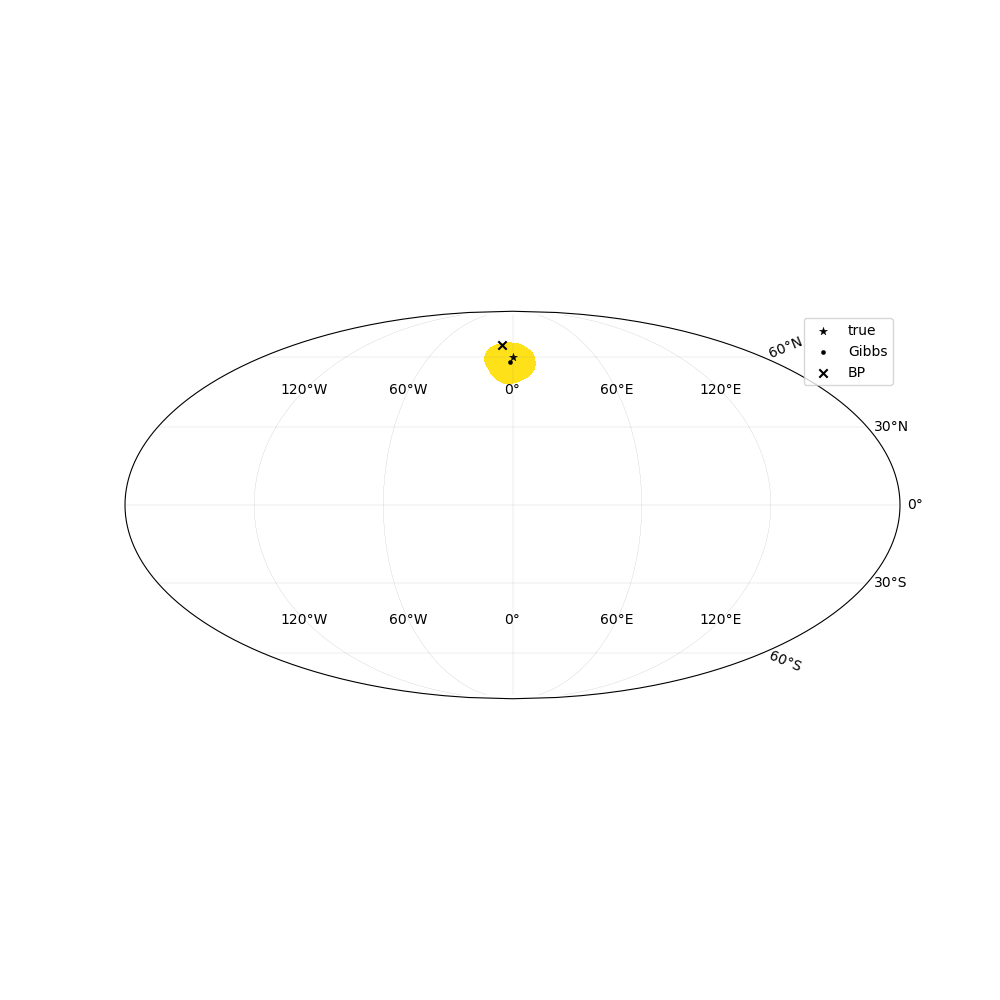} &  \includegraphics[width = 0.4\linewidth,trim={9.5cm 10.3cm 1cm 7.7cm},clip, cframe=gray]{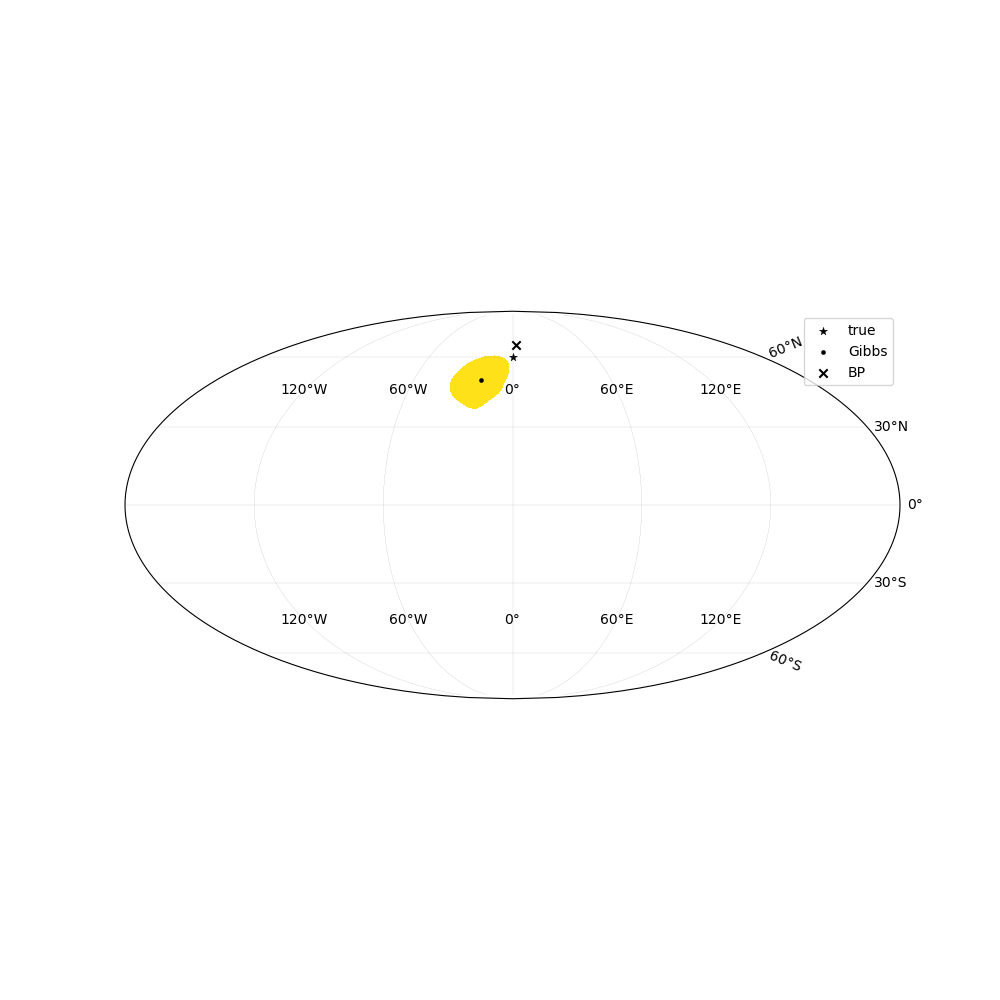}\\ 
            \multicolumn{2}{c}{\footnotesize (c) Source at longitude $0\degree$, latitude $60\degree$N}\\ \vspace{-0.2cm} \\
            \includegraphics[width = 0.4\linewidth, trim={9.5cm 10.3cm 1cm 7.7cm},clip, cframe=gray]{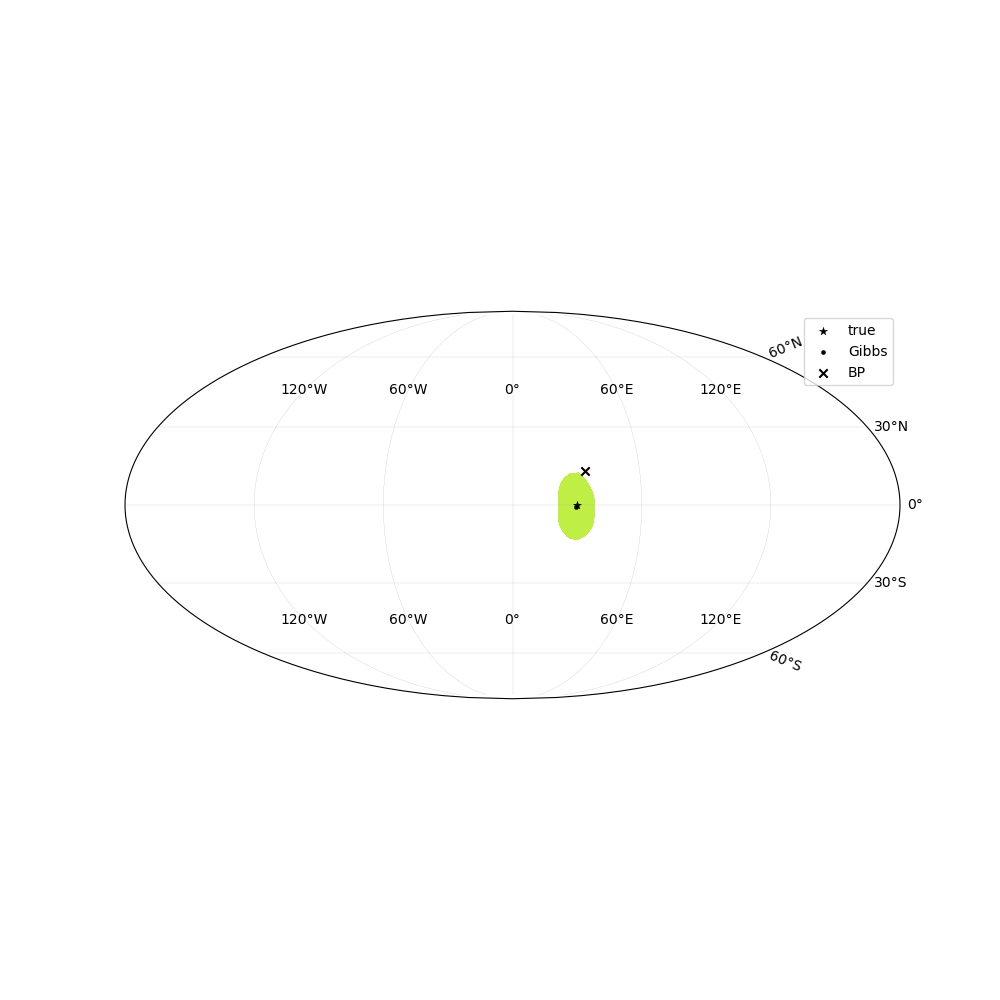} &  \includegraphics[width = 0.4\linewidth,trim={9.5cm 10.3cm 1cm 7.7cm},clip, cframe=gray]{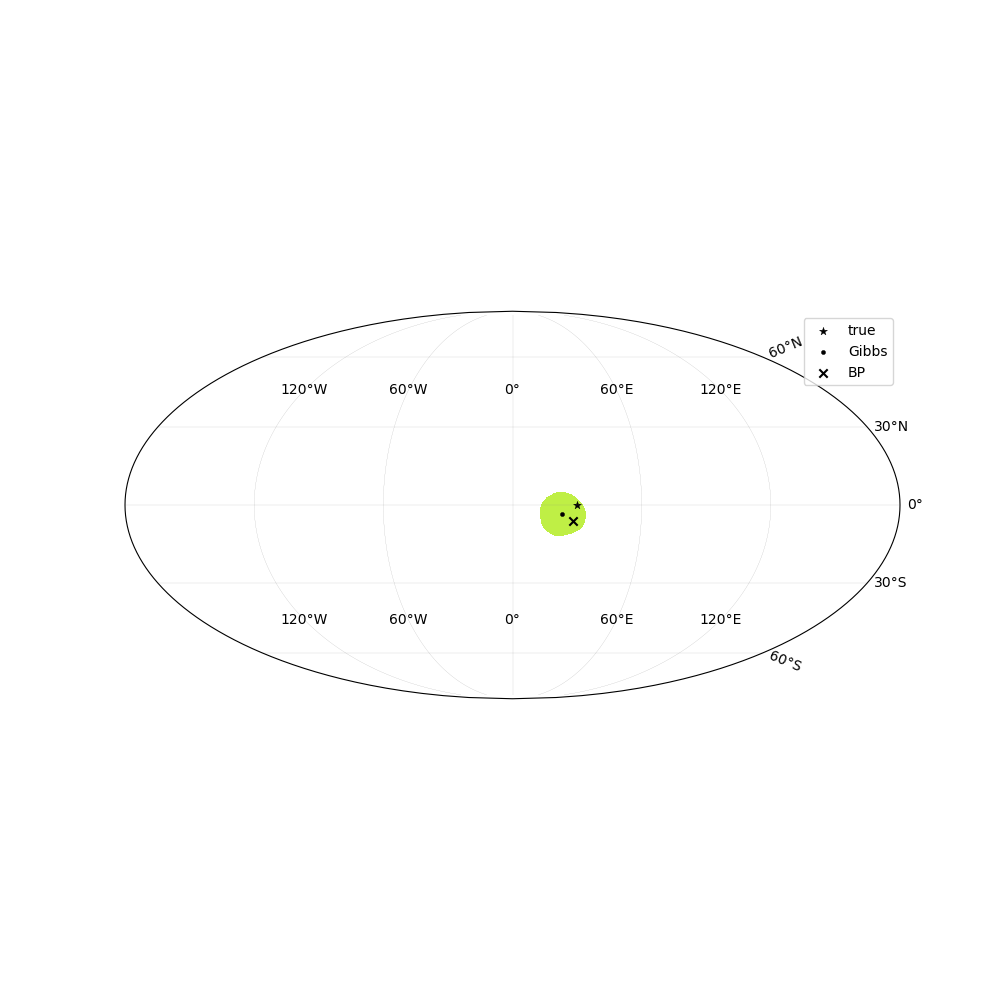}\\
            \multicolumn{2}{c}{\footnotesize (d) Source at longitude $30\degree$E, latitude $0\degree$} \\ \vspace{-0.2cm}\\
            \includegraphics[width = 0.4\linewidth, trim={9.5cm 10.3cm 1cm 7.7cm},clip, cframe=gray]{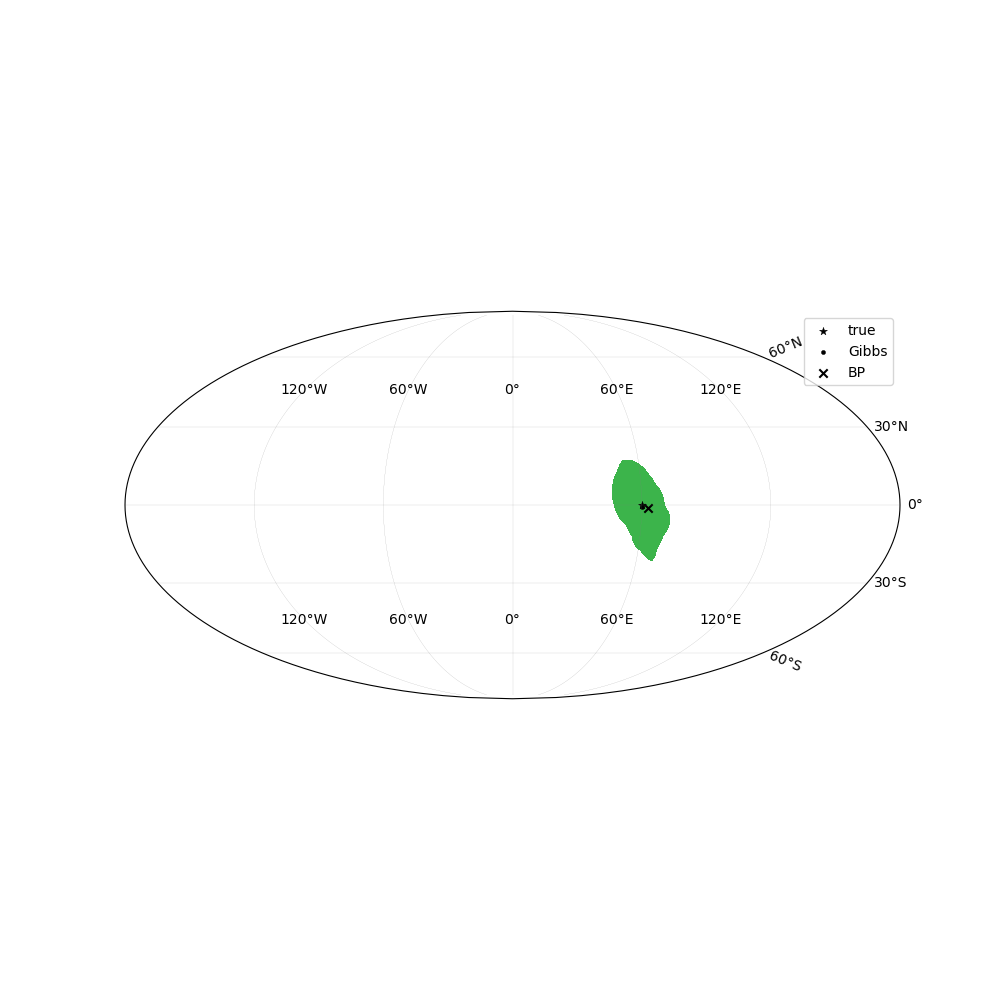} &  \includegraphics[width = 0.4\linewidth,trim={9.5cm 10.3cm 1cm 7.7cm},clip, cframe=gray]{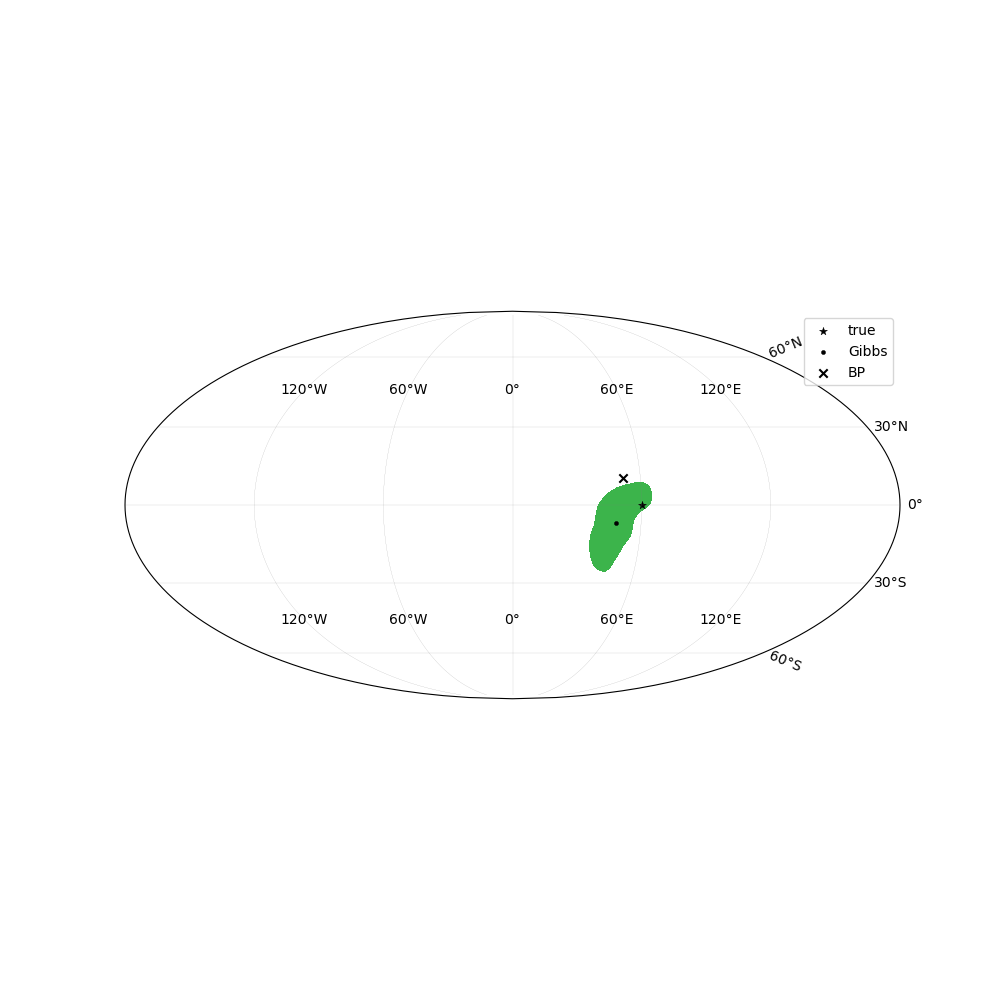}\\
            \multicolumn{2}{c}{\footnotesize (e) Source at longitude $60\degree$E, latitude $0\degree$}
    \end{tabular}
    \caption{\small The $5$th best result and worst results among the obtained distributions are respectively depicted  on the left and right columns  for sources at positions $(0^\circ, 0^\circ)$, $(0^\circ, 30^\circ \text{N})$, $(0^\circ, 60^\circ \text{N})$, $(30^\circ\text{E}, 0^\circ)$, $(60^\circ\text{E}, 0^\circ)$.}
    \label{fig:res_best_worst1}
\end{figure}

\begin{figure}[!ht]
    \centering
    \begin{tabular}{cc}
            \includegraphics[width = 0.4\linewidth, trim={9.5cm 10.3cm 1cm 7.7cm},clip, cframe=gray]{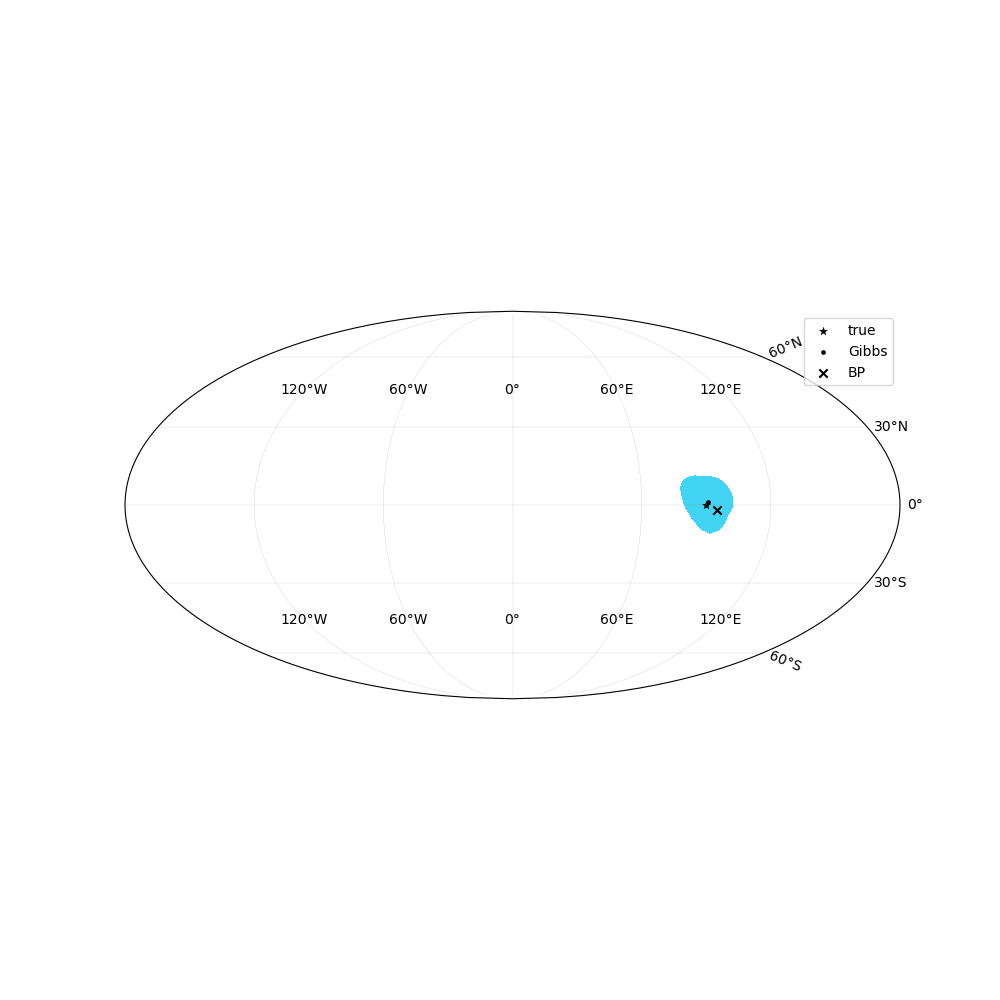} &  \includegraphics[width = 0.4\linewidth,trim={9.5cm 10.3cm 1cm 7.7cm},clip, cframe=gray]{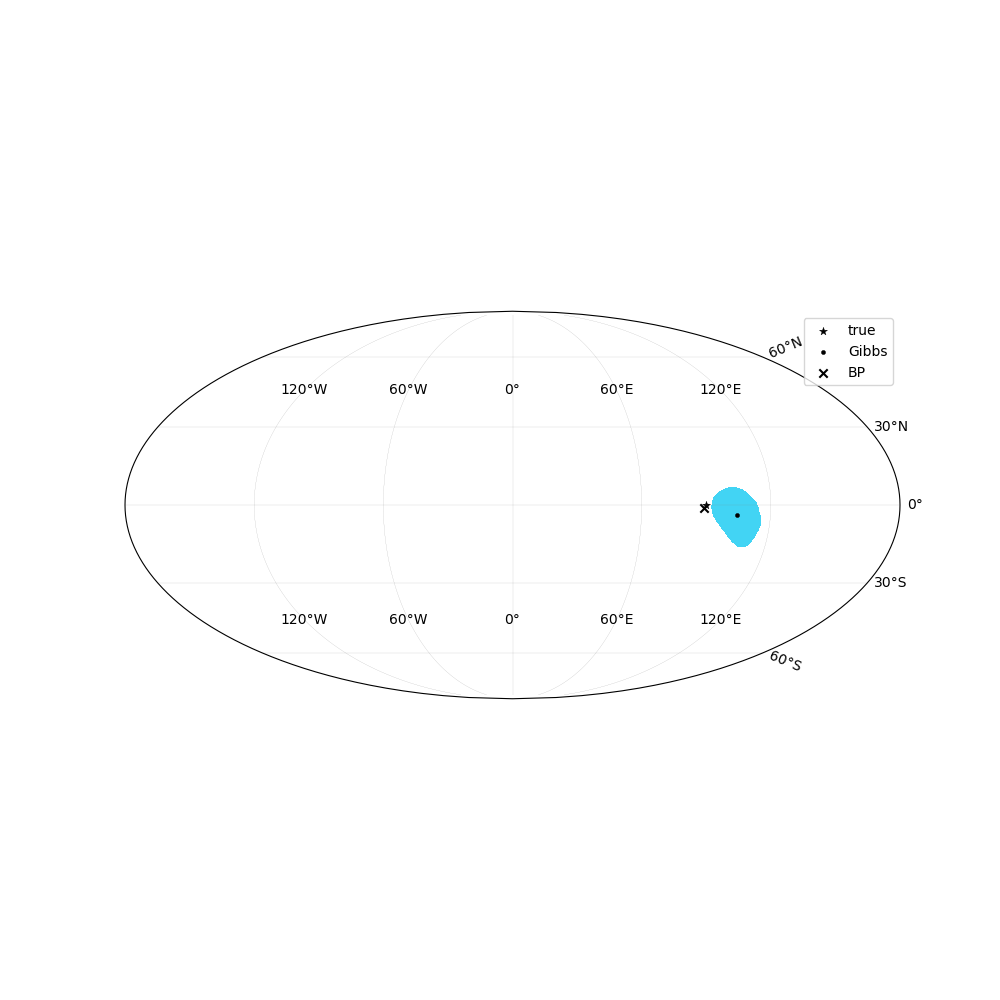}\\ 
            \multicolumn{2}{c}{\footnotesize (a) Source at longitude $90\degree$E, latitude $0\degree$}\\\vspace{-0.2cm}  \\
            \includegraphics[width = 0.4\linewidth, trim={9.5cm 10.3cm 1cm 7.7cm},clip, cframe=gray]{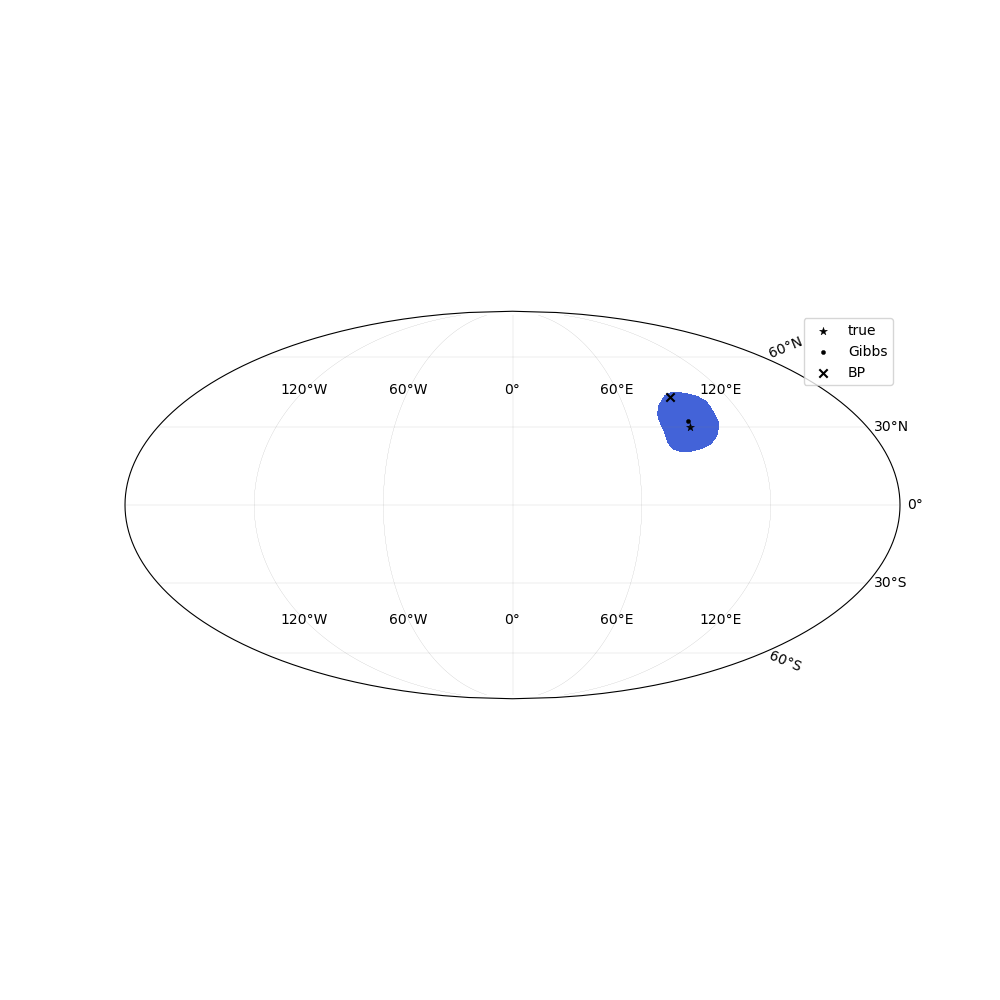} &  \includegraphics[width = 0.4\linewidth,trim={9.5cm 10.3cm 1cm 7.7cm},clip, cframe=gray]{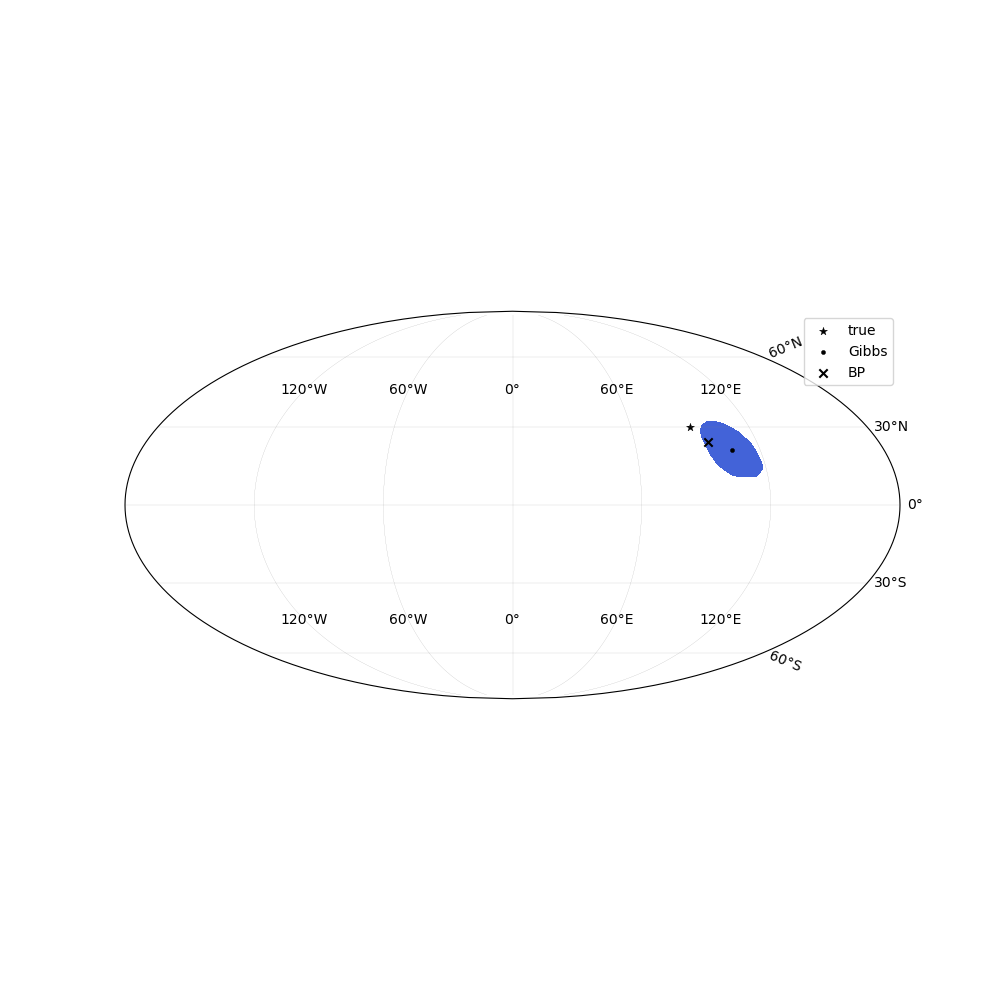}\\
            \multicolumn{2}{c}{\footnotesize (b) Source at longitude $90\degree$E, latitude $30\degree$N}\\ \vspace{-0.2cm} \\
            \includegraphics[width = 0.4\linewidth, trim={9.5cm 10.3cm 1cm 7.7cm},clip, cframe=gray]{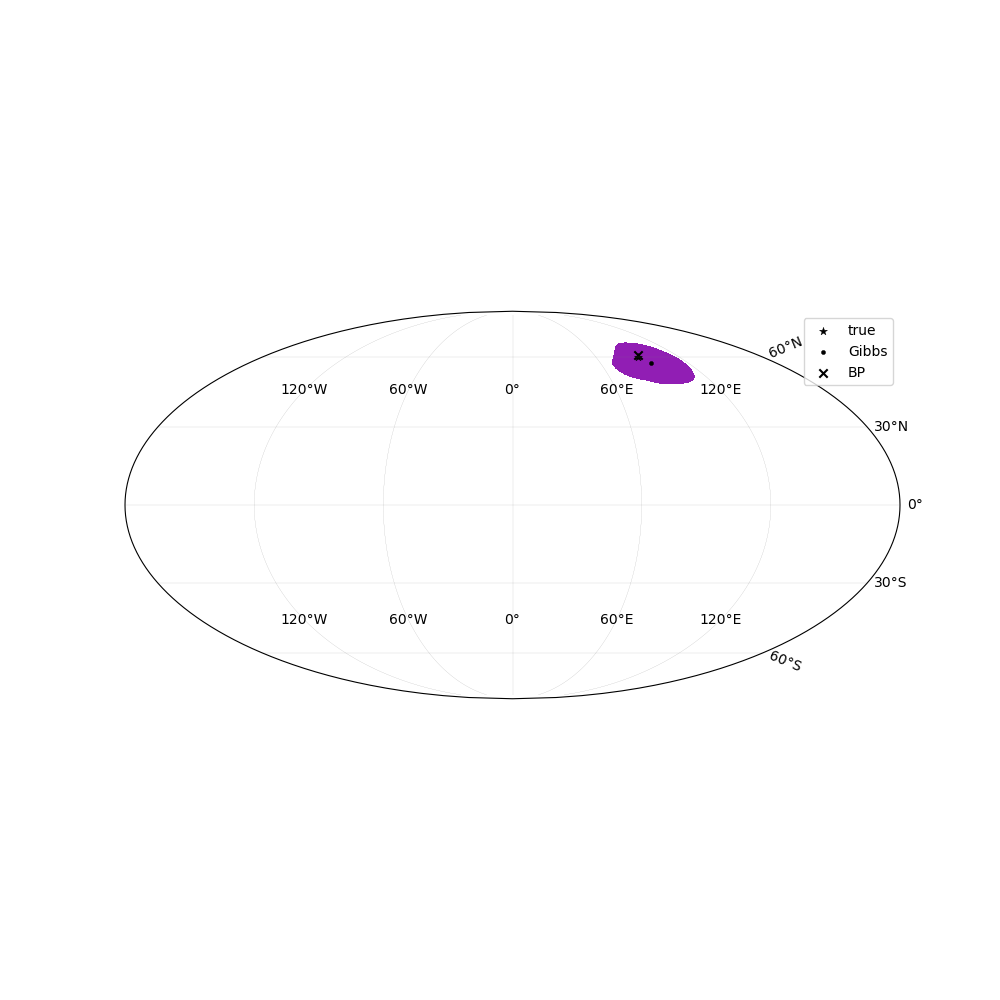} &  \includegraphics[width = 0.4\linewidth,trim={9.5cm 10.3cm 1cm 7.7cm},clip, cframe=gray]{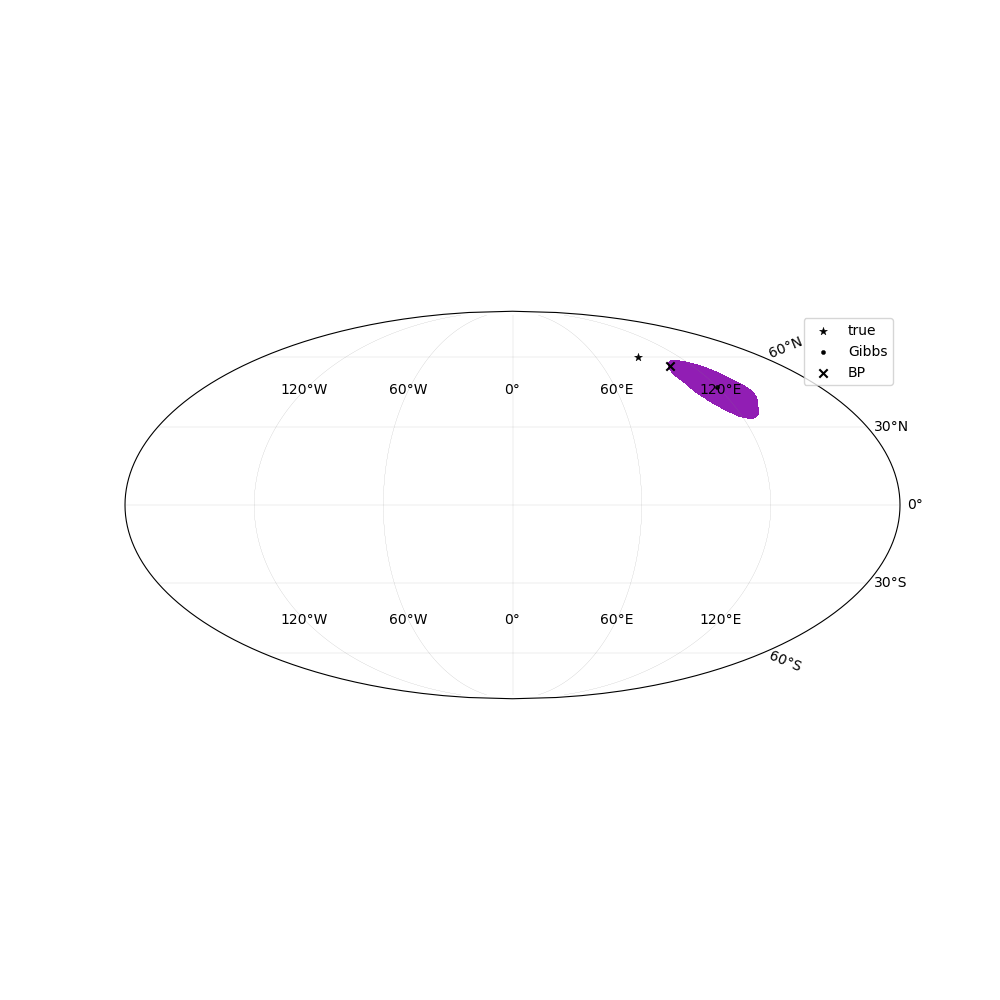}\\ 
            \multicolumn{2}{c}{\footnotesize (c) Source at longitude $90\degree$E, latitude $60\degree$N}\\ \vspace{-0.2cm} \\
            \includegraphics[width = 0.4\linewidth, trim={9.5cm 10.3cm 1cm 7.7cm},clip, cframe=gray]{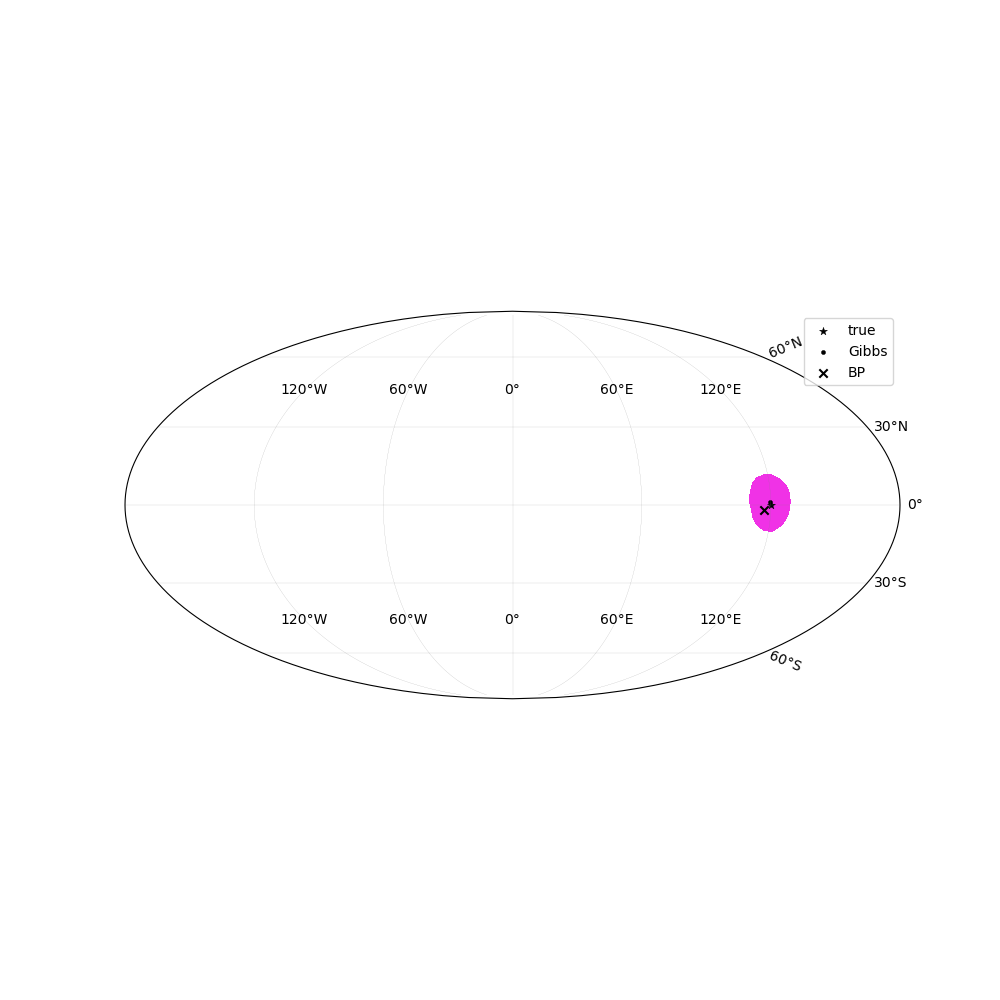} &  \includegraphics[width = 0.4\linewidth,trim={9.5cm 10.3cm 1cm 7.7cm},clip, cframe=gray]{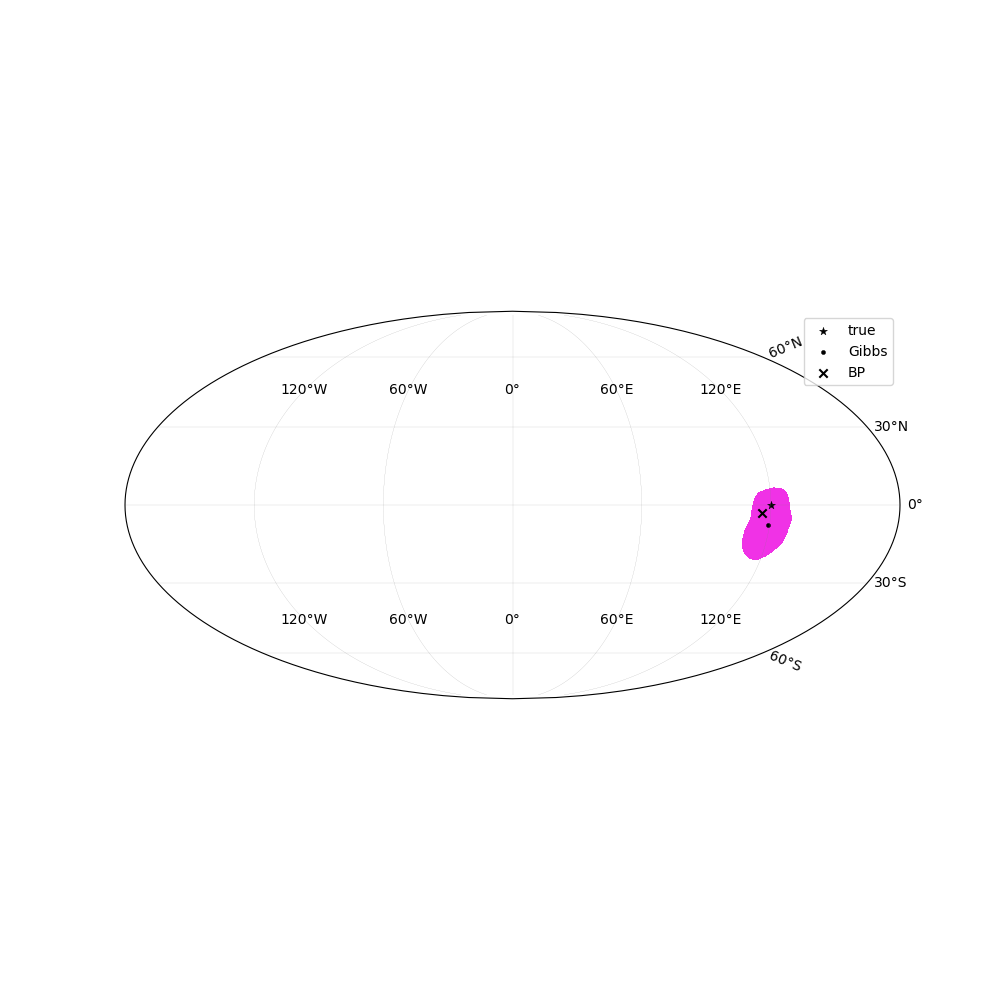}\\
            \multicolumn{2}{c}{\footnotesize (d) Source at longitude $120\degree$E, latitude $0\degree$} \\ \vspace{-0.2cm} \\
            \includegraphics[width = 0.4\linewidth, trim={9.5cm 10.3cm 1cm 7.7cm},clip, cframe=gray]{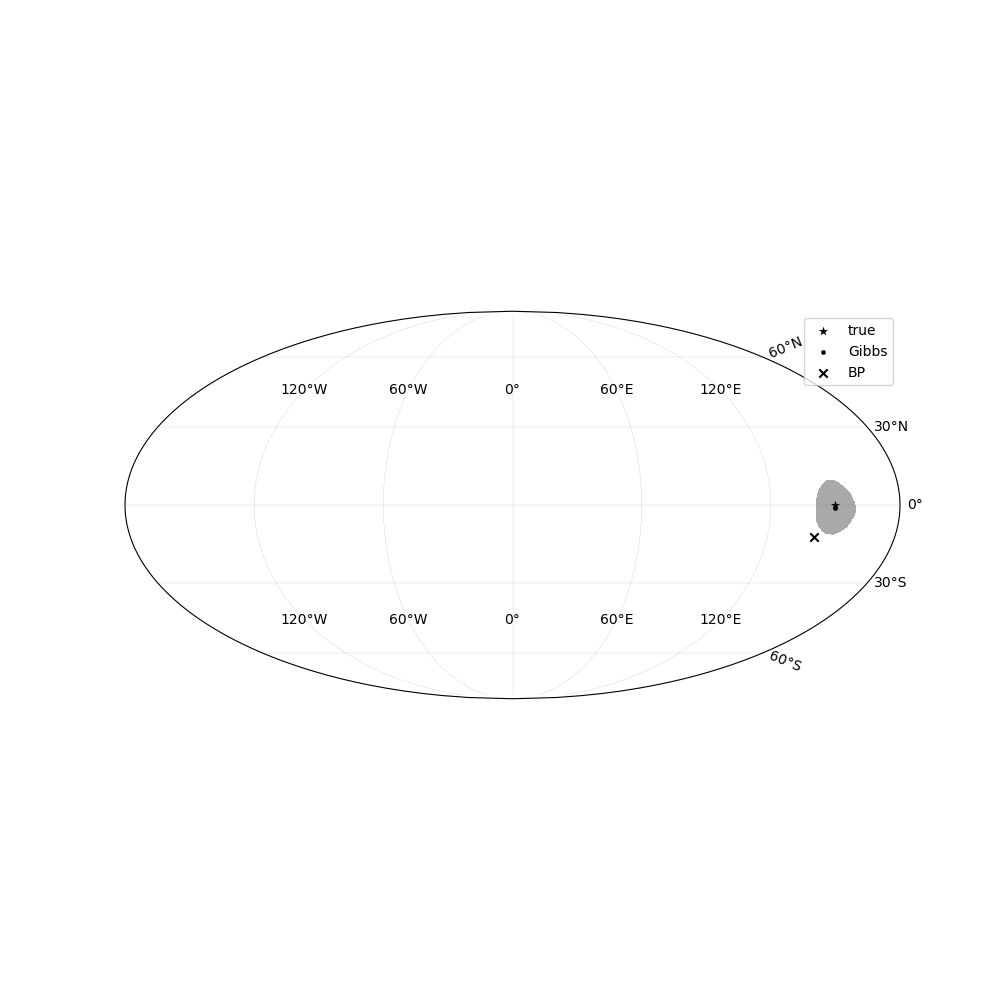} &  \includegraphics[width = 0.4\linewidth,trim={9.5cm 10.3cm 1cm 7.7cm},clip, cframe=gray]{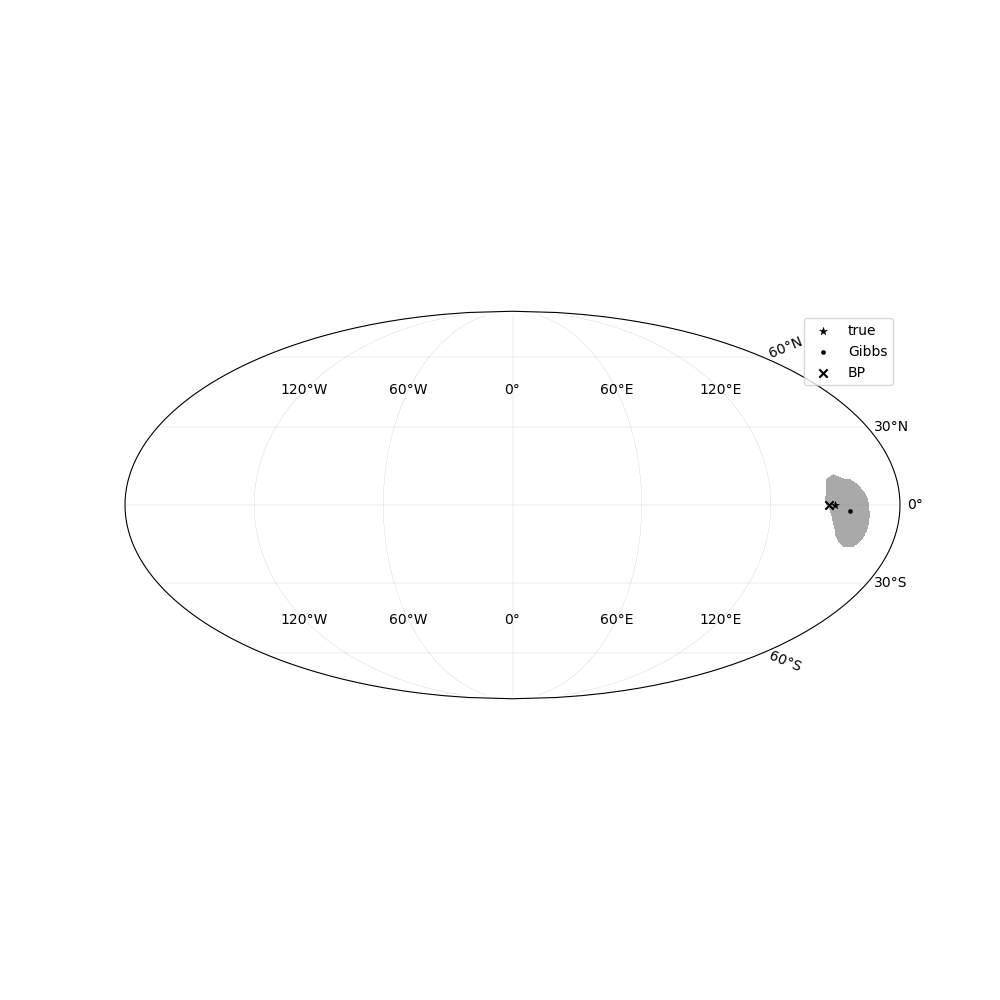}\\
            \multicolumn{2}{c}{\footnotesize (e) Source at longitude $150\degree$E, latitude $0\degree$}
    \end{tabular}
    \caption{\small The $5$th best result and worst results among the obtained distributions are respectively depicted  on the left and right columns  for sources at positions
$(90^\circ\text{E}, 0^\circ)$, $(90^\circ\text{E}, 30^\circ\text{N})$, $(90^\circ\text{E}, 60^\circ\text{N})$, $(120^\circ\text{E}, 0^\circ)$ and  $(150^\circ\text{E}, 0^\circ)$.}
    \label{fig:res_best_worst2}
\end{figure}

%% file: boite_moustache.tex
\begin{figure}
   \centering
   \scalebox{0.6}{ \begin{tikzpicture}
\begin{axis}[
boxplot/draw direction=x,
 xmin = -25, xmax =325,
xlabel={\small Geodesic distance to the true position of the source (in mm)},
height=0.7\textwidth,
boxplot={
    draw position={1/3 + floor(\plotnumofactualtype/2) + 1/3*fpumod(\plotnumofactualtype,2)},
    box extend=0.3
},
every y tick label/.append style={font=\small},
                every x tick label/.append style={font=\scriptsize},
y=2cm,
ytick={0,1,2,...,10},
y tick label as interval,
yticklabels={$(0^\circ; 0^\circ )$, $(0^\circ; 30^\circ \text{N})$, $(0^\circ; 60^\circ \text{N})$, $(30^\circ\text{E}; 0^\circ)$, $(60^\circ\text{E};0^\circ)$, $(90^\circ\text{E}; 0^\circ\text{N})$, $(90^\circ\text{E};30^\circ\text{N})$, $(90^\circ\text{E}; 60^\circ\text{N})$, $(120^\circ\text{E}; 0^\circ)$,  $(150^\circ\text{E}; 0^\circ)$},
y tick label style={
        text width=2.5cm,
        align=right
    },
]

\addplot+[draw=black, fill = rouge]
table[row sep=\\,y index=0] {
data \\ 17.8476865659091 \\	20.2744199564013\\	16.8117602967357\\	75.1172045242435	30.1506973211373\\	43.2652900034360\\	92.9989017028015\\	7.54356604101459\\
46.3472844101245\\	40.1946875677357\\	16.7242214991072\\	20.4328745399533\\	32.1114010662273\\	36.1780266473395\\	53.3379451758881\\	26.4821211463841\\	15.7410328421970\\	69.5507926701454\\	6.59358093999754\\	22.9065145240168\\	16.6633334096726\\	103.131478452024\\	25.0666560984829\\	11.5703913642616\\	7.45127627679186\\	18.8085488762175\\	29.9730913152680\\	39.3939078415334\\	396.764454709000\\	16.1267409295471\\	71.5563713003964\\	37.1484261580483\\	11.9941555978336\\	34.7607945920751\\	23.2650730173453\\	21.2966626881738\\	62.4357795784530\\	23.6286363509005\\	16.0459522232092\\	37.8453611556897\\	33.8290679138293\\	59.8049842592450\\	34.0167018340494\\	27.1078907453940\\	17.6619173564554\\	19.2492620170894\\	38.1957408734951\\	9.87874469865628\\	16.9327003462324\\	53.3124418678542\\
};

\addplot+[draw=black, pattern={north east lines},pattern color=rouge]
table[row sep=\\,y index=0] {
data\\ 45.0037419820022\\	7.40461691752994\\	28.1917173352552\\	165.030344623082\\	18.8759672488002\\	23.4122430159298\\	11.7075489646906\\	23.4122430159298\\	53.8553705738413\\	171.096764336603\\	28.1917173352552\\	49.3510230218981\\	18.8759672488002\\	28.1917173352551\\	33.1092362477460\\	10.4719755119660\\	23.4122430159298\\	35.1112202790635\\	28.1917173352551\\	22.2093370820437\\	0\\	29.6071792824624\\	21.5874986239789\\	33.1092362477460\\	23.4122430159298\\	23.4122430159298\\	35.1112202790634\\	35.1112202790634\\	16.5568903765368\\	42.1879907201141\\	28.1917173352551\\	37.7359826843487\\	45.0037419820022\\	35.1112202790634\\	30.5205263309716\\	23.4122430159298\\	37.7359826843487\\	21.5874986239789\\	384.298244836477\\	28.1917173352551\\	33.5100559618667\\	16.5568903765368\\	33.1092362477460\\	26.1722779011795\\	28.1917173352551\\	45.0037419820022\\	31.8476956372437\\	16.5568903765368\\	16.5568903765368\\	33.1092362477461\\
};

\addplot+[draw=black, fill = orange]
table[row sep=\\,y index=0] {
data\\ 8.36479209797171\\	12.7792161247029\\	41.3699935925147\\	49.6110397352267\\	56.1784548028750\\	46.2492187569638\\	33.7601622870491\\	27.7386576666439\\	49.5719480525715\\	48.0968039597718\\	23.1855061941086\\	7.38566478259065\\	15.4314915987034\\	17.5002915390877\\	39.3865796971637\\	41.7197469796571\\	44.1872958775351\\	26.1032067253242\\	31.6354511667894\\	27.0701010309214\\	17.9985241065526\\	38.7521928804516\\	21.1614191106404\\	19.8567850615008\\	59.5455725896163\\	33.5697466387038\\	38.6834975415034\\	26.0314199819174\\	11.6367977999855\\	18.3727475709000\\	86.3418887624235\\	19.3445289424119\\	90.8586660284894\\	60.1450280027577\\	5.17847617269472\\	18.0790614488898\\	34.2866539579557\\	10.8651479619944\\	27.0031180570988\\	35.9464152994604\\	62.6724586489160\\	16.6863435546754\\	12.9779842440163\\	12.2387577250476\\	33.6062666481188\\	11.1854493233722\\	35.7381035299741\\	19.8129172429248\\	11.6055864379999\\	12.1611738076889\\
};

\addplot+[draw=black, pattern={north east lines},pattern color=orange]
table[row sep=\\,y index=0] {
data\\ 6.94134144090877\\	67.4658494479520\\	25.1167808381091\\	42.4853893022786\\	10.4719755119659\\	11.4293418896634\\	20.9095754012531\\	23.7790994967727\\	6.91142746321078\\	26.5882398247613\\	77.2562877114439\\	58.2185401713824\\	18.7888212127838\\	130.899693899575\\	80.9316925010268\\	10.4318271486073\\	16.3677869276320\\	79.5067689322577\\	33.8769576799679\\	38.2213642551050\\	77.2562877114439\\	60.0475863122395\\	76.1879487086769\\	18.2042851079472\\	27.9344044954452\\	5.23598775598294\\	105.740438638560\\	26.5882398247613\\	37.2323689322799\\	20.9095754012531\\	77.9689470034258\\	31.1849331699898\\	89.1040479822310\\	46.1732057640528\\	6.94134144090877\\	37.2323689322799\\	66.5232390948943\\	6.94134144090877\\	18.2042851079472\\	4.53448402186212\\	300.569414043003\\	17.2731445506455\\	20.9095754012531\\	13.9115458812690\\	21.4477503967600\\	22.6706925169582\\	58.2185401713824\\	5.23598775598294\\	49.2930005732062\\	36.5322475713237\\
};
\addplot+[draw=black, fill = yellow]
table[row sep=\\,y index=0] {
data\\ 57.8760735796463\\	5.49662128000615\\	34.6114670507788\\	20.7807312902063\\	79.2502843546261\\	68.2527388949818\\	19.9895281150954\\	55.4874339972727\\	11.8416528941121\\	71.8638774605851\\	55.2281625693858\\	3.08334639613036\\	45.2922561612739\\	77.0614497285902\\	42.1335557594602\\	61.2080838304866\\	23.5817834225514\\	28.6874644541967\\	16.6270554971333\\	14.9449593496266\\	157.397052753244\\	11.8267697927633\\	34.6252967774846\\	33.6638100546338\\	16.5090727387248\\	41.5688743832151\\	16.9722673832158\\	16.8655328786680\\	23.8117527716099\\	98.4253380628777\\	10.6333998145493\\	26.4583703604438\\	38.0124905000801\\	17.9433188608435\\	60.7023435776758\\	18.1156758651531\\	27.8537298298382\\	49.2562905124559\\	16.0627782463827\\	35.2375287322448\\	30.7085797907451\\	28.6266246039487\\	44.7001244273408\\	65.0461373322854\\	124.644561565885\\	73.8257334159480\\	66.3831661357780\\	9.59444878720388\\	100.169178538624\\ 30.0164916750376\\
};

\addplot+[draw=black, pattern={north east lines},pattern color=yellow]
table[row sep=\\,y index=0] {
data\\52.5382013871996\\	52.4048557352260\\	54.5180532350103\\	14.5814201231783\\	32.2059349965733\\	31.1225449072003\\	21.1262980642503\\	35.5567229480598\\	2.61796895641850\\	29.9007936693199\\	33.9774893004753\\	32.5514444804592\\	11.8468446069553\\	36.7645587933410\\	55.1609800854175\\	29.4639365237597\\	63.1366608762877\\	20.9439510239319\\	38.2407370724532\\	52.4048557352260\\	16.6317386162700\\	37.9296521042584\\	32.2059349965733\\	33.4077100710895\\	53.9543194508501\\	44.6257431148905\\	47.3327593634821\\	45.9813622976323\\	2.61796895641850\\	5.23578838055237\\	40.7574919108032\\	36.0858949092843\\	62.8657994031343\\	47.1761971535104\\	0\\	21.6640934118215\\	9.33912423569526\\	42.1611887417034\\	10.8133079905714\\	48.7483747848116\\	48.5717733259921\\	15.7079632679490\\	26.6193428600583\\	21.5105346711207\\	22.7035850418312\\	70.4278984723760\\	5.83621820049289\\	5.87161168551396\\	68.0975273428087\\	39.1984618623282\\
};

\addplot+[draw=black, fill = lime]
table[row sep=\\,y index=0] {
data\\ 3.08507192916767\\	16.6256316613226\\	27.5490240197627\\	37.3375173006285\\	28.2940128701259\\	14.2058098588475\\	33.4878932434367\\	20.7545490235666\\	8.60012747173755\\	26.4431988302412\\	3.38899534868090\\	4.97552169939129\\	40.9288614488235\\	56.3173467467894\\	16.4440978915187\\	13.5523855127030\\	13.5507227828530\\	2.55815922873336\\	20.9781458090068\\	12.8730622354646\\	18.6095156677195\\	49.2089063619458\\	56.7047845206342\\	8.76203262516993\\	19.9812743058422\\	10.3875031564510\\	40.6480397436122\\	15.2970696898129\\	2.50013849553890\\	11.3903689757428\\	40.3795114607358\\	19.2961454325179\\	35.8307335339125\\	17.3617728212554\\	16.5698234762510\\	37.8662728950324\\	32.9449341534483\\	52.3118215393203\\	23.5109214242798\\	15.0070900987750\\	37.9744866725819\\	6.67043126268163\\	8.96758811359470\\	33.5513954029099\\	21.6554465765752\\	18.7070131387133\\	20.0466919344705\\	7.89822328931697\\	3.82221078972478\\	15.8765151060206\\
};

\addplot+[draw=black, pattern={north east lines},pattern color=lime]
table[row sep=\\,y index=0] {
data\\15.7079632679490\\	197.152514087848\\	26.1799387799149\\	190.296548623730\\	146.523550467763\\	21.5874986239789\\	14.8081057181385\\	37.0221804658497\\	26.6970996870849\\	30.5205263309717\\	21.5874986239789\\	104.151374532178\\	33.1092362477460\\	33.5100559618668\\	48.2224454370888\\	57.5958653158129\\	7.40461691752994\\	41.8879020478639\\	120.427718387609\\	52.2984701778646\\	97.7001727183515\\	18.8759672488002\\	314.396582733743\\	210.511977616687\\	130.899693899575\\	142.799254644841\\	142.261279448150\\	119.193427383389\\	18.8759672488002\\	235.802145294918\\	30.5205263309716\\	107.392867173001\\	116.598873518488\\	44.7183477585598\\	15.7079632679490\\	30.5205263309717\\	137.626424902796\\	167.843844617715\\	33.1092362477461\\	267.101621875797\\	111.441030379272\\	231.139226211349\\	28.1917173352551\\	179.579170383501\\	57.5958653158129\\	51.5334378331756\\	52.2984701778646\\	16.5568903765368\\	71.1640666854127\\	116.987199282376\\
};

\addplot+[draw=black, fill = green]
table[row sep=\\,y index=0] {
data\\ 71.1803921614017\\	85.7321685196060\\	47.7462002599722\\	24.9983677158253\\	18.9169907478793\\	3.26575454325907\\	25.4983080873028\\	7.00796770213695\\	7.12024841359392\\	1.81664239002029\\	24.0765835889839\\	3.69583034781563\\	18.0057298678848\\	31.1604595156968\\	53.5295248802917\\	40.9176790454404\\	74.7838697332589\\	28.9972594605656\\	34.5587784764393\\	27.4461214592081\\	9.05812674748839\\	26.5985117052213\\	26.7039537733018\\	35.8429869892804\\	3.60758851049277\\	7.92284484293623\\	9.76789677810397\\	6.17612058114078\\	30.3630784400641\\	20.0679137452527\\	51.4093069899775\\	318.656894453442\\	5.43539471711223\\	30.8177847793554\\	10.9535960333439\\	11.4965142032602\\	9.83178862735684\\	22.5580520756052\\	46.5454819510324\\	4.61487367744833\\	7.12947204868370\\	10.2622176534876\\	57.2261066532761\\	151.880224954770\\	5.07409105393644\\	18.2624341115545\\	19.4522700287977\\	13.1689935320722\\	14.6017026741424\\	23.8662802606232\\
};

\addplot+[draw=black, pattern={north east lines},pattern color=green]
table[row sep=\\,y index=0] {
data\\ 66.9199353523968\\	448.790562008520\\	15.7079632679490\\	114.618122131117\\	10.4719755119659\\	99.5680157550203\\	117.041288680029\\	94.8086190026878\\	11.7075489646905\\	10.4719755119659\\	15.7079632679490\\	89.3282590139060\\	335.904250786461\\	18.8759672488002\\	20.9439510239320\\	72.8478710196810\\	319.623103561589\\	26.1722779011794\\	10.4719755119659\\	52.2984701778646\\	26.1722779011794\\	37.0221804658496\\	11.7075489646905\\	107.378490284106\\	31.8476956372437\\	10.4719755119659\\	14.8081057181385\\	39.8606804884895\\	59.1418699769789\\	5.23598775598300\\	42.2117698036786\\	232.094300218761\\	31.8476956372437\\	11.7075489646905\\	112.691482672764\\	18.8759672488002\\	10.4719755119659\\	51.7689921614366\\	16.5568903765368\\	16.5568903765368\\	458.373623404445\\	131.224921154738\\	61.2415578728684\\	43.1687964004090\\	114.127025401483\\	15.7079632679490\\	31.4159265358979\\	10.4719755119660\\	46.8016049324653\\	83.9350018287327\\
};

\addplot+[draw=black, fill = cyan]
table[row sep=\\,y index=0] {
data\\ 37.9999028599668\\	42.1147158174728	8.21945989086650	41.2143771887633	22.7851518042842\\	40.3120329604964\\	32.7031498641379\\	73.1837209098526\\	34.5158286734517\\	64.6901029970610\\	40.8640260738098\\	42.6191584563822\\	14.2822978649020\\	90.6283206469935\\	22.5159873835235\\	77.1343782150999\\	78.8197406974341\\	76.7771168575787\\	52.5400698085974\\	7.24842885674955\\	6.29374368034279\\	27.8942363951455\\	39.3715882717647\\	3.79751582270896\\	5.91456141924369\\	51.6658269854177\\	26.4418201554463\\	93.1725190569736\\	23.4850396025811\\	17.1819349197493\\	36.5290468973049\\	37.6536602360443\\	22.5955337183969\\	41.8023911341901\\	73.2110706288878\\	45.7582222889063\\	36.0453478227165\\	17.5081313171337\\	6.56929350297788\\	8.93171039872410\\	21.7276423329340\\	50.3566290930370\\	49.0248839387211\\	104.077362039168\\	19.5480874762030\\	13.3184733331751\\	22.5088062078608\\	41.4859920448876\\	69.7736796591281\\	225.983820896493\\
};

\addplot+[draw=black, pattern={north east lines},pattern color=cyan]
table[row sep=\\,y index=0] {
data\\ 97.7001727183515\\	44.7183477585598\\	252.148131989320\\	130.748746284708\\	83.9350018287327\\	49.3510230218981\\	91.2502690805091\\	54.6423457788904\\	21.5874986239789\\	14.8081057181385\\	7.40461691752990\\	372.320826170111\\	28.1917173352552\\	33.1092362477460\\	101.236250820685\\	78.7101586880945\\	7.40461691752990\\	63.0464533698898\\	28.1917173352552\\	28.1917173352552\\	66.1819928352158\\	48.2224454370888\\	114.160407106547\\	18.8759672488002\\	43.1687964004091\\	11.7075489646906\\	7.40461691752990\\	15.7079632679489\\	26.6970996870850\\	30.5205263309717\\	26.6970996870849\\	16.5568903765368\\	26.6970996870849\\	37.0221804658497\\	79.7344063737342\\	11.7075489646906\\	11.7075489646906\\	16.5568903765368\\	174.884450485488\\	92.8670571762352\\	33.1092362477460\\	117.326385482395\\	15.7079632679490\\	98.4284606057058\\	47.4115038538249\\	188.757043136496\\	49.3510230218981\\	22.2093370820437\\	57.8304556244468\\	224.734655195632\\
};

\addplot+[draw=black, fill = blue]
table[row sep=\\,y index=0] {
data\\ 61.6607403441666\\	56.0173260858944\\	64.6663879869471\\	11.5404575371515\\	6.25852164791630\\	117.710016437900\\	87.6435330242342\\	11.7696248814538\\	31.0973611610510\\	43.5045336304551\\	82.7368456895071\\	54.2213370811453\\	34.2640363639380\\	23.0538286492779\\	28.4437100597596\\	12.0891075517123\\	83.2668479313671\\	10.4983271781504\\	27.1920310084123\\	87.4963187817090\\	62.7351745556062\\	84.6583063697821\\	35.4483999245823\\	41.6333477966772\\	14.6800307966630\\	17.4096459388191\\	75.3781942964016\\	43.3168589604538\\	55.3205627935435\\	13.5922640591176\\	27.0429858467984\\	362.848403804129\\	52.9022260846683\\	25.8250115888446\\	58.8296707715171\\	4.20392990258474\\	25.9249949608186\\	77.9423167621094\\	51.1235616294762\\	18.6531450681048\\	90.3206347879728\\	29.7155657911296\\	34.8918987703610\\	17.2207243757469\\	36.7050191581089\\	84.0633957811601\\	13.1142935292406\\	41.3822965768005\\	107.100494334923\\	160.962219893501\\
};

\addplot+[draw=black, pattern={north east lines},pattern color=blue]
table[row sep=\\,y index=0] {
data\\ 82.7549099183822\\	336.295625293625\\	326.325229282931\\	40.5890099313541\\	157.851003417443\\	26.5882398247613\\	27.7771309092199\\	63.3950672985160\\	106.336459043438\\	25.1167808381091\\	148.424498223724\\	79.8851444264748\\	27.9344044954452\\	169.406956938021\\	233.064971550649\\	54.9644733157578\\	13.7918762816238\\	61.3833754120792\\	37.8285544415841\\	103.418760658603\\	139.052118393773\\	167.960188315784\\	54.9706909471350\\	13.7918762816238\\	264.639121714727\\	67.7310491037251\\	233.064971550649\\	395.664679276749\\	17.0558412428236\\	46.0673071728279\\	127.216932086580\\	186.946284320922\\	42.0604175009492\\	36.8234332894275\\	102.313089104155\\	32.3215189951053\\	24.1938037406229\\	25.1749233062124\\	38.0930757113969\\	164.522889876594\\	42.0604175009492\\	31.7207124092905\\	27.8361973826922\\	120.427718387609\\	10.4318271486072\\	27.9344044954452\\	18.1370724615213\\	80.2506749341491\\	24.1938037406229\\	137.049166432153\\
};

\addplot+[draw=black, fill = purple]
table[row sep=\\,y index=0] {
data\\ 35.9259608657734\\	139.302103333145\\	46.1657887237722\\	21.9005716543829\\	97.9777561356618\\	25.9004154770660\\	46.4124393669856\\	100.043700311792\\	62.6236147256231\\	3.98155805740055\\	12.7061405406453\\	6.41263583124889\\	185.702267122774\\	33.9570119097917\\	41.0950970131371\\	31.9853647390615\\	33.5115333907801\\	19.2167961405542\\	66.7071598975536\\	23.1402062383185\\	48.6153017018232\\	58.6389802767605\\	39.7714771041644\\	54.1063624470022\\	115.193556070185\\	56.5158999145858\\	18.4458793494848\\	25.3267206303316\\	41.4426633654287\\	291.517229887412\\	31.1502014852527\\	27.7883355149650\\	42.7472169809356\\	61.4741649514791\\	78.9416926338072\\	75.7615911031196\\	52.1722357936122\\	52.5133990742162\\	65.1347123945233\\	20.3625324781717\\	43.4293429285480\\	24.6212919745030\\	366.556615066670\\	39.2475208103638\\	23.2671141781662\\	104.382682880569\\	5.10514227075778\\	28.0744683219543\\	42.4754059578311\\	88.3559241367327\\
};

\addplot+[draw=black, pattern={north east lines},pattern color=purple]
table[row sep=\\,y index=0] {
data\\ 33.9791240173579\\	11.8468446069553\\	36.7645587933410\\	358.757416202612\\	59.8924067438659\\	58.7443751981871\\	17.0650198694735\\	72.8816329704071\\	283.519327362176\\	64.2605822321868\\	36.9438802569219\\	58.7443751981871\\	113.893342563460\\	18.6098256413158\\	11.5636886604490\\	14.2773290882864\\	13.0868538375809\\	73.5657483826615\\	123.658943325562\\	65.0683554129293\\	26.7327937582667\\	34.7044548258728\\	64.7503700387363\\	49.4921982575780\\	46.4504179988284\\	199.731565470624\\	7.34830146954562\\	18.6098256413158\\	28.5344901798943\\	219.015302845258\\	39.2988291900568\\	27.0748602579101\\	305.269765016517\\	48.4458003959547\\	20.8178784130436\\	236.188694576609\\	7.46022754140711\\	151.921589099471\\	86.6458534737217\\	40.4458898265226\\	5.23578838055234\\	22.5317876900811\\	348.708328155715\\	186.898004027273\\	15.7025793117932\\	88.5882997588790\\	5.23578838055237\\	38.3563517484639\\	52.8756629967669\\	191.688012409554\\
};

\addplot+[draw=black, fill = magenta]
table[row sep=\\,y index=0] {
data\\ 12.5045902622969\\	7.30693369134546\\	3.40238298306985\\	19.8699886133386\\	34.1571214542759\\	3.82772390814262\\	12.8116019310367\\	29.2013529282663\\	7.43693659194691\\	14.7759403790546\\	14.4529736317034\\	8.55089414641274\\	32.5164178442274\\	18.5302428002397\\	8.05969330746260\\	27.3624389584664\\	13.4093089359156\\	13.7372956364897\\	30.7520881422558\\	35.7591906247905\\	38.0314196781934\\	13.5665207368795\\	20.4958166393975\\	35.7024455775094\\	39.2438814203969\\	18.5970792659796\\	226.153997548390\\	5.75084916605376\\	11.1812593679616\\	13.6171550603314\\	40.4304793320123\\	19.4363350687756\\	10.0473375120127\\	15.5021733925976\\	29.6031391001049\\	25.3110256472001\\	11.7070206827737\\	26.6489704910234\\	75.5893366584908\\	34.5567622682607\\	8.22363840986737\\	9.80663948944120\\	2.02634030853306\\	10.8857606824206\\	11.4193850989138\\	1.54640874508091\\	20.6476923046677\\	10.6265882673206\\	58.6818881202357\\	32.7106436710740\\
};

\addplot+[draw=black, pattern={north east lines},pattern color=magenta]
table[row sep=\\,y index=0] {
data\\ 161.565790413638\\	102.854553966494\\	11.7075489646906\\	265.468451276087\\	26.1722779011794\\	142.600534200720\\	40.8636899092491\\	55.5806555847932\\	59.6065929248420\\	72.8478710196810\\	37.0004969789167\\	371.436110421904\\	30.5205263309716\\	18.8759672488002\\	23.4122430159298\\	157.030086900228\\	29.6071792824624\\	161.565790413638\\	7.40461691752992\\	52.2984701778647\\	45.0037419820023\\	22.2093370820437\\	52.2984701778646\\	33.1092362477460\\	26.1722779011794\\	463.398635090303\\	461.172744481012\\	18.8759672488002\\	61.2415578728684\\	22.2093370820437\\	124.013704513427\\	272.414118208399\\	277.541780557941\\	7.40461691752992\\	22.2093370820437\\	11.7075489646906\\	23.4122430159298\\	302.102611095236\\	16.5568903765368\\	133.025398369956\\	15.7079632679490\\	190.833585803658\\	18.8759672488002\\	267.035375555132\\	30.5205263309716\\	15.7079632679490\\	23.4122430159298\\	22.2093370820437\\	42.1879907201142\\	65.5139633954495\\
};

\addplot+[draw=black, fill = grey]
table[row sep=\\,y index=0] {
data\\ 5.54111662095934\\	17.7743211865849\\	46.3474385643343\\	20.7286764575332\\	29.1957613083140\\	5.12350996628788\\	14.1884217136522\\	20.4536774682740\\	32.1146737421076\\	21.0224741514570\\	4.15279799545199\\	7.63410002953774\\	20.7327267081344\\	17.7296680700672\\	2.50506875902215\\	29.8774612183582\\	44.5594407451632\\	10.7056103314504\\	9.48238804669349\\	37.4305563867126\\	33.4514515367277\\	27.3081342119348\\	16.0953908979999\\	32.7867895788508\\	12.6647270986175\\	40.2134662827369\\	29.2883773272829\\	7.76041546324637\\	8.51302719990266\\	5.34261543037898\\	14.3772761284225\\	37.3712852816315\\	6.45845222136193\\	7.08422407448814\\	12.6669490019154\\	7.24011350267445\\	24.9480944622517\\	12.3994741557594\\	12.6559083599477\\	26.0814524729732\\	15.9776204855912\\	22.1837963768272\\	14.8888843307140\\	26.3104345512970\\	18.8072652221146\\	19.6284824079834\\	40.5182880613174\\	5.69470212725450\\	18.9254224421854\\	8.21449432057439\\
};

\addplot+[draw=black, pattern={north east lines},pattern color=grey]
table[row sep=\\,y index=0] {
data\\ 75.3437095445525\\	71.1640666854127\\	74.8544210918118\\	69.8263675622138\\	63.6859148627404\\	21.5874986239789\\	79.2189871189314\\	91.2502690805091\\	77.1585333438193\\	65.5139633954493\\	94.2385542853851\\	54.6423457788904\\	57.5958653158129\\	5.23598775598300\\	37.0004969789168\\	18.8759672488001\\	77.1585333438193\\	26.6970996870851\\	28.1917173352553\\	15.7079632679489\\	35.1112202790635\\	26.1722779011794\\	48.2640628759401\\	89.3282590139060\\	33.5100559618668\\	30.5205263309717\\	22.2093370820437\\	53.3863577156416\\	77.1585333438193\\	96.4715585810402\\	77.1585333438193\\	93.4188166594500\\	59.6739832641255\\	29.6071792824625\\	63.6859148627404\\	64.7376465271518\\	94.2385542853852\\	70.1449811056714\\	33.5100559618669\\	74.0332505148523\\	63.1998973865014\\	29.6071792824625\\	26.1799387799149\\	63.0464533698898\\	66.1819928352158\\	68.8549545588807\\	26.1722779011794\\	37.0004969789168\\	11.7075489646906\\	69.8263675622138\\
};
\end{axis}
\end{tikzpicture}}
\vspace{-0.2cm}
    \caption{\small Box plots summarizing the distributions of the sets of distances obtained according to the position of the source. Filled boxes correspond to results of Gibbs sampler, boxes with hatches to BP results. }
    \label{fig:boite_moustache}
\end{figure}

%% file: boite_moustache_far_v2.tex
\begin{figure}
   \centering
   \scalebox{0.6}{ \begin{tikzpicture}
\begin{axis}[
boxplot/draw direction=x,
 xmin = -25, xmax =420,
xlabel={\small Geodesic distance to the true position of the source (in mm)},
height=0.7\textwidth,
boxplot={
    draw position={1/3 + floor(\plotnumofactualtype/2) + 1/3*fpumod(\plotnumofactualtype,2)},
    box extend=0.3
},
every y tick label/.append style={font=\small},
                every x tick label/.append style={font=\scriptsize},
y=2cm,
ytick={0,1,2,...,10},
y tick label as interval,
yticklabels={$(0^\circ; 0^\circ )\,\,$, $(120^\circ\text{E}; 0^\circ)\,\,$, $(0^\circ; 0^\circ )\,\,$, $(30^\circ\text{E}; 0^\circ)\,\,$},
y tick label style={
        text width=2.5cm,
        align=right
    },
]

\addplot+[draw=black, fill = blue]
table[row sep=\\,y index=0] {
data \\ 34.6730702144731\\	24.2579452818496\\	49.0556602654015\\	131.053381768897\\	28.7264976713387\\	29.1398404975946\\	84.6669315313817\\	11.1039712216001\\	38.8041463153037\\	17.9910797972625\\	31.4213026692819\\	9.86283742966451\\	30.1017630182829\\	33.5567571422857\\	51.6266827835314\\	41.3373146596786\\	142.759216628799\\	73.7396016613715\\	3.10423013748651\\	72.8424014199848\\	29.5199970572408\\	106.584215894403\\	15.9690253363776\\	65.4926187472589\\	23.5846628654715\\	11.4173556758119\\	286.002754833589\\	14.2620779545738\\	42.6906956289010\\	32.2433862430424\\	66.1568407198337\\	50.5552906445682\\	23.4659761096798\\	67.4209948367295\\	71.1988593149840\\	88.8559601787557\\	98.9382512600811\\	29.1356981802352\\	47.8989471443807\\	15.6579998156977\\	39.6334343967177\\	63.9133667804824\\	29.6220705023695\\	48.6724464743038\\	34.0904698865575\\	21.7591237604091\\	33.0570129195660\\	54.0838142079138\\	38.9509946376837\\	112.268397260798\\
};
\addplot+[draw=black, pattern={north east lines},pattern color=blue]
table[row sep=\\,y index=0] {
data\\ 37.7359826843487\\	7.40461691752994\\	18.8759672488002\\	329.867228626928\\	33.1092362477460\\	23.4122430159298\\	38.1114010350564\\	23.4122430159298\\	135.615841558792\\	14.8081057181385\\	209.768366117569\\	42.1879907201141\\	110.074711731272\\	28.1917173352551\\	33.1092362477460\\	10.4719755119660\\	23.4122430159298\\	31.8476956372437\\	28.1917173352551\\	119.193427383389\\	33.1092362477460\\	105.786238005092\\	135.615841558792\\	138.457884930740\\	23.4122430159298\\	23.4122430159298\\	335.125501189929\\	130.595371264953\\	16.5568903765368\\	22.2093370820437\\	111.844033842604\\	15.7079632679490\\	45.0037419820022\\	335.125501189929\\	130.595371264953\\	23.4122430159298\\	37.7359826843487\\	21.5874986239789\\	106.709456478457\\	120.605788097450\\	33.5100559618667\\	16.5568903765368\\	33.1092362477460\\	26.1722779011795\\	28.1917173352551\\	130.595371264953\\	31.8476956372437\\	16.5568903765368\\	16.5568903765368\\	39.8606804884895\\
};
\addplot+[draw=black, fill = blue]
table[row sep=\\,y index=0] {
data\\ 11.3432860084836\\	15.9315595805889\\	17.5481100211098\\	37.0756096854832\\	5.24554549915337\\	27.9487514250835\\	15.9468129624264\\	54.4669296217514\\	36.0905431486623\\	358.549722545085\\	16.9336818039213\\	3.65448027452226\\	10.8320082613256\\	20.6534493710504\\	10.9784126414032\\	19.1711890693794\\	75.2340730168908\\	20.0068800536730\\	35.1987365347881\\	8.08007041010338\\	173.496259092743\\	18.7914725550252\\	28.8670871935356\\	39.3382241223925\\	48.2376478596987\\	10.0868436794793\\	147.823036128587\\	24.1617177294949\\	16.4483477916673\\	21.5204012832637\\	44.6916667439645\\	14.1167102427716\\	23.5100060713321\\	25.9627678363027\\	15.9989112683223\\	9.17339137358421\\	419.000357627660\\	11.5713409703877\\	10.0091270972618\\	37.4006595385180\\	424.990582460928\\	280.102167383091\\	5.08544678020854\\	16.2204901184189\\	373.158238985965\\	16.6202448130001\\	14.8436047273102\\	12.1470517059151\\	62.5604877035358\\	404.818221705846\\
};
\addplot+[draw=black, pattern={north east lines},pattern color=blue]
table[row sep=\\,y index=0] {
data\\ 16.5568903765368\\	18.8759672488002\\	15.7079632679489\\	104.224625637084\\	81.2010479265839\\	73.8590750336796\\	73.8590750336796\\	15.7079632679489\\	92.2653602388143\\	310.555069083416\\	30.5205263309716\\	28.1917173352551\\	84.9491631534459\\	99.5680157550202\\	22.2093370820437\\	88.8439890474663\\	30.5205263309716\\	163.687024775434\\	10.4719755119660\\	11.7075489646906\\	298.327532196929\\	176.169845182627\\	15.7079632679489\\	16.5568903765368\\	397.785750696556\\	96.1362670323170\\	77.6209114365857\\	33.5100559618668\\	15.7079632679489\\	336.192898977082\\	15.7079632679489\\	20.9439510239319\\	295.689208427413\\	92.2653602388143\\	161.565790413638\\	331.119830336901\\	293.649243772534\\	77.6209114365857\\	96.1362670323170\\	20.9439510239319\\	305.875160656355\\	408.649676704648\\	10.4719755119660\\	88.5303452777311\\	299.784094555382\\	16.5568903765368\\	77.6209114365857\\	21.5874986239788\\	20.9439510239319\\	305.780895234661\\
};

\addplot+[draw=black, fill = rouge]
table[row sep=\\,y index=0] {
data\\ 91.5714604416135\\	30.2744821177726\\	67.1226598966820\\	102.244442161064\\	120.047934074804\\	113.071640392519\\	126.491549808931\\	32.5077202750500\\	48.7196611047154\\	47.9822741764229\\	77.1426794647797\\	78.1649222148969\\	46.7257679264703\\	13.5207589811407\\	37.3856012404715\\	32.2671110903146\\	34.8869239066353\\	56.8684942804019\\	118.041952936904\\	87.3061457050163\\	16.8611764551320\\	134.062534760381\\	27.2259587915921\\	85.5539006555793\\	62.0122437735875\\	17.9585755375157\\	78.7057184165781\\	88.5669464236047\\	25.8610472427471\\	45.4767157565709\\	91.4035895374635\\	57.7423456935268\\	69.1488106484751\\	75.4072885436631\\	76.4427614585843\\	120.274021102063\\	70.0255550318679\\	102.488817856026\\	74.9742251288123\\	90.1882082680019\\	74.0862142920875\\	82.9681358856351\\	25.9226861971608\\	92.0182111828157\\	137.669815911343\\	1.81730003532951\\	23.8733672128360\\	79.2800564606913\\	9.46769462592345\\	126.284607636346\\
};

\addplot+[draw=black, pattern={north east lines},pattern color=rouge]
table[row sep=\\,y index=0] {
data\\ 159.903167019104\\	136.229333048260\\	141.461317650171\\	155.830109553297\\	115.304768240098\\	141.371669411541\\	43.1687964004091\\	23.4122430159298\\	130.997645246327\\	169.369729476033\\	146.693567159164\\	131.291059267042\\	78.5398163397448\\	106.709456478457\\	147.379033972345\\	154.782551070661\\	131.778625691826\\	152.173011954564\\	151.843644923507\\	125.766291434949\\	168.717485994130\\	125.663706143592\\	109.145514910928\\	127.295120919839\\	23.4122430159298\\	140.650980557667\\	130.997645246327\\	120.857529659225\\	153.156834985491\\	151.926054068295\\	128.203641385066\\	125.663706143592\\	45.0037419820022\\	120.857529659225\\	122.137865841606\\	142.799254644841\\	37.7359826843487\\	159.045780248216\\	136.976229500680\\	136.229333048260\\  33.5100559618667\\	164.205650138839\\	33.1092362477460\\	157.395879594049\\	145.699261109365\\	144.563814685308\\	42.1879907201141\\	16.5568903765368\\	157.395879594049\\	141.461317650171\\
};

\addplot+[draw=black, fill = rouge]
table[row sep=\\,y index=0] {
data\\ 30.5489312000304\\	60.4406962069015\\	13.9495346097000\\	127.751795230877\\	243.225805403651\\	162.848761736845\\	20.4485803837409\\	17.9379502120434\\	62.3455087556410\\	12.7603323833614\\	22.5851309088685\\	107.390086061045\\	76.5371846135770\\	147.010228136207\\	27.0327763372629\\	59.0002673111933\\	26.9343475359590\\	71.8832307477286\\	407.445645165700\\	17.7041773183226\\	28.1869479730722\\	370.130653914383\\	62.8305431041133\\	16.9563387097395\\	25.0140097585134\\	42.3576255303681\\	56.1103359096896\\	273.718219631635\\	9.01128227365035\\	23.1181292850631\\	69.4499174007124\\	39.0400327013125\\	46.0650900519419\\	140.589872662347\\	5.41422421402507\\	299.899269161789\\	42.9968349849749\\	94.2372814096563\\	56.0691219004271\\	32.6424251048194\\	182.885748706137\\	26.3544828580610\\	15.3205324229545\\	79.7216986193603\\	16.2515108010375\\	12.4061240635519\\	46.5343418065120\\	33.1031220990136\\	29.8216358485268\\	313.996304525123\\
};

\addplot+[draw=black, pattern={north east lines},pattern color=rouge]
table[row sep=\\,y index=0] {
data\\ 47.1238898038469\\	38.1114010350564\\	51.5334378331756\\	30.5205263309716\\	322.573606710081\\	351.965567748051\\	7.40461691752990\\	11.7075489646906\\	316.592946925297\\	248.020573698092\\	78.5398163397448\\	374.398682758408\\	23.4122430159298\\	49.6524725360303\\	53.3863577156416\\	21.5874986239789\\	23.4122430159298\\	338.891818613870\\	312.520040404230\\	234.123526794493\\	351.471474180629\\	392.748048850621\\	28.1917173352551\\	11.7075489646906\\	336.889178959789\\	31.8476956372437\\	35.1112202790635\\	168.087203509413\\	42.2117698036786\\	166.378331762101\\	16.5568903765368\\	35.1112202790635\\	11.7075489646906\\	240.846277757465\\	11.7075489646906\\	310.555069083415\\	11.7075489646906\\	51.5334378331756\\	373.025980916868\\	227.888243365235\\	224.326015095576\\	390.390229263066\\	26.1722779011794\\	168.361650978415\\	35.1112202790635\\	314.688946345388\\	42.1879907201141\\	30.5205263309716\\	234.642175981356\\	53.3863577156416\\
};

\end{axis}
\end{tikzpicture}}
\vspace{-0.2cm}
    \caption{\small Box plots summarizing the distributions of the sets of distances obtained according to the position of the two sources. Filled boxes correspond to results of Gibbs sampler, boxes with hatches to BP results. }
    \label{fig:boite_moustache_2sources}
\end{figure}